\documentclass[sigconf]{acmart}

\AtBeginDocument{%
  }

\copyrightyear{2025}
\acmYear{2025}
\setcopyright{rightsretained}
\acmConference[C\&C '25]{Creativity and Cognition}{June 23--25, 2025}{Virtual, United Kingdom}
\acmBooktitle{Creativity and Cognition (C\&C '25), June 23--25, 2025, Virtual, United Kingdom}\acmDOI{10.1145/3698061.3726907}
\acmISBN{979-8-4007-1289-0/2025/06}
\acmISBN{978-1-4503-XXXX-X/18/06}

\usepackage{acmart-taps}

\usepackage{subfig}
\usepackage{tabularx}
\usepackage{caption}
\usepackage{array}
\usepackage{multirow}
\usepackage{colortbl}

    \newenvironment{myquote}{\begin{quote}\leftskip-14pt\rightskip-14pt}{\end{quote}}
        \newcommand*{\participant}[1]{{\small{\fontfamily{cmss}\selectfont{(#1)}}}}
        \newcommand*{\quoted}[1]{{\small{\fontfamily{cmss}\selectfont{#1}}}}
        \newcommand{\squote}[2]{\begin{myquote}\quoted{#2 \participant{#1}}\end{myquote}}

    \newenvironment{myquote2}{}{}

\begin{document}

\title[Thoughtful, Confused, or Untrustworthy: How Text Presentation Influences\\ Perceptions of AI Writing Tools]{Thoughtful, Confused, or Untrustworthy: How Text Presentation Influences Perceptions of AI Writing Tools}

\author{David Zhou}
\affiliation{%
 \institution{University of Illinois Urbana-Champaign}
 \city{Urbana}
 \state{Illinois}
 \country{USA}}

\author{John R. Gallagher}
\affiliation{%
 \institution{University of Illinois Urbana-Champaign}
 \city{Urbana}
 \state{Illinois}
 \country{USA}}

\author{Sarah Sterman}
\affiliation{%
 \institution{University of Illinois Urbana-Champaign}
 \city{Urbana}
 \state{Illinois}
 \country{USA}}

\renewcommand{\shortauthors}{Zhou et al.}

\begin{abstract}
  AI writing tools have been shown to dramatically change the way people write, yet the effects of AI text presentation are not well understood nor always intentionally designed. Although text presentation in existing large language model interfaces is linked to the speed of the underlying model, text presentation speed can impact perceptions of AI systems, potentially influencing whether AI suggestions are accepted or rejected. In this paper, we analyze the effects of varying text generation speed in creative and professional writing scenarios on an online platform ($n=297$). We find that speed is correlated with perceived humanness and trustworthiness of the AI tool, as well as the perceived quality of the generated text. We discuss its implications on creative and writing processes, along with future steps in the intentional design of AI writing tool interfaces.

\end{abstract}


\begin{teaserfigure}
  \includegraphics[width=\textwidth]{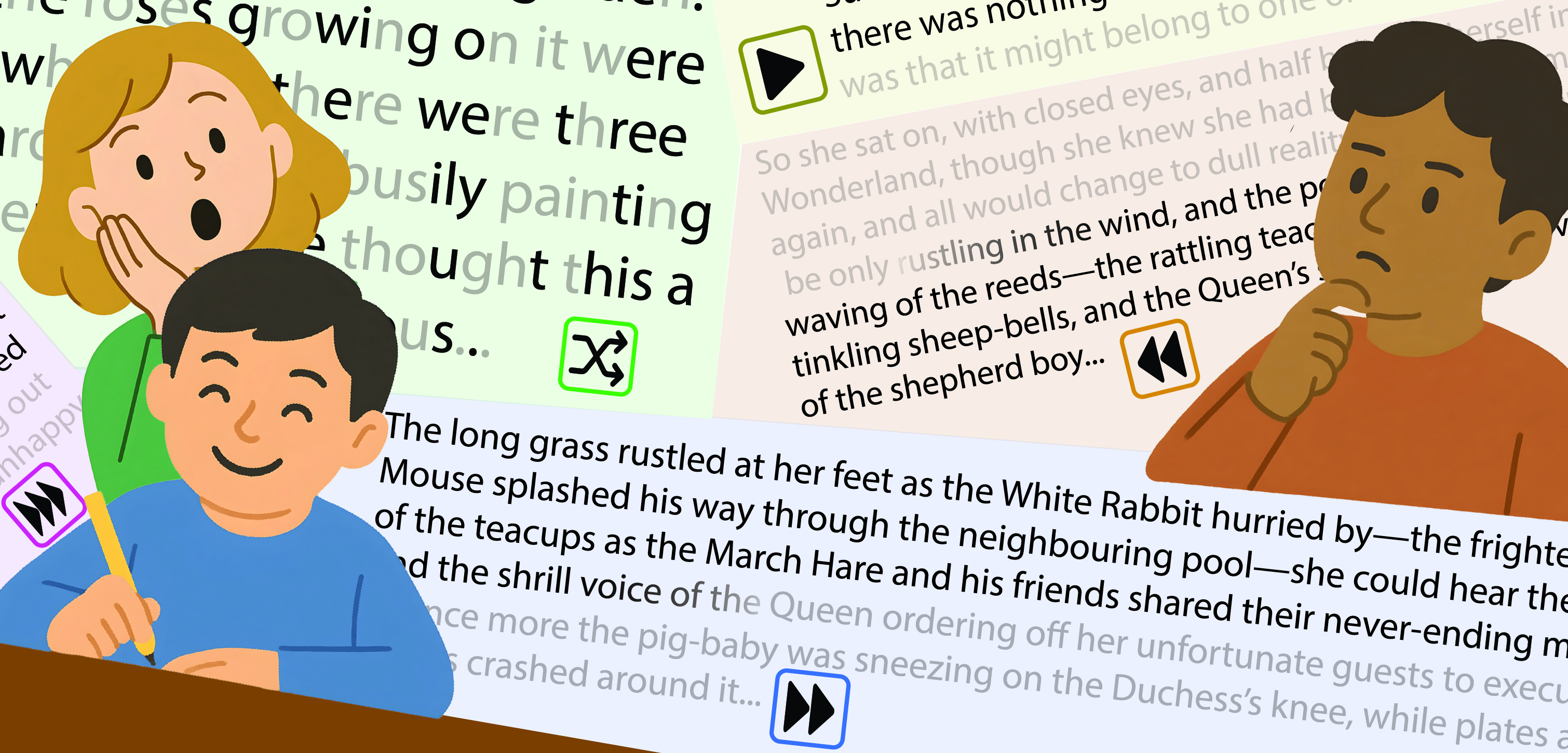}
  \caption{Text presentation style is a key design element of AI writing tools. This paper explores the possible impacts of five text presentation speeds (i.e. streaming speeds) on the perceptions of tools and outputs. See Table \ref{tab:my_label} for a more precise example of each speed. Figure text adapted from Alice’s Adventures in Wonderland (Lewis Carroll). Clip art figures were generated using DALL·E.}
  \Description{Three users are reacting to text. On the left, a smiling user in a blue shirt writes with a pencil while another user in a green shirt appears surprised. To the right, a user in an orange shirt appears to be thinking, with a hand on their chin. The background is filled with excerpts, each in different colored panels, from Alice's Adventures in Wonderland (Lewis Carroll). Several playback buttons are overlaid on the text to describe the speed and order of the text appearance.}
  \label{fig:teaser}
\end{teaserfigure}

\maketitle

\enlargethispage{12pt}
\section{Introduction}

Generative AI tools provide fluent text to humans in many contexts, from creative practice to professional work. The decisions a writer makes about the generated content---whether to use or reject a sentence, believe or disbelieve a claim, or whether to even consider the text generation---depend on their perceptions of the quality and trustworthiness of the text and model \cite{singh22stolen, clark18creative}.

How generated text appears on the screen---whether the AI generation is ``typed'' \cite{ChatGPT, MetaAI, AnthropicClaude, GoogleGemini}, shown as grey in-line text completions \cite{wu18smart}, or appended without animation or distinguishing features \cite{lee22coauthor, zhou24aillude}---has no objective effect on the content. Yet, in many contexts, information speed influences the perceptions of both content and its expressor. For example, in spoken language, speakers with moderate to slightly faster speech rates are perceived as more competent and socially attractive \cite{street83influence, smith75effects}. However, speaking too fast carries an impression of anxiety, impatience, and lack of empathy \cite{schraeder19public, martinuzzi22why, ni19do}. 

While writing processes are highly idiosyncratic \cite{prior13writing}, word processing tools have standardized many parts \cite{kirschenbaum16track}. One metric that captures some of the effects of this digital environment is \textit{typing speed}. Typing speed has a profound impact on users’ perceptions of the writing process, with novices being associated with slow typing speeds and experts with fast typing speeds \cite{sears93investigating}. In synchronous co-writing, inserting and deleting words might be interpreted as struggling with the thought process, even though rewriting is common in creative writing \cite{wang17why}.

In this paper, we study whether similar effects on perception exist in AI text presentation.
For example, text appearance speed is commonly coupled to generation speed to improve perceived instantaneity \cite{agarwal23llm, saumya24how}. Yet text that appears quickly may signal high confidence and efficiency, encouraging users to accept the completion without much scrutiny. Conversely, slower text generation might suggest deliberation, prompting users to become more critical. Another choice is how human-like a tool appears: AI writing tools have been designed to employ human-like elements such as character-by-character text appearance to simulate typing \cite{jakesch23cowriting} and avatars \cite{cahill21supporting} to evoke human-like perception.
AI tools are frequently framed as conversational agents, being deliberately anthropomorphized, thereby leading people to relate to them as collaborators in addition to software tools \cite{khadpe20conceptual, tian24designing}.

While seemingly innocuous, these design choices can produce unintentional effects in how the tool and its output are perceived.
Moreover, \textit{perceptions} of quality and trustworthiness can influence decisions made on the generated content, which impacts how we reflect on words, ideas, and structure. As our relationship to our tools shapes what we think and do \cite{dalsgaard17instruments}, understanding these relations is essential to understanding the effects that generative AI has on our writing and thinking processes.

Using these approaches as a starting point, we investigate how the speed and style of text presentation affect user perceptions of the AI content in two contexts: creative writing and professional correspondence. 
This research is guided by the general question: \textit{How do decisions about the presentation of AI text affect user perception of AI tools and their outputs?} We operationalize this through the following specific questions: 

\begin{enumerate}

\item \textbf{RQ1:} How does the speed and order of AI text completion appearance affect:
\begin{enumerate}
    \item {reading comfort}; 
    \item perceived quality of AI text;
    \item how users perceive anthropomorphic attributes of the system;
    \item users' attitude towards adoption.
\end{enumerate}

\item \textbf{RQ2:} Does the genre of the writing task (creative writing vs professional writing) influence users' perceptions? 
\end{enumerate}

We conducted an online experiment in which participants were instructed to imagine themselves as writers co-writing with AI tools that each present text differently. 
Our quantitative results show that the \textit{medium} text appearance speed---600 wpm, or slightly faster than average reading speed \cite{rayner16so}---is the most comfortable to read (\textbf{RQ1a}) and results in the highest perceived quality of the AI text (\textbf{RQ1b}). We find the \textit{slow} and \textit{medium} speeds to appear more human-like (\textbf{RQ1c}), that \textit{backwards} and \textit{random} appear less trustworthy (\textbf{RQ1c}) and do not find evidence of influence on other attributes (\textbf{RQ1c}) or attitude towards adoption (\textbf{RQ1d}).
We do not find evidence that genre affects preferred speed of the tool (\textbf{RQ2}). 

In our qualitative analysis, we find that users attribute human-like qualities to text presentation variants, read along with text generation and prefer the \textit{medium} speed, and have divisive opinions when considering both genre and text appearance style. Finally, we discuss implications for writing and creative processes, signaling the need to mitigate unintentional manipulations while improving awareness that such perceptual effects exist.


\enlargethispage{12pt}

\section{Related Work}

\subsection{Unintended Effects of Writing Tool Interaction}
    Every user interface consists of design choices that influence how a user relates to a tool. Whether intentional or not, these choices manifest in the user experience, work processes, and attitudes, even going beyond the task itself to affecting mood, self-conceptualization, and community dynamics.

    Early word processors were touted to automate the mechanical aspects of drafting and editing, improving ease-of-use and leading to more revisions on the level of words and sentences compared to writing by hand \cite{collier83word, dave10drafting}. Perhaps an unknown consequence at the time of design, however, was that novice writers tended to make more lexical edits \cite{sommers80revision} while experienced writers tended to make global changes \cite{hill91revising}. The advent of AI writing tools has created a similar circumstance where designing for convenience has created concerns about its potential downstream effects, such as questions related to psychological ownership, control, and agency \cite{lee22coauthor, zhou23creative, biermann22from, zhou24aillude, singh22stolen, yuan22wordcraft, gero19metaphoria}. AI writing tools add, summarize, rewrite, restructure, and suggest new ideas \cite{clark18creative, singh22stolen, skjuve23user, gabriel15inkwell, sudowrite, roemmele18automated}. When we consider the latent influence writing tools exert on writing cognition \cite{dalsgaard17instruments, flower81cognitive}, it is essential to negotiate how these tools affect creativity and composition.

    Past work has already revealed unanticipated impacts. Text completions can unknowingly influence one's personal stance, opinions, and writing \cite{jakesch23cowriting}.
    As creative writing is deeply personal \cite{murray91all}---a process of making meaning through the lens of the self \cite{flower80rhetorical}---it is important to understand how intentional and unexpected effects of common design choices impact written expression, as authenticity \cite{gero23social} and personal strategies \cite{biermann22from} are valued when writing tools are used. 
    
\subsection{Text Appearance Speed}

The available design choices in the field of writing tools are still being explored. Lee et al. describe a design space for intelligent and interactive writing tools \cite{lee2024design}.  In this framework, ``System-Output Type'' describes what the tool produces for the user, categorized into \textit{analysis}, \textit{generation}, or \textit{proposal}.  In this work, we look specifically at a subset of \textit{generation}: \textit{text completion interfaces}, which take pre-existing user text and add direct followups. Examples of this interaction include AI co-writing research systems \cite{lee22coauthor, roemmele15creative, yuan22wordcraft, zhou24aillude} and sentence completions in text editors and email programs.

Text completions in email programs, text editors, and mobile interfaces are displayed fast enough to appear immediate.
In Gmail's Smart Compose, in-line text suggestions are provided in real time, designed to respond within 100ms for the user to not notice any delays \cite{wu18smart}. Large language model interfaces including ChatGPT's Chat interface \cite{ChatGPT}, its API (configurable \cite{ChatGPTAPIStreaming}), and others \cite{MetaAI, AnthropicClaude, GoogleGemini} produce responses token-by-token, which allows partial responses to appear before the entire response is generated. Current LLM performance metrics prioritize speed using metrics such as time to first token, tokens per second, and latency \cite{microsoft23how, nvidia24metrics, liu24andes, dong24large}. While these design choices and metrics optimize for perceived system responsiveness, their agility and immediacy may do little to encourage critical evaluation.

Understanding human reading and writing speeds offers useful context for evaluating text appearance in AI systems. 
Early work in human-computer interaction had investigated the relation between text presentation rate on VDUs (visual display units) and reading comprehension, finding that a medium presentation speed (30 cps) resulted in higher reading comprehension rates than both the slowest (15 cps) and fastest (960 cps) conditions, but not the slightly faster condition (120 cps) \cite{tombaugh85effect}.
Research indicates that ``college-educated adults who are considered good readers'' read between two to four hundred words per minute \cite{rayner16so}, or about 4.44 tokens to 8.89 tokens per second \cite{openai24what}. Typing, on the other hand, is slower, with speeds rarely in excess of 120 wpm \cite{ayres05on} (\(\sim\) 2.67 tokens per second). 
Large language model interfaces produce and show text orders of magnitude faster \cite{conde24speed, llmleaderboard}.
AI-generated text is optimized for speed with little consideration for other effects.

Historically, interface design has targeted approximately one second to support rapid cycles of user-system interaction \cite{card91information, nielsen94usability} due to elementary cognitive operations occurring around that range \cite{newell90unified}.
This indicates the need for UI updates to be timed to a clock rather than as an indirect effect of computation speed \cite{nielsen93response}. For example, animations such as text scrolling between two anchors in a viewport can be too fast for user comprehension if the visual update was tied to computation speed. The problem of too fast text generation has been identified in the task scheduling domain \cite{liu24andes}, but not yet been explored in HCI.

\subsection{Assessing Perceptions of AI Writing Tools}

We suggest that understanding how users perceive AI text appearance (``typing'') is critical for understanding its effects on the creativity and writing process.

Modern AI systems are often perceived as agentic to varying degrees---more than other forms of technology but less than humans \cite{vanneste24artificial}. \textit{Perceived agency}, or having the capacity to think, plan, and act, is classically a necessary component of trust \cite{mayer95integrative, vanneste24artificial}, along with expectations of positive intentions \cite{rousseau98introduction, vanneste24artificial}. As AI systems are designed to operate with some level of autonomy, trust (i.e. ``willingness to rely'') is a relevant factor to consider \cite{rousseau98introduction}. The degree of trust, agency, and respect for an AI system, such as a writing tool, substantially influences a user's expectations for and engagement with the tool, especially when considering that anthropomorphism can influence trust \cite{leong19robot}, even making ``fallible'' information appear reliable \cite{maeda24when}.

Some design decisions seem to be made to evoke agentic interaction through metaphor. For example, the ellipses waiting symbol is the visual language used by texting applications as one waits while a friend types. Conversely, a throbber (``spinning loading icon'') would evoke waiting on a software system. Hwang et al. found that typing indicators led users to interpret delays as related to their texting partner's cognitive processes \cite{hwang19when}. Similarly, writers often made assumptions about the writing processes of others (e.g. inserting and deleting words being interpreted as struggle) when watching others write \cite{wang17why}, and watching writing replays ``humanized'' the author \cite{carrera22watch}. Many current AI tools use a flashing text caret, despite the fact that there is no ``typing'' happening behind the scenes; there is no mind considering what to say. 
We might wonder whether such a design decision changes how the user perceives the AI: as a helpful, trustworthy, respectful or disruptive ``collaborator'' or ``tool''.

Research in other domains provides useful context for understanding the relation between perceptions and text appearance rate. 
Literature in speech perception has shown that speech rate \cite{smith75effects}, tone \cite{wu18smart}, and accent \cite{woolridge24do} influence the listener's perception of the speaker, even though these qualities do not alter the spoken content. Fast speech increases the speaker's perceived competence and attractiveness \cite{street83influence, smith75effects}, while medium speech was attributed with the highest level of benevolence, relative to fast and slow \cite{smith75effects}.

Perceptions of AI writing tools differ when considering personal writing values, tasks, and genre \cite{gero23social}. To understand the effects of genre, including its socially situated norms and conventions of practice, we employ in our study the concept of \textit{genre systems}: intermediate links between institutional structural properties and individual communication \cite{berkenkotter01genre}. Within a genre, participants (e.g. writers) make a recognizable action or movement, which is followed by recognizable responses by others \cite{bazerman94systems, karsten14writing}. Bazerman provides an example: legal writing is more than a grouping of similar documents; it includes the various statements for court rulings,  correspondence, forms, and appeals for laws, letters, etc, each governed by past norms and conventions, and each all governing future texts within the genre \cite{bazerman94systems}. 

In this work, we use the concept of genre systems to align our study to real-world writing situations and to investigate the specific effects of contrasting genres. We choose creative and professional writing as suitable genres. These represent common writing tasks and are commonly studied in the writing tool domain (creative: \cite{biermann22from, clark18creative, gabriel15inkwell, gero19metaphoria, gero23social, ippolito22creative, lee22coauthor, kim23cells, lee2024design, roemmele15creative, roemmele18automated, schmitt21characterchat, singh22stolen, yuan22wordcraft, zhou24aillude}; professional: \cite{gero22sparks, kim23cells, lee2024design, weber23structured, hui18introassist, hui23lettersmith, wanbsganss22modeling}), in which we situate this work.



\section{Study Design}
To measure the influence of text appearance and perceptions, we test five presentation styles representing a range of text appearance speeds, each italicized for clarity: \textit{slow}, \textit{medium}, \textit{fast}, \textit{backwards}, and \textit{random}; and two genres, \textit{creative} and \textit{professional}. See Sec. \ref{sec:independentVariables} for implementation details and Table \ref{tab:my_label} for an example of each style.
These hypotheses were finalized one month before commencing data collection.

To guide our study, we formulate eight hypotheses around \textbf{RQ1} and \textbf{RQ2} and the presentation conditions. These are grounded in a mixture of related prior work in both HCI \cite{card91information, nielsen94usability, khurana24why, vanneste24artificial, lee92trust, saisubramanian21understanding} and writing studies \cite{mavrakakis21writing, robson22reflective}, as well as the design goals and intuition developed from experience in the field.
As the space of literature in specifically AI text presentation is underexplored, we took hypothesis formulation as an open-ended and participatory exploration, inspired by existing problem structuring methods \cite{rosenhead96whats, goel92structure}.%

\subsection{Independent Variables}
\label{sec:independentVariables}

\begin{table*}
    \centering
    \begin{tabular}{|c|c|c|p{8.5cm}|}
    \hline
        \textbf{Speed} & \textbf{WPM} & \textbf{Character Display Order} & \textbf{Example text completion after 0.5 sec}.  \cr
        \hline
        Slow & 160 & forward sequentially & \textcolor{purple}{The quic}\textbf{|}\cr
        \hline
        Medium & 600 & forward sequentially &  \textcolor{purple}{The quick brown fox jumps over}\textbf{|}\cr
        \hline 
        Fast & 6,000 & forward sequentially & \textcolor{purple}{The quick brown fox jumps over the lazy dog.}\textbf{|}\cr
        \hline
        Random & 600 & random insertion &  \textcolor{purple}{h irow}\textbf{|}\textcolor{purple}{fxer hedog}\cr 
        \hline
        Backwards & 600 & backwards sequentially &  \textbf{|}\textcolor{purple}{er the lazy dog}.\cr
        \hline
    \end{tabular}
    \caption{Parameters and example of each text appearance style. ``|'' indicates the cursor position. AI text was presented to the users in red to distinguish from initial text. Note that for the \textit{fast} continuation, the text generation would have finished in 0.07 seconds. For reference, the example string is 44 characters long, including spaces: \textit{"The quick brown fox jumps over the lazy dog."}}
    \Description{This table shows the words per minute (WPM), appearance order, and an example for each speed with the phrase "The quick brown fox jumps over the lazy dog." Slow appears at 160 WPM forward sequentially and the example is "The quic". Medium appears at 600 WPM forward sequentially and the example is "The quick brown fox jumps over". Fast appears at 6,000 WPM forward sequentially and the example shows "The quick brown fox jumps over the lazy dog."; it finishes appearing in 0.07 seconds. Random appears at 600 WPM randomly and the example shows "h irowfxer hedog". Backwards appears at 600 WPM backwards and the example shows "er the lazy dog."}
    \label{tab:my_label}
\end{table*}

The experiment has two independent variables: text presentation styles and genre systems.

\subsubsection{Text Presentation Styles} (categorical, within-subjects). Table \ref{tab:my_label} shows an example of each presentation style. 
    \begin{enumerate}
        \item \textbf{Fast:} Characters are displayed, in order, with 1.67ms delay between insertions; 6,000 words per minute (\(\sim\)133 tokens per second \cite{openai24what}). ``Fast'' approximates potential real-world LLM generation speed, e.g. GPT-4o mini (133.1 tokens per second \cite{llmleaderboard}).
        \item \textbf{Medium:} Characters are displayed, in order, with 16.67ms delay between insertions, 600 words per minute (\(\sim\)13 tokens per second); slightly faster than average reading speed \cite{rayner16so}.
        \item \textbf{Slow:} Characters are displayed, in order, with 62.5ms latency; approximately 160 words per minute (\(\sim\)3.6 tokens per second), or slightly faster than human typing speed \cite{ayres05on}.
        \item \textbf{Backwards:} Characters are displayed backwards (starting with the final character) with 16.67ms latency, approximately 600 words per minute (\(\sim\)13 tokens per second). ``Backwards'' was designed as a drastically non-anthropomorphic text appearance style, although intermediate generation can still be read.
        \item \textbf{Random:} Characters were displayed, in random order, with 16.67ms latency via \texttt{insertion-sort}; approximately 600 words per minute (\(\sim\)13 tokens per second). ``Random'' was also designed as a drastically non-anthropomorphic text appearance style, where intermediate generation is difficult to read.
    \end{enumerate}
    To select these typing speeds, we ran a series of pilot tests to gauge the experiences of the speeds.  Our initial intuition was to match human typing speeds, having 160 WPM as our ``fast'' value and 40 WPM as our ``slow'' speed.  However, participants felt all of these were extremely slow when waiting to read text, as opposed to typing oneself. 
    Thus, we revised our speed anchors to a broader range of real-world experiences, such as typical LLM text completion speeds during AI co-writing (for "Fast"), and an upper bound of human typing speed (for "Slow"). Older studies on VDUs found that medium presentations (30 cps and 120 cps, corresponding to 300 wpm and 1,200 wpm, assuming that a word is on average five characters long, plus a space) led to the highest comprehension rates \cite{tombaugh85effect}. Since our study does not evaluate \textit{reading comprehension} like Tombaugh et al.'s study \cite{tombaugh85effect}, and acknowledging that users can skim at much higher speeds (2x--4x compared to average reading speeds) \cite{rayner16so}, we set our \textit{medium} presentation to 600 wpm.
    
    ``Backwards'' and ``random'' were designed to be distinctly \textit{non-human} ways of text-entry and reading, and serve to contrast typical text appearance and anthropocentric interaction design. We do not expect ``backwards'' and ``random'' to improve usability---rather the opposite---but included these styles to study the effects of non-anthropocentric design.

    The experience of reading the ``backwards'' presentation is meant to be jarring and disruptive.  Leonardo da Vinci's right-to-left script (or \textit{mirror-writing}) was hypothesized to appear enigmatic and to reduce the readability of his texts to the untrained eye \cite{bambach03leonardo}. Although a key differentiator between our ``backwards'' text appearance and da Vinci's mirror script is that ``backwards'' produces a left-to-right readable version after fully appearing, \textit{the experience of reading the ``backwards` generation echoes the cognitive disruption of reading mirror script, deteriorating readability.} ``Random'' text appearance references the visual shorthand for depicting complex computer data streaming and commands (e.g. the visual cue for ``hacking'' in movies), which is stylized to appear machinic and difficult to read for dramatic effect---distinctly non-human.
    
    \subsubsection{Genre System} (categorical, within-subjects). A genre system consists of the intersecting genres that facilitate a specific kind of work \cite{gross17technical, bazerman94systems, berkenkotter01genre}. 
    For each genre, writers were instructed to imagine that they were writing for a specific purpose: 

    \begin{enumerate}
    \item \textbf{Creative Writing:} Participants were asked to imagine themselves writing a fictional story for a blog for a YA audience. The specific scenarios include writing portions of a story introduction, setting exposition, dialogue between two characters, verse for a chant, and a romance.
    
    \item \textbf{Professional Communication:} Participants were asked to imagine themselves writing for a tourist agency. The specific scenarios include writing portions of a brochure, a voice-over script, a business email, an employee training module, and a project proposal. 

    \end{enumerate}

    The initial writing and completions for each scenario were generated together using GPT-4, revised for the study (e.g. removing references to real entities), and split into initial text (presented in black) and AI completion (presented in red). The full instructions, prompts, scenarios, and survey instruments are provided in the supplemental materials.

\subsection{Dependent Variables}
\label{sec:dependentVariables}
Feelings of comfort, text quality, human-like interaction, and attitude towards adopting the AI tool were measured using eight five-point Likert scale survey questions. 
Each of the eight dependent variables correspond with the eight hypotheses described in the following section, and the groupings correspond to subquestions in \textbf{RQ1}. Questions and statements were informed by previous literature \cite{liu23modeling, novikova18rankME, draxler24ai}, adapted to the topic of our study (i.e. text appearance) and iterated on during pilot testing.

\begin{enumerate}
    \item \textit{Perceptions of overall reading comfort.} Measured by a single item, corresponding to $\mathbf{H_1}$.
    \item \textit{Perceptions of overall text quality.} Measured by a single item, corresponding to $\mathbf{H_2}$.
    \item \textit{Perceptions of human-like interaction.} Measured by three items, corresponding to $\mathbf{H_3}, \mathbf{H_5}, \mathbf{H_6}$.
    \item \textit{Attitude towards adoption.} Measured by three items, corresponding to $\mathbf{H_4}, \mathbf{H_7}, \mathbf{H_8}$.
\end{enumerate}

\subsection{Hypotheses}
\label{sec:hypotheses}

\textbf{Hypothesis 1---Comfort:} The \textit{medium} condition will be considered more comfortable than \textit{slow, fast, backwards, and random}. This will not vary by genre. 
\textit{Rationale.}
Here we assume that comfort is primarily mediated by lack of disruption to reading, and therefore words appearing in the standard order, at the same speed as or slightly faster than the user's reading speed will create the least disruption \cite{card91information, nielsen94usability}.

\textbf{Hypothesis 2---Quality:} There will be no variance between the judged quality of the text. This will not vary by genre. 
\textit{Rationale.}
Since the text content is randomly assigned between presentation conditions, there should be no effect from the actual quality of the content.

\textbf{Hypothesis 3---Human-like.} \textit{Backwards} and \textit{random} appearance will appear least human-like; \textit{fast}, \textit{medium}, and \textit{slow}, will appear most human-like.
\textit{Rationale.} AI technology is typically perceived somewhere between human and inanimate technology \cite{vanneste24artificial, lee92trust}. Since humans do not type by inserting characters backwards or randomly, we expect that the forward presenting styles appear most human-like.

\textbf{Hypothesis 4---Purpose:} In the professional genre, the \textit{fast} condition will be judged more like it was created for this purpose.  In the creative condition, \textit{slow} and \textit{medium} will be judged more like it was created for this purpose. 
\textit{Rationale.}
Here we assume that efficiency is the primary value of the professional case, and therefore a fast presentation will be viewed as most appropriate.  In the creative case, efficiency may take a backseat to values like reflection, uniqueness, or engagement, which may be supported by a slower interaction with the text. 

\textbf{Hypothesis 5---Trustworthiness:} \textit{Backwards} and \textit{random} presentation will be perceived as less trustworthy compared to text presentations that display text in order. Feelings of trust will not vary by genre.
\textit{Rationale.} Khurana et al. found that trust in LLM interfaces can depend on the presentation of its output, specifically by breaking down instructions step-by-step \cite{khurana24why}. Since the \textit{random} or \textit{backwards} outputs will seem jarring and cannot be read until it fully appears, we expect a decrease in trust for those presentation styles \cite{saisubramanian21understanding}. 

\textbf{Hypothesis 6---Respect:} For professional writing, users will find the fast continuation more respectful; for creative writing, users will find the slow continuation more respectful. The \textit{backwards} and \textit{random} presentations will not be perceived as respectful compared to the other presentation types. 
\textit{Rationale.} Since writing in a professional setting is often constrained by time, we hypothesize that fast presentation will be perceived as more respectful. Conversely, since creative writing is often felt as reflective and slow \cite{mavrakakis21writing, robson22reflective}, we expect that the slower presentation style will be perceived as more respectful.

\textbf{Hypothesis 7---Liked:} We hypothesize that users will prefer the \textit{fast} and \textit{medium} speed over the rest.
\textit{Rationale.} Following past user interface design practices \cite{card91information, nielsen94usability}, users will prefer a more responsive text appearance over ones that take significantly longer to fully appear because faster speeds will be less disruptive to user flow. 

\textbf{Hypothesis 8---Would Use:} 
Users would use the \textit{fast} speed for professional writing and either prefer 
the \textit{medium} and \textit{slow} speeds or be unwilling to use AI text completions for creative writing in general. 
\textit{Rationale.} We believe that the typical time constraints in professional writing will lead to greater preference for the \textit{fast} speed and the slower speeds will be preferable for creative writing due to the deep reflection that often takes place. However, we expect that a user's general attitude towards generative AI and creative tasks may override whether they would use the AI tool or not, particularly if their views on AI being used for creativity are negative.



\section{Methods}

To study the effects of text presentation speed and order on perceptions of an AI system, we deployed an interactive survey in which participants would imagine themselves as a writer completing a particular task, then trigger an automatic AI completion on a pre-written text. We used fixed texts and generations to remove the variability (and potential confound) of content and generation output.
Our data analysis plan and hypotheses were finalized one month prior to data collection. After data collection and analysis, we ran additional nonparametric tests to confirm our original findings.

\subsection{Experiment Flow}
\label{sec:procedure}
\begin{figure*}
    \centering
    \includegraphics[width=0.9\linewidth]{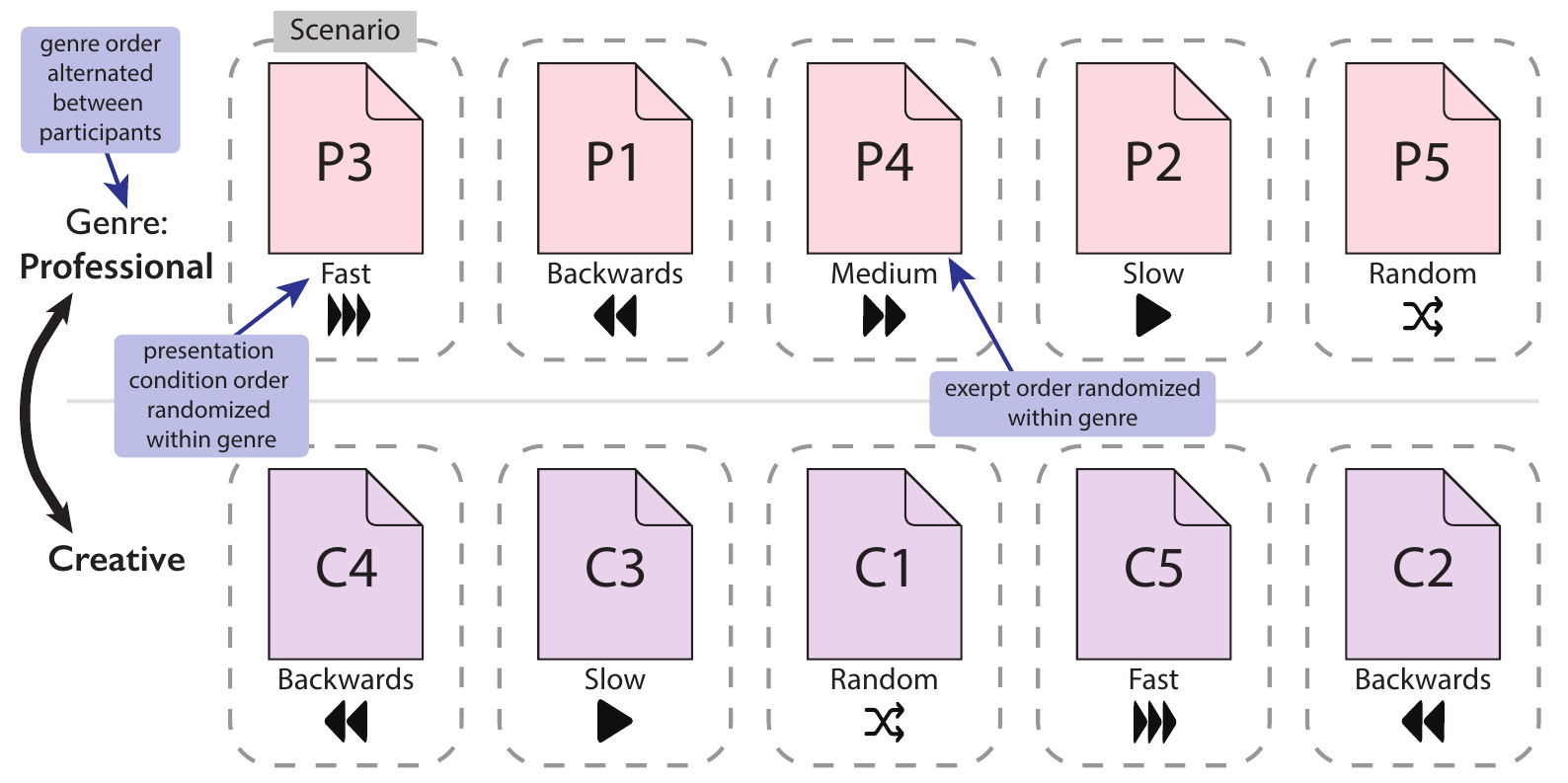}
    \caption{This figure represents one example study order. For each participant, all marked conditions were randomized. Participants viewed ten scenarios, five in a row for one genre (in this case, beginning with \textit{professional}), and the remaining five in a row for the second genre (here, \textit{creative}). The genre order alternated between participants, and within each genre, the sequence of each scenario was randomized for each participant. Each of the five speeds was shown twice (once for each genre), and the sequence of speeds was randomized as well. For each scenario, the questionnaire appears after the AI completion fully appears.}
    \label{fig:study-setup}
    \Description{This figure shows a hypothetical sequence of scenarios. There are five boxes denoting professional scenarios on top, and five boxes denoting creative scenarios on the bottom. Each scenario contains text and a speed, which are represented in smaller boxes inside each scenario. In the figure, the order of scenarios is P3 Fast (P denotes professional), P1 Backwards, P4 Medium, P2 Slow, P5 Random, C4 Backwards (C denotes creative), C3 Slow, C1 Random, C5 Fast, C2 Backwards. The genre order alternates between participants and the sequence of scenarios within each genre and speeds are randomized.}
\end{figure*}

The study consisted of three main steps. Figure \ref{fig:study-setup} shows a possible participation flow of the study, demonstrating the grouping by genre and categories of randomization.

\begin{enumerate}
    \item Participants were presented with two task genres, creative and professional, in counterbalanced order. Within each task,
    participants would view five writing scenarios and imagine themselves as a writer performing that task:
    \begin{enumerate}
        \item Each scenario was assigned one of the five appearance conditions that determined how its AI continuation would be presented. The order of the five scenario texts and five text appearance styles were independently randomized within each genre. Each scenario was shown to a participant exactly once. Each appearance style was shown to a participant exactly once per genre (i.e. twice across the experiment).
        \item Each scenario displayed a writing prompt, a non-editable Quill.js text editor \cite{quill} with initial writing pre-filled to establish the imagined writing context, and a questionnaire that appeared after the AI text was added.
        \item Participants were instructed to first read the prompt, then the initial text (presented in black), and then request an AI continuation by clicking a button.  The new AI text would be ``typed'' in a red color to distinguish it from the initial text. 
        \item After the AI continuation fully appeared, participants completed the questionnaire before continuing to the next scenario. 
    \end{enumerate} 
    \item Participants were asked an open-ended question, instructing them to reflect on each text appearance style and any differences they perceived between genre.
    \item Participants completed an exit questionnaire about past experience with AI tools, attitudes towards AI, prior writing experience, and demographics.
\end{enumerate}

\subsection{Participants}

\begin{figure*}%
\centering%
\subfloat[\centering Desktop interface displaying a creative scenario.]{{\includegraphics[width=0.71\textwidth]{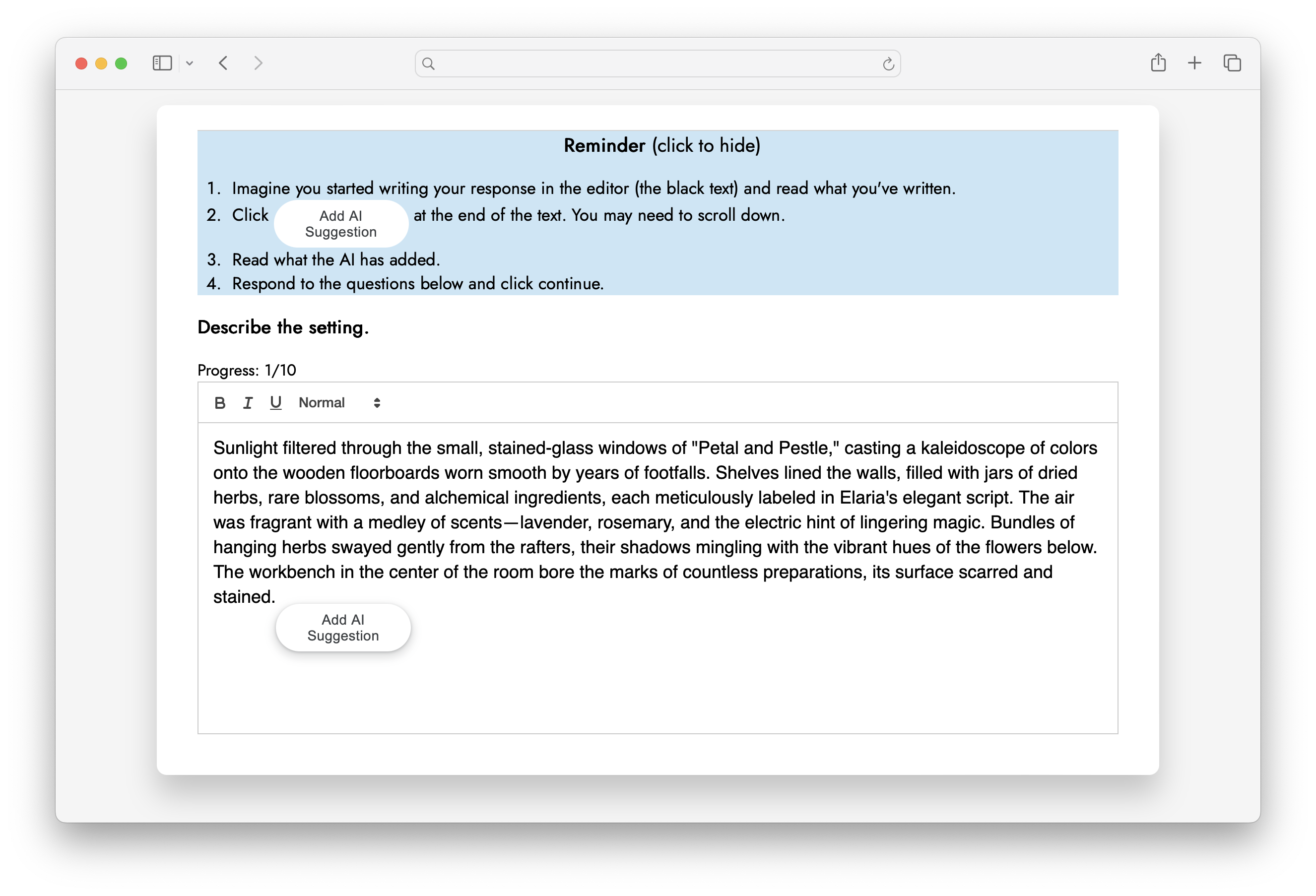}}}%
\quad%
\subfloat[\centering Mobile interface displaying a professional scenario.]{{\includegraphics[width=0.26\textwidth]{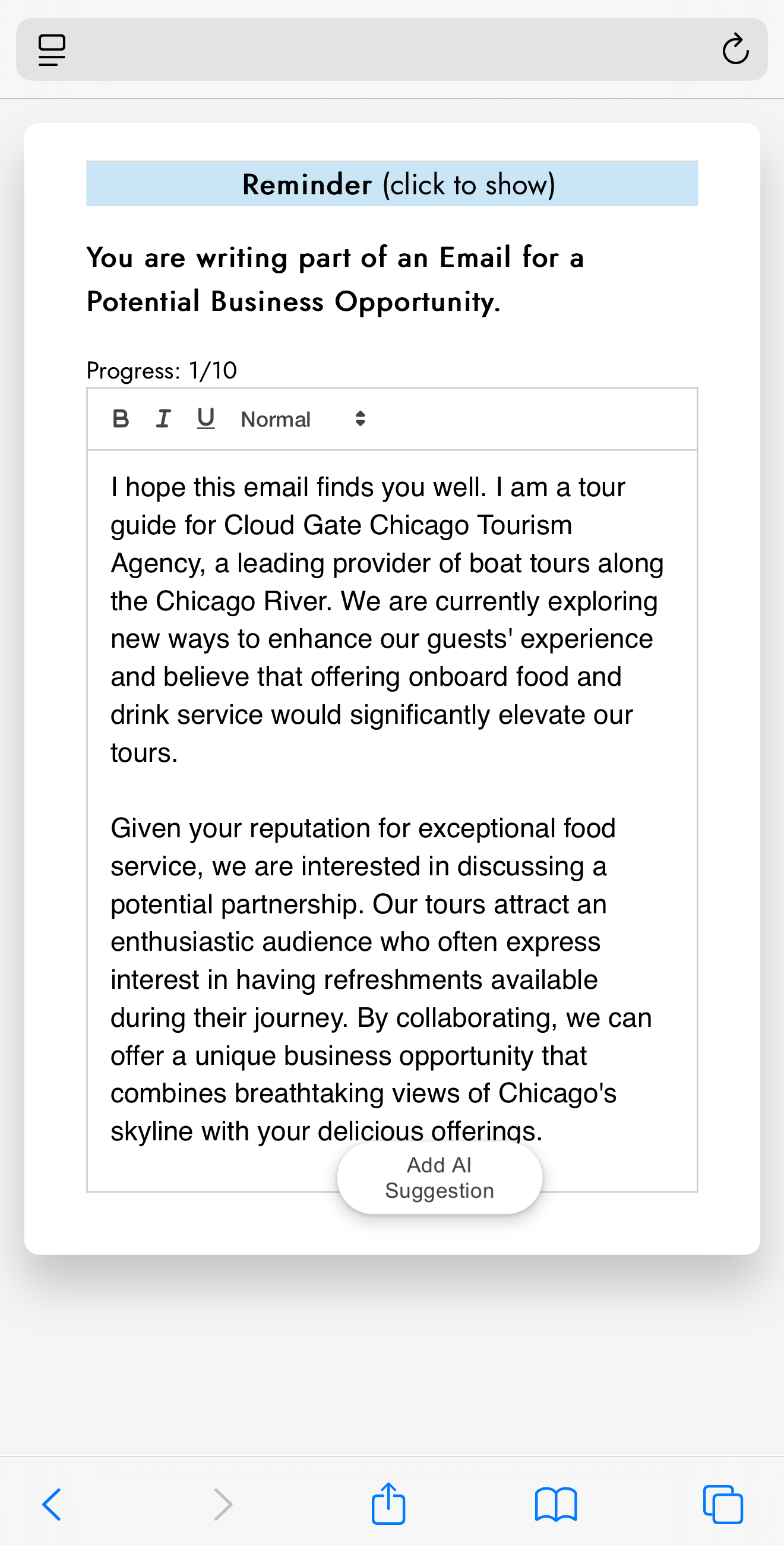}}}%
\\%
\subfloat[\centering Desktop interface displaying the creative scenario with the AI completion shown in red.]{{\includegraphics[width=0.71\textwidth]{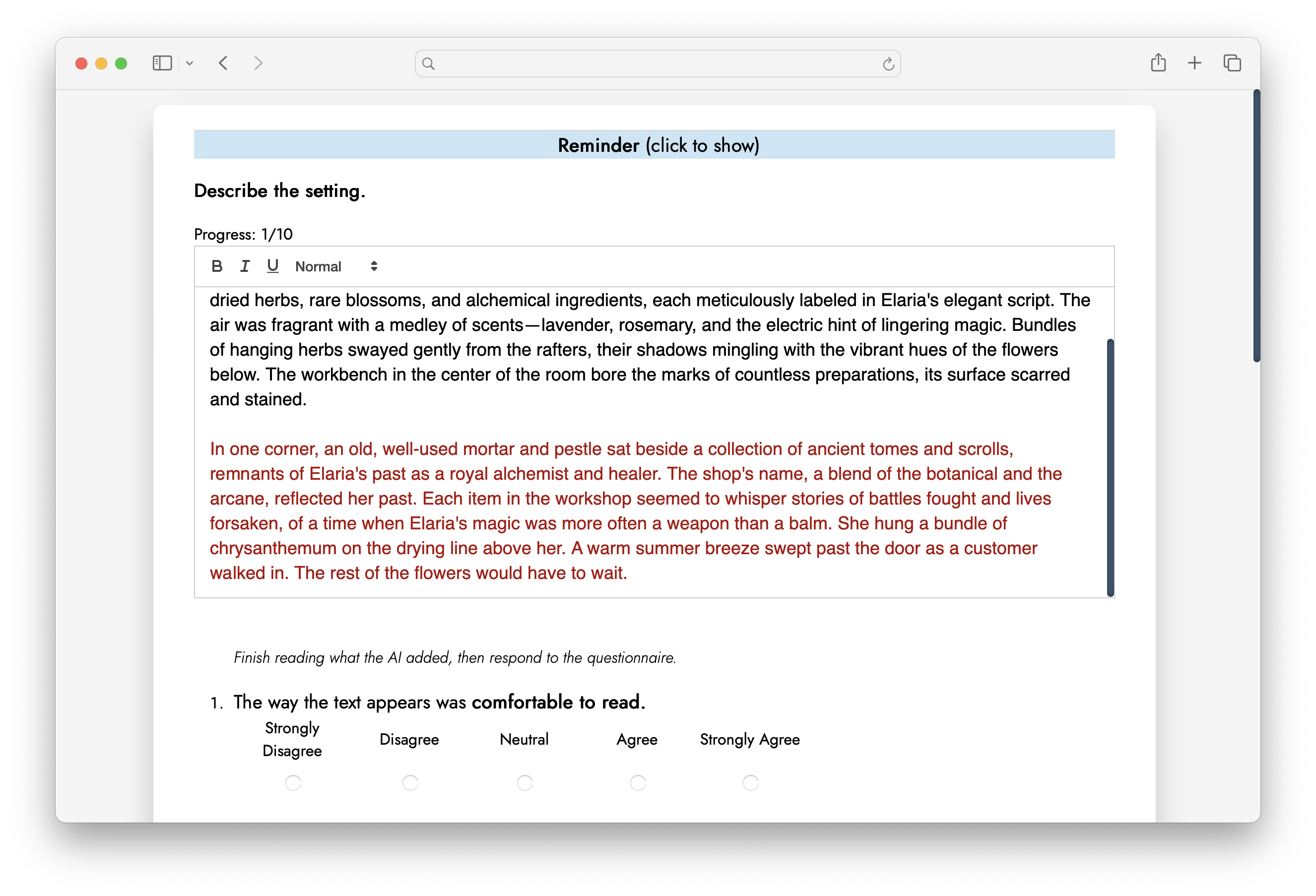}}}%
\quad%
\subfloat[\centering Mobile interface with AI completion, scrolled to show the survey questions.]{{\includegraphics[width=0.26\textwidth]{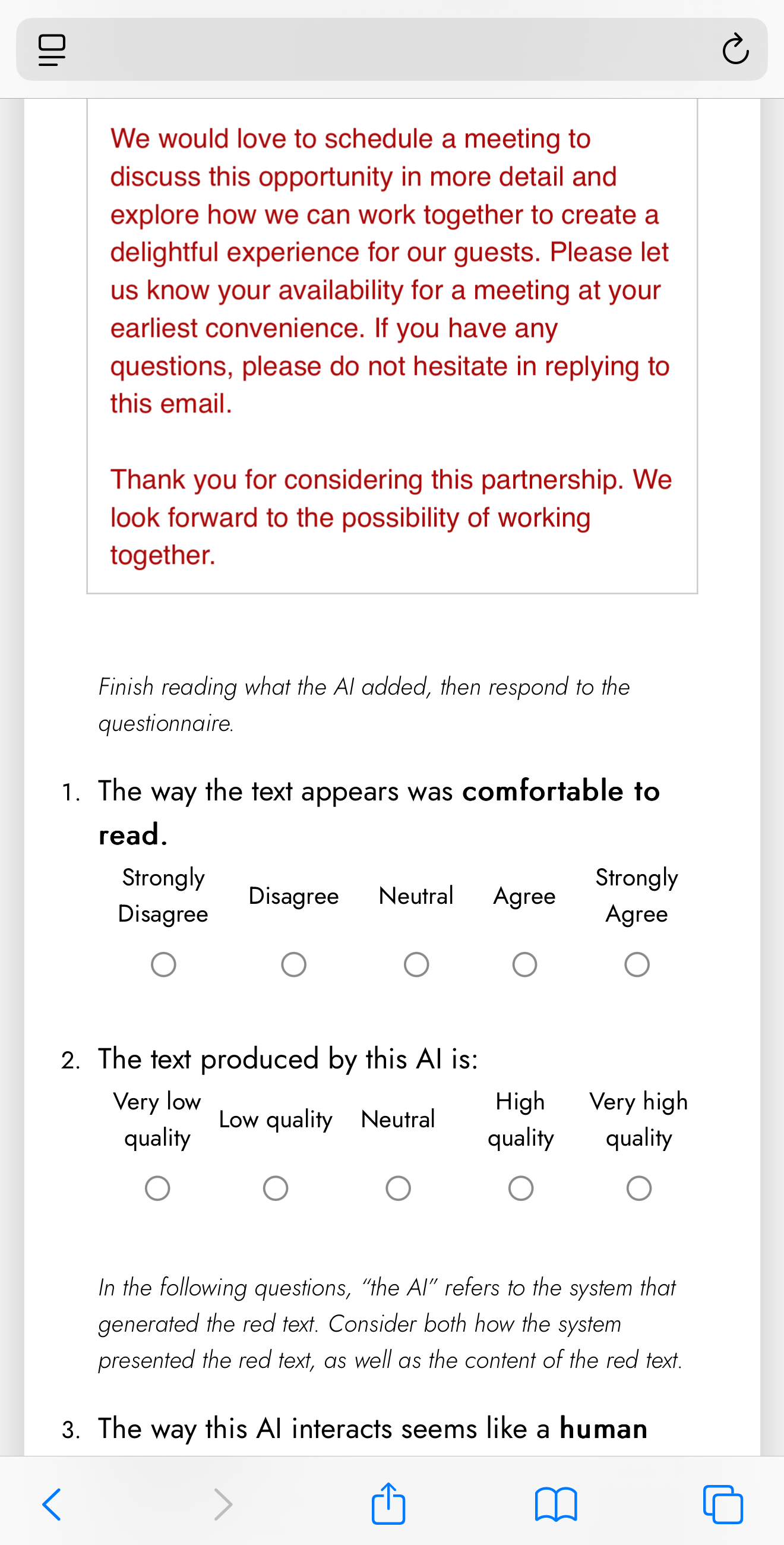}}}%
\caption{Survey interface. The viewable text area dynamically resizes according to the width of the display. On mobile, users must scroll down to the end of the text in order to see the ``Add AI Suggestion'' button. AI text displays in red after the user presses ``Add AI Suggestion.'' The survey appears after the AI text fully appears.}

    \Description{The upper two images show the survey interface on (left) a desktop browser and (right) a mobile browser. In the image with the desktop browser, there is instructional text titled "Reminder," which describes the series of steps a participant must complete to continue to the next scenario. The image with the mobile browser hides these instructions. Both images show a text editor with initial text presented in black. Both interfaces show the "Add AI Suggestion" button. The Reminder window shows regardless of desktop and mobile on the 1st scenario. The bottom two images show the interface after a participant presses "Add AI Suggestion." The AI text is appended to the editor and is presented in red. Portions of the survey are shown in both images.}
    \label{fig:typing-speed-interface}
\end{figure*}

We recruited 301 participants\footnote{Using \textit{G*Power 3} \cite{gpower3}, we computed the required sample size for ANOVA with main effects and interactions, resulting in a sample size of 297. We used the default effect size 0.25 (following Yatani \cite{yatani16effect}, suggesting Cohen's interpretation of a \textit{medium} effect size (0.25), and after initial pilot tests)}, a significance threshold of $\alpha=\frac{0.05}{8}=0.00625$ due to multiple hypotheses testing after Bonferroni correction, and power of 0.8., of whom 231 were from Prolific\footnote{https://www.prolific.com/} and 70 from mailing lists at our institution and word-of-mouth. All participants were at least 18 years old and had high proficiency in English. Prolific participants were compensated at the Prolific-recommended \$12 USD hourly rate (average completion time \textasciitilde 19 minutes); non-Prolific participants were entered into a drawing with four \$15 USD gift cards. After removing participants that did not complete the entire study or provided irrelevant content in the open-ended question, 297  participants remained. 
Participants were permitted to complete the survey with a computer or mobile device (Figure \ref{fig:typing-speed-interface}).
Our study was declared exempt by our institution's IRB. 

The average participant age was 34 years. Fifty-nine percent held at least a Bachelor's degree; 28\% were previously paid to write; 91\% had previously used an AI writing assistant. Across all participants, 87\% took the study on a computer while the remaining 13\% completed it on a mobile device. Forty-nine percent identified as female; 46\% male; 3\% self-described; 1\% chose not to disclose (due to rounding, these ratios add to 99\%). 
We checked for any effect of device type or recruitment location on our hypotheses and were unable to observe any effects.  We therefore use all the data for all subsequent analysis; however, we make no causal claims about the potential effect or lack thereof from modality.

\subsection{Analysis Methods}
\subsubsection{Quantitative Analysis Methods}
\label{sec:quant_method}
Before conducting any statistical analyses, we mapped all (five-point) Likert scale responses onto an ordinal scale ranging from -2 (strongly disagree) to 2 (strongly agree). Normality and homoscedasticity were checked with diagnostic plots. Histograms of response values appeared approximately normal, though with a left skew ($[-0.5, -0.8]$). We generally observed departure along the tails of Q--Q plots. Residuals did not vary between groups. As the skewness values fall between $\pm 1$, we proceed with the analysis \cite{george16ibm}.

We conducted independent multi-way ANOVAs with \textit{text appearance style} and \textit{genre} as the independent variables, along with their interaction term. We report the effect size $\eta^2_p$ next to the $F$ and $p$-value for each hypothesis test. Since we tested eight separate hypotheses, we applied Bonferroni correction, adjusting the significance threshold to $\alpha=\frac{0.05}{8}=0.00625$. Note: Bonferroni adjustment is conservative and may increase Type II (false negative) error. We conducted Tukey's HSD for significant ANOVAs to analyze pairwise differences within each genre and appearance style condition. We controlled for FWER at $\alpha=0.05$. We analyzed by genre, as our research questions are motivated by the potential contrast caused by conventionalized social motives \cite{miller84genre}.

Although Likert data is nonparametric, ANOVA is robust against deviations to non-normality \cite{norman10likert}, and has been used to analyze Likert data in HCI \cite{yurrita23disentangling}. Nonetheless, to support our original parametric analyses, we also performed Align Rank Transform (ART) \cite{wobbrock11aligned} before re-running ANOVAs and then re-performed post-hoc analysis using ART-c \cite{elkin21uist}. We noted an additional significant multiple comparison result for $\mathbf{H_4}$. This deviation did not change the interpretation of our conclusion for $\mathbf{H_4}$ (see Sec. \ref{sec:results_attitude_adoption}). All other results matched the results of the parametric analyses.

\subsubsection{Qualitative Analysis Methods}
We performed reflexive thematic analysis \cite{braun2006TA} on the open-ended response of the exit survey. As perceptions and metaphors of text appearance speed are underexplored, inductive coding allows us to independently build up our understanding of perceptions from the data.

Two authors independently open-coded 50 of the free response answers, after which saturation was achieved. The authors then discussed their codes to share interpretations, determine alignment, and reach consensus. Both coders grouped these responses into clusters. The first author open-coded the remaining 247 responses using the clusters as guidance. These codes and clusters were discussed and revised to identify key themes both related to and outside of the hypotheses, extending our analysis to external factors and other topics.

\subsubsection{Mitigating Potential Content Quality Confound}
In our study, we sought to isolate the effects of text appearance speed from text content. Text appearance speeds and texts were independently randomized. Because individual participants each saw specific speed-text pairs, we derived themes from multiple participants with different speed-text pairs in the qualitative analysis to ensure that individual claims do not arise solely from text content.

We evaluated Flesch-Kincaid scores \cite{kincaid75derivation} for each text as a measure of reading ease and as a proxy for text style consistency. We did not find notable differences in readability scores between texts within either genre. The professional texts scored within the college- and college graduate-level range, which is within expectation for professional writing, while the creative texts scored between the sixth-grade- to ninth-grade- level range, suitable for a YA audience\footnote{As a reference, J. K. Rowling's \textit{Harry Potter} series also scored within the sixth-grade- to ninth-grade- level range \cite{readableflesch}.}. 

We chose a within-subjects study design to mitigate noise from participant background and experience. In our results, we found strong effects from text appearance speed. 
Replication work at scale would be valuable future work, as this is a preliminary study.

\section{Results}
We find sufficient evidence to support hypotheses $\mathbf{H_1}$, $\mathbf{H_3}$, and $\mathbf{H_5}$, and reject $\mathbf{H_2}$, where we initially hypothesized that presentation style would have no effect on perception of quality. Quantitative and qualitative results are grouped by \textbf{RQ1} subquestion.  We use quantitative results to accept or reject hypotheses, and qualitative data to deepen and contextualize our understanding of the effects. 
For clarity, \underline{dependent variables} are underlined, and \textit{independent variables} are italicized. 

\aptLtoX[graphic=no,type=html]{\begin{table*}
\centering
\begin{tabular}{|l|ccc|} 
\hline
\multirow{2}{*}{} & \multicolumn{3}{c|}{\textbf{Genre}}\cr 
\cline{2-4}
 & \textbf{Significance} & \textbf{Professional} & \textbf{Creative}  \cr 
\hline
$\mathbf{H_1}:$ Comfort & $p=0.505$ & $\begin{matrix}Mean = 0.496 \\ SD = 1.330 \end{matrix}$ & $\begin{matrix}Mean = 0.467 \\ SD = 1.293 \end{matrix}$ \cr
\rowcolor[rgb]{0.925,0.925,0.925} $\mathbf{H_2}:$ Quality & $p=0.002$*** & $\begin{matrix}Mean = 0.837 \\ SD = 0.845 \end{matrix}$ \textbf{a} & $\begin{matrix}Mean = 0.741 \\ SD = 0.868 \end{matrix}$ \textbf{b} \cr
$\mathbf{H_3}:$ Human-like & $p=0.358$ & $\begin{matrix}Mean = 0.466 \\ SD = 1.184 \end{matrix}$ & $\begin{matrix}Mean = 0.428 \\ SD = 1.136 \end{matrix}$  \cr
\rowcolor[rgb]{0.925,0.925,0.925} $\mathbf{H_4}:$ Purpose & $p<0.001$*** & $\begin{matrix}Mean = 0.838 \\ SD = 1.007 \end{matrix}$ \textbf{a} & $\begin{matrix}Mean = 0.686 \\ SD = 1.020 \end{matrix}$ \textbf{b} \cr
$\mathbf{H_5}:$ Trustworthiness & $p<0.001$*** & $\begin{matrix}Mean = 0.714 \\ SD = 0.951 \end{matrix}$ \textbf{a} & $\begin{matrix}Mean = 0.587 \\ SD = 0.942 \end{matrix}$ \textbf{b} \cr
\rowcolor[rgb]{0.925,0.925,0.925} $\mathbf{H_6}:$ Respect & $p=0.309$ & $\begin{matrix}Mean = 0.690 \\ SD = 0.921 \end{matrix}$ & $\begin{matrix}Mean = 0.656 \\ SD = 0.939 \end{matrix}$  \cr
$\mathbf{H_7}:$ Liked & $p=0.073$ & $\begin{matrix}Mean = 0.562 \\ SD = 1.044 \end{matrix}$ & $\begin{matrix}Mean = 0.495 \\ SD = 1.064 \end{matrix}$  \cr
\rowcolor[rgb]{0.925,0.925,0.925} $\mathbf{H_8}:$ Would Use & $p<0.001$*** & $\begin{matrix}Mean = 0.548 \\ SD = 1.164 \end{matrix}$ \textbf{a} & $\begin{matrix}Mean = 0.402 \\ SD = 1.222 \end{matrix}$ \textbf{b} \cr
\hline
\end{tabular}
\caption{Hypothesis testing results from our study. Asterisks (***) refer to the presence of a significant effect ($p < 0.00625$). Note significance does not directly imply a confirmed hypothesis. For brevity, the result of all pair-wise comparisons conducted for each significant ANOVA is shown using Compact Letter Display \cite{piepho18letters}, i.e. \textbf{a}, \textbf{b}, \textbf{c}. Means which do not share any letter are significantly different. Means which share a letter are not significantly different from each other. Letters rank variables in descending mean order. }
\end{table*}
\setcounter{table}{1}\begin{table*}
\centering
\begin{tabular}{|l|cccccc|} 
\hline
\multirow{2}{*}{}  & \multicolumn{6}{c|}{\textbf{Text Presentation Style}} \\ 
\cline{2-7}
 & \textbf{Significance} & \textbf{Slow} & \textbf{Medium} & \textbf{Fast} & \textbf{Backwards} & \textbf{Random} \\ 
\hline
$\mathbf{H_1}$  & $p<0.001$*** & $\begin{matrix}Mean = 0.874 \\ SD = 1.022 \end{matrix}$ \textbf{b} & $\begin{matrix}Mean = 1.088 \\ SD = 0.847 \end{matrix}$ \textbf{a} & $\begin{matrix}Mean = 0.848 \\ SD = 0.939 \end{matrix}$ \textbf{b} & $\begin{matrix}Mean = -0.096 \\ SD = 1.427 \end{matrix}$ \textbf{c} & $\begin{matrix}Mean = -0.305 \\ SD = 1.521 \end{matrix}$ \textbf{c} \\
\rowcolor[rgb]{0.925,0.925,0.925} $\mathbf{H_2}$  & $p<0.001$*** & $\begin{matrix}Mean = 0.892 \\ SD = 0.814 \end{matrix}$ \textbf{b} & $\begin{matrix}Mean = 0.956 \\ SD = 0.787 \end{matrix}$ \textbf{a} & $\begin{matrix}Mean = 0.816 \\ SD = 0.866 \end{matrix}$ \textbf{b} & $\begin{matrix}Mean = 0.668 \\ SD = 0.863 \end{matrix}$ \textbf{c} & $\begin{matrix}Mean = 0.611 \\ SD = 0.906 \end{matrix}$ \textbf{c} \\
$\mathbf{H_3}$  & $p<0.001$*** & $\begin{matrix}Mean = 0.754 \\ SD = 0.996 \end{matrix}$ \textbf{a} & $\begin{matrix}Mean = 0.781 \\ SD = 0.949 \end{matrix}$ \textbf{a} & $\begin{matrix}Mean = 0.532 \\ SD = 1.045 \end{matrix}$ \textbf{b} & $\begin{matrix}Mean = 0.162 \\ SD = 1.230 \end{matrix}$ \textbf{c} & $\begin{matrix}Mean = 0.007 \\ SD = 1.324 \end{matrix}$ \textbf{c} \\
\rowcolor[rgb]{0.925,0.925,0.925} $\mathbf{H_4}$  & $p<0.001$*** & $\begin{matrix}Mean = 0.869 \\ SD = 0.965 \end{matrix}$ \textbf{a} & $\begin{matrix}Mean = 0.981 \\ SD = 0.887 \end{matrix}$ \textbf{a} & $\begin{matrix}Mean = 0.854 \\ SD = 0.946 \end{matrix}$ \textbf{a} & $\begin{matrix}Mean = 0.591 \\ SD = 1.072 \end{matrix}$ \textbf{b} & $\begin{matrix}Mean = 0.515 \\ SD = 1.119 \end{matrix}$ \textbf{b} \\
$\mathbf{H_5}$  & $p<0.001$*** & $\begin{matrix}Mean = 0.808 \\ SD = 0.896 \end{matrix}$ \textbf{a} & $\begin{matrix}Mean = 0.845 \\ SD = 0.861 \end{matrix}$ \textbf{a} & $\begin{matrix}Mean = 0.714 \\ SD = 0.897 \end{matrix}$ \textbf{a} & $\begin{matrix}Mean = 0.492 \\ SD = 0.977 \end{matrix}$ \textbf{b} & $\begin{matrix}Mean = 0.392 \\ SD = 1.012 \end{matrix}$ \textbf{b} \\
\rowcolor[rgb]{0.925,0.925,0.925} $\mathbf{H_6}$  & $p<0.001$*** & $\begin{matrix}Mean = 0.776 \\ SD = 0.891 \end{matrix}$ \textbf{a} & $\begin{matrix}Mean = 0.825 \\ SD = 0.841 \end{matrix}$ \textbf{a} & $\begin{matrix}Mean = 0.739 \\ SD = 0.887 \end{matrix}$ \textbf{a} & $\begin{matrix}Mean = 0.571 \\ SD = 0.957 \end{matrix}$ \textbf{b} & $\begin{matrix}Mean = 0.455 \\ SD = 1.012 \end{matrix}$ \textbf{b} \\
$\mathbf{H_7}$  & $p<0.001$*** & $\begin{matrix}Mean = 0.677 \\ SD = 1.009 \end{matrix}$ \textbf{a} & $\begin{matrix}Mean = 0.791 \\ SD = 0.907 \end{matrix}$ \textbf{a} & $\begin{matrix}Mean = 0.742 \\ SD = 0.936 \end{matrix}$ \textbf{a} & $\begin{matrix}Mean = 0.274 \\ SD = 1.084 \end{matrix}$ \textbf{b} & $\begin{matrix}Mean = 0.158 \\ SD = 1.156 \end{matrix}$ \textbf{b} \\
\rowcolor[rgb]{0.925,0.925,0.925} $\mathbf{H_8}$  & $p<0.001$*** & $\begin{matrix}Mean = 0.626 \\ SD = 1.151 \end{matrix}$ \textbf{a} & $\begin{matrix}Mean = 0.763 \\ SD = 1.063 \end{matrix}$ \textbf{a} & $\begin{matrix}Mean = 0.646 \\ SD = 1.108 \end{matrix}$ \textbf{a} & $\begin{matrix}Mean = 0.236 \\ SD = 1.244 \end{matrix}$ \textbf{b} & $\begin{matrix}Mean = 0.104 \\ SD = 1.264 \end{matrix}$ \textbf{b} \\
\hline
\end{tabular}
\caption{\textit{(cont.)}}
\end{table*}}{
\begin{table*}
\centering
\begin{tabular}{|l|ccc|} 
\hline
\multirow{2}{*}{} & \multicolumn{3}{c|}{\textbf{Genre}}\\ 
\cline{2-4}
 & \textbf{Significance} & \textbf{Professional} & \textbf{Creative}  \\ 
\hline
$\mathbf{H_1}:$ Comfort & $p=0.505$ & $\begin{matrix}Mean = 0.496 \\ SD = 1.330 \end{matrix}$ & $\begin{matrix}Mean = 0.467 \\ SD = 1.293 \end{matrix}$ \\
\rowcolor[rgb]{0.925,0.925,0.925} $\mathbf{H_2}:$ Quality & $p=0.002$*** & $\begin{matrix}Mean = 0.837 \\ SD = 0.845 \end{matrix}$ \textbf{a} & $\begin{matrix}Mean = 0.741 \\ SD = 0.868 \end{matrix}$ \textbf{b} \\
$\mathbf{H_3}:$ Human-like & $p=0.358$ & $\begin{matrix}Mean = 0.466 \\ SD = 1.184 \end{matrix}$ & $\begin{matrix}Mean = 0.428 \\ SD = 1.136 \end{matrix}$  \\
\rowcolor[rgb]{0.925,0.925,0.925} $\mathbf{H_4}:$ Purpose & $p<0.001$*** & $\begin{matrix}Mean = 0.838 \\ SD = 1.007 \end{matrix}$ \textbf{a} & $\begin{matrix}Mean = 0.686 \\ SD = 1.020 \end{matrix}$ \textbf{b} \\
$\mathbf{H_5}:$ Trustworthiness & $p<0.001$*** & $\begin{matrix}Mean = 0.714 \\ SD = 0.951 \end{matrix}$ \textbf{a} & $\begin{matrix}Mean = 0.587 \\ SD = 0.942 \end{matrix}$ \textbf{b} \\
\rowcolor[rgb]{0.925,0.925,0.925} $\mathbf{H_6}:$ Respect & $p=0.309$ & $\begin{matrix}Mean = 0.690 \\ SD = 0.921 \end{matrix}$ & $\begin{matrix}Mean = 0.656 \\ SD = 0.939 \end{matrix}$  \\
$\mathbf{H_7}:$ Liked & $p=0.073$ & $\begin{matrix}Mean = 0.562 \\ SD = 1.044 \end{matrix}$ & $\begin{matrix}Mean = 0.495 \\ SD = 1.064 \end{matrix}$  \\
\rowcolor[rgb]{0.925,0.925,0.925} $\mathbf{H_8}:$ Would Use & $p<0.001$*** & $\begin{matrix}Mean = 0.548 \\ SD = 1.164 \end{matrix}$ \textbf{a} & $\begin{matrix}Mean = 0.402 \\ SD = 1.222 \end{matrix}$ \textbf{b} \\
\hline
\end{tabular}
\caption{Hypothesis testing results from our study. Asterisks (***) refer to the presence of a significant effect ($p < 0.00625$). Note significance does not directly imply a confirmed hypothesis. For brevity, the result of all pair-wise comparisons conducted for each significant ANOVA is shown using Compact Letter Display \cite{piepho18letters}, i.e. \textbf{a}, \textbf{b}, \textbf{c}. Means which do not share any letter are significantly different. Means which share a letter are not significantly different from each other. Letters rank variables in descending mean order.}
\Description{This table shows the results of each ANOVA. *** indicates a significant ANOVA result and bolded letters are rankings from a to c indicating relative user score (using CLD). a is ranked highest; b is second highest; c is lowest. Text presentation style was significant for H1 and Medium is ranked a, Slow and Fast are ranked b, and Backwards and Random are ranked c. Genre was significant for H2 and professional was ranked a and creative was ranked b; text presentation style was significant for H2 and Medium is ranked a, Slow and Fast are ranked b, and Backwards and Random are ranked c. Text presentation style was significant for H3 and Medium and Slow are ranked a, Fast is ranked b, and Backwards and Random are ranked c. Genre was significant for H4 and professional was ranked a and creative was ranked b; text presentation style was significant for H4 and Slow, Medium, Fast are ranked a, and Backwards and Random are ranked b. Genre was significant for H5 and professional was ranked a and creative was ranked b; text presentation style was significant for H5 and Slow, Medium, Fast are ranked a, and Backwards and Random are ranked b. Text presentation style was significant for H6 and Slow, Medium, Fast are ranked a, and Backwards and Random are ranked b. Text presentation style was significant for H7 and Slow, Medium, Fast are ranked a, and Backwards and Random are ranked b. Genre was significant for H8 and professional was ranked a and creative was ranked b; text presentation style was significant for H8 and Slow, Medium, Fast are ranked a, and Backwards and Random are ranked b.}
\addtocounter{table}{-1}
\centering
\begin{tabular}{|l|cccccc|} 
\hline
\multirow{2}{*}{}  & \multicolumn{6}{c|}{\textbf{Text Presentation Style}} \\ 
\cline{2-7}
 & \textbf{Significance} & \textbf{Slow} & \textbf{Medium} & \textbf{Fast} & \textbf{Backwards} & \textbf{Random} \\ 
\hline
$\mathbf{H_1}$  & $p<0.001$*** & $\begin{matrix}Mean = 0.874 \\ SD = 1.022 \end{matrix}$ \textbf{b} & $\begin{matrix}Mean = 1.088 \\ SD = 0.847 \end{matrix}$ \textbf{a} & $\begin{matrix}Mean = 0.848 \\ SD = 0.939 \end{matrix}$ \textbf{b} & $\begin{matrix}Mean = -0.096 \\ SD = 1.427 \end{matrix}$ \textbf{c} & $\begin{matrix}Mean = -0.305 \\ SD = 1.521 \end{matrix}$ \textbf{c} \\
\rowcolor[rgb]{0.925,0.925,0.925} $\mathbf{H_2}$  & $p<0.001$*** & $\begin{matrix}Mean = 0.892 \\ SD = 0.814 \end{matrix}$ \textbf{b} & $\begin{matrix}Mean = 0.956 \\ SD = 0.787 \end{matrix}$ \textbf{a} & $\begin{matrix}Mean = 0.816 \\ SD = 0.866 \end{matrix}$ \textbf{b} & $\begin{matrix}Mean = 0.668 \\ SD = 0.863 \end{matrix}$ \textbf{c} & $\begin{matrix}Mean = 0.611 \\ SD = 0.906 \end{matrix}$ \textbf{c} \\
$\mathbf{H_3}$  & $p<0.001$*** & $\begin{matrix}Mean = 0.754 \\ SD = 0.996 \end{matrix}$ \textbf{a} & $\begin{matrix}Mean = 0.781 \\ SD = 0.949 \end{matrix}$ \textbf{a} & $\begin{matrix}Mean = 0.532 \\ SD = 1.045 \end{matrix}$ \textbf{b} & $\begin{matrix}Mean = 0.162 \\ SD = 1.230 \end{matrix}$ \textbf{c} & $\begin{matrix}Mean = 0.007 \\ SD = 1.324 \end{matrix}$ \textbf{c} \\
\rowcolor[rgb]{0.925,0.925,0.925} $\mathbf{H_4}$  & $p<0.001$*** & $\begin{matrix}Mean = 0.869 \\ SD = 0.965 \end{matrix}$ \textbf{a} & $\begin{matrix}Mean = 0.981 \\ SD = 0.887 \end{matrix}$ \textbf{a} & $\begin{matrix}Mean = 0.854 \\ SD = 0.946 \end{matrix}$ \textbf{a} & $\begin{matrix}Mean = 0.591 \\ SD = 1.072 \end{matrix}$ \textbf{b} & $\begin{matrix}Mean = 0.515 \\ SD = 1.119 \end{matrix}$ \textbf{b} \\
$\mathbf{H_5}$  & $p<0.001$*** & $\begin{matrix}Mean = 0.808 \\ SD = 0.896 \end{matrix}$ \textbf{a} & $\begin{matrix}Mean = 0.845 \\ SD = 0.861 \end{matrix}$ \textbf{a} & $\begin{matrix}Mean = 0.714 \\ SD = 0.897 \end{matrix}$ \textbf{a} & $\begin{matrix}Mean = 0.492 \\ SD = 0.977 \end{matrix}$ \textbf{b} & $\begin{matrix}Mean = 0.392 \\ SD = 1.012 \end{matrix}$ \textbf{b} \\
\rowcolor[rgb]{0.925,0.925,0.925} $\mathbf{H_6}$  & $p<0.001$*** & $\begin{matrix}Mean = 0.776 \\ SD = 0.891 \end{matrix}$ \textbf{a} & $\begin{matrix}Mean = 0.825 \\ SD = 0.841 \end{matrix}$ \textbf{a} & $\begin{matrix}Mean = 0.739 \\ SD = 0.887 \end{matrix}$ \textbf{a} & $\begin{matrix}Mean = 0.571 \\ SD = 0.957 \end{matrix}$ \textbf{b} & $\begin{matrix}Mean = 0.455 \\ SD = 1.012 \end{matrix}$ \textbf{b} \\
$\mathbf{H_7}$  & $p<0.001$*** & $\begin{matrix}Mean = 0.677 \\ SD = 1.009 \end{matrix}$ \textbf{a} & $\begin{matrix}Mean = 0.791 \\ SD = 0.907 \end{matrix}$ \textbf{a} & $\begin{matrix}Mean = 0.742 \\ SD = 0.936 \end{matrix}$ \textbf{a} & $\begin{matrix}Mean = 0.274 \\ SD = 1.084 \end{matrix}$ \textbf{b} & $\begin{matrix}Mean = 0.158 \\ SD = 1.156 \end{matrix}$ \textbf{b} \\
\rowcolor[rgb]{0.925,0.925,0.925} $\mathbf{H_8}$  & $p<0.001$*** & $\begin{matrix}Mean = 0.626 \\ SD = 1.151 \end{matrix}$ \textbf{a} & $\begin{matrix}Mean = 0.763 \\ SD = 1.063 \end{matrix}$ \textbf{a} & $\begin{matrix}Mean = 0.646 \\ SD = 1.108 \end{matrix}$ \textbf{a} & $\begin{matrix}Mean = 0.236 \\ SD = 1.244 \end{matrix}$ \textbf{b} & $\begin{matrix}Mean = 0.104 \\ SD = 1.264 \end{matrix}$ \textbf{b} \\
\hline
\end{tabular}
\caption{\textit{(cont.)}}
\end{table*}
\label{tab:quantresults}}

\subsection{Reading Comfort}
We found that text appearance speed influences reading comfort.
\textit{Medium} was the most \underline{comfortable to read}. We found a main effect in \textit{presentation style} ($\mathbf{H_1}: F_{4,2965}=170.61, p<0.001, \eta^2_p=0.19$). Post-hoc analysis revealed that the most comfortable speed was \textit{medium}, followed by \textit{fast} and \textit{slow}, then \textit{backwards} and \textit{random}. 

To contextualize the connection between \textit{appearance speed} and \underline{comfort}, participants described how they followed along with the generated text. The most frequent behavior was \textbf{reading along with the generation} where people preferred the speeds that \textbf{matched their reading speed} and let them ``keep up'' with the AI:

\squote {P186}  {I prefer the slow and medium AI tools, mostly because I could read along as it appeared.}

\squote {P274}  {I like the slow one the best. It was easy to read and keep up with. It flowed the smoothest.}

Yet \textit{which} option most closely matched their reading speed varied between participants:

\squote {P5}{The medium version felt best when I was following along with the text in real time, since it most closely matched my reading speed.}

Matching comprehension speed was important for usability, echoing past insights suggesting that coupling computation and display speed can be overwhelming \cite{nielsen93response}. Similar findings for optimizing reading comprehension and comfort have been reported from research in closed captioning speed \cite{szarkowska18viewers, jensema96closed} and implemented in closed captioning standards \cite{fcc14closed, bbc24subtitle}.

For faster readers, \textbf{maintaining attention} was a problem for presentations that were too slow:

\squote{P273}   {I prefer medium or fast --- I am a fast reader so I lose focus when the pace is slow and have to reread to make sure I remember everything about the passage.}

Slower readers had the inverse problem, where a fast presentation felt \textbf{overwhelming}:

\squote{P22}{[Medium] was just fast enough to keep up with my reading while also allow myself to slowly take in what was being written without too much information all at once with fast.}

And some participants expressed experiencing both overwhelm and difficulty with attention:

\squote{P186} {I preferred the medium version. The fast version was very overwhelming, and if it was too slow it was hard to pay attention. The backwards version was very distracting.}

Minor mismatches were considered \textbf{annoying}:

\squote{P155}  {I strongly disliked the very slow version as it was slower than I read and thus highly annoying.}

However, when presented with the non-standard order conditions \textit{backwards} and \textit{random}, participants who tried to read along were far more than annoyed---they found it \textbf{jarring} and \textbf{frustrating}:

\squote{P84}{I liked the medium or fast version because the slow took to long and backwards and random were jarring.}

\squote{P94}{Slow, medium, fast were the ones I will prefer over backwards and random as it was irritating to read and I had to scroll through the content to read it from start.}

Some participants expressed physical discomfort with the backwards and random presentations: 

\squote{P209}{I really didn't like the text that came too fast, random, and backwards; it felt painful on the eyes to read.}

\squote{P226}{The backwards type makes it hard to read. I hated the rando type, it hurt my eyes.}

Understanding that participants read along with the AI generation helps explain both the preference for forward presentations and the \textit{medium} speed.
Similar to reading speed considerations in closed captioning \cite{szarkowska18viewers, jensema96closed, fcc14closed, bbc24subtitle}, our findings suggest that text appearance speed influences comfort and fluency in keeping up with the text. Mismatches between text appearance and reading speed were frustrating and uncomfortable for users.

\subsection{Quality of the AI Text}
\textit{Medium} evoked the highest \underline{perceived AI text quality}. While we did not expect perceptions of \underline{AI text quality} to be influenced by either independent variable, we found small but significant effects corresponding to \textit{presentation style} ($\mathbf{H_2}: F_{4,2965}=17.70, p<0.001, \eta^2_p=0.02$) and \textit{genre} ($F_{1,2968}=9.59, p<0.001, \eta^2_p<0.01$). Post-hoc analysis revealed that the speed leading to the highest perceived quality was \textit{medium}, followed by \textit{fast} and \textit{slow}, then \textit{backwards} and \textit{random}.

Before conducting the qualitative analysis, we confirmed statistically there was no relation between participant-reported quality judgments and text scenario. 
Qualitatively, \textbf{quality} judgments were attributed to text appearance speed.  
The qualitative responses go beyond the single quality rating given for each scenario, for instance attributing beliefs about \textbf{accuracy}, \textbf{detail}, and \textbf{depth}:

\squote{P39} {the slow forward one is fun because it gives you an illusion that the computer is using more power to generate a response leading to an assumption of an accurate response}

\squote{P202} {The slower versions were too detailed, while the faster versions often lacked depth.}

\squote{P220}{The slow tool was useful for detailed exploration but sometimes felt too laborious for quicker needs. The fast tool was great for rapid brainstorming but sometimes sacrificed nuance. }

Each of these participants saw different sets of excerpts for these speeds. The constant factor here is not literal ``detail'' or ``depth'' of the text, but rather the presentation speed, reflecting the statistical analysis---the overall perception of ``medium''  as higher quality.

Participants also felt the random speed to be more creative than the other speeds:

\squote{P174}{For creative writing, I found the random version intriguing, as it added unexpected twists that sparked new ideas, though it sometimes disrupted the flow.}

\squote{P106} {the random had a lot more creativity to it.}

\squote{P216} {for very specific creative writing projects, the "random" setting could be interesting and inspiring.}

Again, these participants saw a different creative scenario for the random speed (P174 saw \textit{the beginning of the story}; P106 saw \textit{the romance}; P216 saw \textit{the setting}), yet attributed ``unexpected twists'' or having ``more creativity'' to the content.

Overall, we see the perceptual influences of text appearance speed modify judgments of textual quality. Text quality influences how text generations are assessed and used, for example influencing the extent of their revision \cite{roemmele18automated, zhou24aillude} and the overall treatment, evaluation, and usage of AI text \cite{ippolito22creative}. We suggest that \textit{perceptions of quality} similarly influence writing judgments and process.

\subsection{Anthropomorphic Attributes of the System}

Three survey questions contributed to this analysis. 
We found sufficient evidence to show that \textit{slow} and \textit{medium} were the most \textbf{human-like}, which was also reflected in the qualitative analysis.

\textit{Slow} and \textit{medium} were perceived as most \underline{human-like}. Here, we found a main effect in \textit{presentation style} only ($\mathbf{H_3}: F_{4,2965}=57.97, p<0.001, \eta^2_p=0.07$). Post-hoc analysis revealed that the \textit{slow} speed ($p=0.006$) and \textit{medium} speed ($p=0.001$) were seen as more human-like than the \textit{fast} speed, while \textit{fast} appeared more human-like than both \textit{backwards} ($p<0.001$) and \textit{random} ($p<0.001$). No significant difference was found between \textit{medium} and \textit{slow} ($p=0.999$), or \textit{backwards} and \textit{random} ($p=0.119$). We found no substantial effect from \textit{genre system}.

The qualitative results contextualize these findings and show additional attributions of human-like behavior. 
\textbf{Thoughtfulness} was often attributed to the \textit{slow} and \textit{medium} presentations: 

\squote{P202} {I found the medium version of the AI tools to be the most effective. It balanced speed with thoughtful content generation}

\squote{P220} {[Medium] balanced efficiency with depth, allowing for thoughtful responses without overwhelming speed.} 

Here, ``balance'' evoked an interpretation of thoughtfulness; an example of metaphor as hermeneutic \cite{sheehan99metaphor}. The effects of believing anthropomorphism have been linked to greater perceptions of accuracy and reduced risk when a user receives information from an LLM \cite{cohn24believing}. Thoughtfulness is a distinctively human trait, which we see connected to text appearance speed.

Speed also affected participants' perceptions of \textbf{humanness} and \textbf{collaboration}: 

\squote{P94} {[Slow] seemed more human and more collaborative than other versions. I liked the fast and backwards tool the least, as it seemed robotic.}

\squote{P3} {The non-intuitive ways of generating text (especially random, and to a lesser extent, backwards) are difficult and uncomfortable to read, and do not seem to be a natural human-collaborator-like way to write.}

P94 contrasted human and robotic based on presentation, and both participants linked ``human'' and ``collaborative''. 
Collaboration was connected with the feeling of watching someone typing in real time: 

\squote{P110} {I liked the AI tools that outputted the writing assignment slowly word by word the way a human would type it.}

\squote{P115} {The slow felt the most like it was a collaborative writing as it looked as though someone was typing the words out.}

Research has shown that people apply social expectations to computers \cite{nass00machines, turkle84second}, even though computers are by nature asocial machines. AI writing tools produce words, are interactive, and fulfill a function traditionally performed by humans. The combination of these factors form sufficient bases for individuals to cue ``humanness'' \cite{nass00machines}.

The forward presentations were perceived as more \underline{trustworthy}. Here we found that both \textit{genre} ($\mathbf{H_5}: F_{1,2968}=13.89, p<0.001, \eta^2_p<0.01$) and \textit{presentation style} ($\mathbf{H_5}: F_{4,2965}=27.21, p<0.001, \eta^2_p=0.03$) individually had significant effects. The post-hoc analysis revealed that \textit{slow}, \textit{medium}, and \textit{fast} were perceived as more trustworthy over \textit{random} and \textit{backwards} (all $p<0.001$). 

We found comments on trustworthiness to support the quantitative findings. For instance, P56 described \textit{backwards} and \textit{random} as seemingly \textbf{erroring}:
\squote{P56}{The text that showed up backwards or random was very off-putting and it almost even seemed like it was erroring at first.}

One participant grouped \textit{fast} together with \textit{backwards} and \textit{random}:

\squote{P112}{The fast, backwards and random AI did instill a sense of mistrust.}

Another had included \textit{slow}:

\squote{P222}{I hated the slow, random, and backwards because they felt very unnatural in the delivery. The others felt natural and more trusted.}

And, for one participant, even both \textit{fast} and \textit{slow} were regarded as less trustworthy, describing them as \textbf{clunky} and \textbf{robotic}:

\squote{P193}{Medium worked perfect for both the professional and creative scenarios. The others seemed clunky or robotic making it seem untrustworthy.}

Overall, participants’ trust judgments aligned with how natural the text appeared. Anthropomorphic cues---such as forward, human-like presentation---increased apparent trustworthiness, corresponding to research on dishonest anthropomorphism \cite{leong19robot, maeda24when}, while non-forward styles were deemed unusual, and therefore less trustworthy.

We did not find \textit{fast} to be perceived as more \underline{respectful} for \textit{professional}, or \textit{slow} to be perceived as more \underline{respectful} for \textit{creative}. While we found \textit{speed} to be significant ($\mathbf{H_6}: F_{4,2965}=16.90, p<0.001, \eta^2_p=0.02$), we did not find evidence that \textit{fast} was perceived as more \underline{respectful} than \textit{slow} for the professional genre ($p=1.000$) or \textit{slow} perceived as more respectful than \textit{fast} for creative ($p=0.919$). Qualitatively, we did not find evidence of any speed appearing more respectful to the user.

In sum, we see attribution of features to the text to be caused not by content, but by presentation speed: thoughtfulness, quality, creativity, humanness, and collaboration.  The qualitative results here provide evidence that speed is one factor influencing these anthropocentric perceptions.

\subsection{Attitude Towards Adoption}
\label{sec:results_attitude_adoption}
Three questions contributed to an analysis of participants' attitudes towards potentially adopting the AI tools into their practice. 
We did not find statistical evidence supporting our hypothesized effects for particular speeds. The free response answers echo this lack of unity. Instead, participants expressed \textbf{strong speed preferences for both genres}, but varied in which they preferred.

We did not find \textit{fast} to be judged more \underline{appropriate} for \textit{professional}, or \textit{slow} or \textit{medium} to be judged more \underline{appropriate} for \textit{creative}. While we found that both \textit{genre} ($F_{1,2968}=17.40, p<0.001, \eta^2_p<0.01$) and presentation style ($\mathbf{H_4}: F_{4,2965}=23.55, p<0.001, \eta^2_p=0.03$) individually had significant effects, post-hoc analysis did not\footnote{\textit{Medium} was rated higher than \textit{fast} in the ARC-c post-hoc comparisons.} reveal differences between the forward-presenting styles. We did not find evidence that \textit{fast} was judged higher on \underline{purpose} in the professional scenarios over \textit{medium} or \textit{slow}; neither did we find evidence that \textit{slow} and \textit{medium} were judged higher than \textit{fast} in the creative scenarios.

We did not find \textit{fast} or \textit{medium} being \underline{liked} more than \textit{slow}. While we found \textit{presentation style} having a significant effect ($\mathbf{H_7}: F_{4,2965}=40.08, p<0.001, \eta^2_p=0.06$), we did not find that \textit{fast} was \underline{liked} over \textit{slow} ($p=0.803$) or \textit{medium} was liked over \textit{slow} ($p=0.302$).

Lastly, we did not find participants hypothetically \underline{using} \textit{fast} more than others for \textit{professional} or \textit{slow} more than others for \textit{creative}. While we found that both \textit{genre} ($\mathbf{H_8}: F_{1,2968}=11.66, p<0.001, \eta^2_p<0.01$) and \textit{presentation style} ($\mathbf{H_8}: F_{4,2965}=36.01, p<0.001, \eta^2_p=0.05$) had significant effect, we did not find participants preferring the \textit{fast} speed over \textit{medium} ($p=1.000$) over \textit{slow} ($p=0.982$) for professional writing, or preferring \textit{medium} over \textit{fast} ($p=0.739$) or \textit{slow} preferred over \textit{fast} ($p=1.000$).

First, it is important to note that some participants were against using AI in writing at all, regardless of genre. 
Qualitatively, we noted that some participants outright rejected AI for any sort of writing because writing and expression are innately human. Others felt AI was permissible in professional writing, where the writing is generally not as focused on self-reflection and identity expression, and not justified in creative writing. Yet, others still permit using AI for creative writing to generate many ideas, and disallow AI in professional communication as it might convey error.

Opinions were divided over the best presentation types for professional and creative writing. Regardless of preference, the participants' choices were primarily driven by their \textbf{writing values}.  

Participants who preferred faster appearance speeds in the creative writing scenarios felt that \textbf{\textit{fast} offered spontaneity.} 

\squote{P107} {Creative writing thrived with the fast version, allowing for more spontaneity in ideas.}

\squote{P285} {AI tools that offered complex, intricate, and emotionally charged additions were favored for creative writing; the "fast" version proved to be especially useful.}

On the other hand, participants who preferred slower appearance speeds in the creative writing scenarios felt that \textbf{\textit{slow} was the most thoughtful in eliciting creativity.} Many participants viewed creative writing as a slower and reflective practice, desiring intentionality in the creative task:
\squote{P69} {A quick speed could be used in informative writing but the creative writing made it seem at least a little more thoughtful when slowed down.}
\squote{P86} {slow and medium feel the most thought out for creative writing }

The primary value in professional writing for participants was \textbf{efficiency}.
\textbf{\textit{Fast} resulted in the lowest waiting time} for complete text continuations to be shown and was perceived to be the \textbf{most clear:} 

\squote{P128} {The fast tool provided quick and clear enhancements.}

\squote{P174} {I preferred the medium and fast versions for professional writing due to their efficiency and clarity.}

Finally, a smaller portion of participants felt that \textit{slow} was preferred for professional writing due to its perceived \textbf{clarity and thoroughness.} Effective professional communication relies on concision and completeness, which \textit{slow} was perceived to offer: 

\squote{P285} {Clarity and precision were essential for professional writing, so the "medium" and "slow" versions were better suited.}

\squote{P224} {The slow tool was beneficial for thorough, in-depth responses, especially in professional contexts, but felt cumbersome for creative tasks.}

In each of these cases, the participants prefer a speed that they perceive as matching their values. Sometimes that perception is grounded in objective features of the condition, such as \textit{fast} providing greater efficiency, and othertimes dependent on implicit perceptual features, such as the thoughtfulness of \textit{slow}. Writers have diverse preferences for AI writing tools, arising from differences in values and processes \cite{gero22sparks, gero19metaphoria, ippolito22creative, biermann22from}. We extend these findings and show that the diversity of writing practices and values leads to a wide range of preferences for text appearance speed.



\section{Discussion}
Our results show that text appearance speed influences how users perceive generated outputs and relate to the tool. We discuss the implications of the perceptual effects stemming from AI interface design, as well as the impact on writing and creative processes.

\subsection{Implications for Intentional and Context Specific Design of AI Systems}

Text appearance speed is neither a neutral nor an inconsequential decision, influencing users to modify their judgments of system quality and trustworthiness, therefore influencing how generated text is used. 
In high-stakes situations, or even in frequent low-stakes situations, these decisions may have significant impacts on decision-making during writing. 
Passive acceptance of computer-generated writing continuations---a form of automation-induced complacency \cite{parasuraman93performance}---leads to more predictable and suboptimal writing outcomes \cite{arnold20predictive} and loss of agency and psychological ownership \cite{draxler24ai}.

Past work has emphasized the value of flexible LLM writing interfaces that can be configured to task and preference \cite{kim23cells}. In some video games, particularly RPGs such as Pokémon\footnote{https://bulbapedia.bulbagarden.net/wiki/Options} and Final Fantasy\footnote{https://strategywiki.org/wiki/Final\_Fantasy/Interface \\ Accessed on 16 April 2025}, players can choose the text appearance speed that they prefer. We similarly suggest giving users control over text generation appearance. For those who prefer to read along, configurability will enable them to curate a comfortable and appropriate experience. For example, when a task might require maximum efficiency, appearance speed could be tied to computation speed, and when a task may benefit from deeper consideration, a slow speed may be appropriate.  

While this study specifically investigates text presentation style, we see it as an example of how seemingly innocuous or foundational design decisions can have unanticipated perceptual effects. Writing support interfaces will have many such design choices. On a high level, the choice of interaction metaphor, such as tool or agent, not only influences how users relate to these systems \cite{lee2024design} through mechanisms like character-by-character text with random pauses \cite{jakesch23cowriting} and use of first-person pronouns \cite{peng20exploring}, but also influences how users make sense of and use the feedback provided to them. 
Words like ``reason'', ``plan'', and ``collaborate'' are innocuously used as metaphors (for users) and design lenses (for designers) to make sense of AI tools, but this terminology is uncritical \cite{sarkar23enough, agre14toward}, leading to unintentional harm \cite{sarkar23enough} and imprecise interaction effects.

We encourage the HCI and AI communities to consider all of the possible effects their design choices will have on user perceptions, and make intentional choices for the safety and benefit of people using these systems. For example, the choice to anthropomorphize computer systems can induce trust-forming behaviors in users and carry unintentional risks, such as making fallible information appear trustworthy \cite{maeda24when}, sometimes intentionally so \cite{leong19robot}.
An AI creative writing tool might be designed to backspace its generation randomly to simulate human typing, but this anthropomorphic design can unintentionally promote acceptance of its output by appearing thoughtful and trustworthy. 
Misconceptions of AI limit users' abilities to effectively use and critique these tools \cite{long20what}.
When design choices manipulate our sense of quality and trust, and change how a tool is used and understood, it is important that we are aware of their consequences.

\subsection{Perceptual Influences on Writing Cognition and Creative Process} 
In recent studies on writing assistants, Flower and Hayes' cognitive process model of writing \cite{flower81cognitive} has been used as a framework to explain the influence of writing tools on writing processes \cite{gero22design, zhang23visar, zhou24aillude}.
Our observed perceptual effects have the potential to reshape how writers engage with the writing task, influencing how writing is organized, evaluated, and revised. Tools that evoke human-like interaction, such as the \textit{medium} speed, might elicit tangible influence on personal goal setting.  

Creative writing is shaped by personal perspective and lived experience \cite{murray91all, flower81cognitive}, and is by nature a creative act of discovery and meaning-making. 
This process is essential to the character of the written outcome \cite{flower80rhetorical}. Design elements that potentially change how users reflect on their writing and the writing that is produced by writing tools can influence what is ultimately expressed by the author.

Writing is not only thinking, but also feeling \cite{brand87why, niles14randomized}, and creative writing is deeply personal and open-ended \cite{hunt88self}.
There are no predetermined paths to creative goals, nor are there explicit definitions of acceptable output \cite{amabile83social}. Therefore, the creative process is meaningful in itself. Tools play a crucial role in this process \cite{dalsgaard17instruments}, thereby influencing the experience and the outcome of creative exploration.
Tools that appear to generate higher quality text can support writing engagement and flow, but can also be disengaging if the suggestions are perceived as too good \cite{gero19metaphoria}.

The qualitative results suggest that text appearance styles evoke attribution of human-like qualities, such as ``thoughtfulness,'' ``emotional,'' and ``spontaneity.'' In creativity tasks, suggestions that appear to be intentional, thoughtful, and authentic are valued and appealing. Appearing thoughtful might foster emotional engagement, but also increase the likelihood that it is accepted. Writers care about agency and personal expression \cite{biermann22from, gero23social}, making it imperative for designers and users to understand that writing tools influence authenticity.

Text generators can influence personal stance and writing outcome \cite{jakesch23cowriting}. Significant effort in machine learning research is dedicated to improving language model output, such as by debiasing word embeddings \cite{bolukbasi16man} and injecting corpus-level constraints to follow a particular distribution \cite{zhao17men}, in order to prevent bias amplification and influence on one's writing and stance. Our study builds upon these findings, showing that perceptual factors and interaction design similarly affect how writers reflect and express themselves.

\subsection{Limitations and Future Work}
In this section, we acknowledge the scope of our study and reflect on limitations and tensions that may be further investigated in future work.

For our study, text suggestions are inserted in-line with human-written text. Other studies have used similar writing scenarios \cite{lee22coauthor, jakesch23cowriting, zhou24aillude}, and writing tools in email clients show suggested text in a similar manner. However, others show suggestions as off-page annotations \cite{hai22beyond} and lists \cite{sudowrite}; others adopt a chat interface \cite{schmitt21characterchat, GoogleGemini, ChatGPT}. Each of these paradigms ascribe different values, perceptions, and metaphors, which may limit the generality of our findings. 

Additionally, users participated in our study by imagining writing rather than directly writing in the interface. We made this choice to limit variability, as participants would unlikely write the same text and receive the same text generations. While some participants referred to the initial text as ``my writing'' and ``what I wrote,'' \textit{imagining writing} does not create the same experience as \textit{writing}.  Therefore there may be additional effects when the user is reflecting on their own ideas and writing their own text. 

For each of our speeds, characters appeared one by one. We suggest that future research explores token-by-token appearance, similar to some commercial tools.
Previous work has implemented timing inconsistency between character insertions to simulate co-writing \cite{jakesch23cowriting}, deletions during generation to correct mistakes \cite{cundy24sequencematch}, and speculative sampling that utilizes a faster but less capable draft model combined with a rejection module that ``rewrites'' outputted text \cite{chen23accelerating}. Human writing process involves series of additions and deletions (i.e. revision \cite{flower81cognitive}), so ``rewriting'' by an LLM may be interpreted as ``thinking'' or even reflection-in-action \cite{schon1983reflective}---distinctively human-like notions.

Finally, we acknowledge limitations in participant recruitment and participation. We recruited people from our institution and from Prolific, with the requirement of fluency in English. While we attempt to have broad reach, it would be difficult to analyze within and between specific subgroups, or recruit less technology-interested groups or potential participants who are not interested in using AI writing tools at all.



\section{Conclusion}
This paper presents a user study investigating how \textit{text appearance styles} and \textit{genres} affect perceptions of writing tools and their generated output. We find that text appearance speed plays a role in reading comfort, perceived writing quality, trustworthiness, and humanness, but we do not find evidence of a consistent interaction between text appearance speed and genre system. 
Qualitatively, we find that users read along with text generation, attribute human-like qualities to presentation speeds, and express varied preferences depending on genre. Our work provides insights on the effects of design considerations for intelligent writing tools and addresses some of the concerns that unintentional design has created. We hope this work contributes to a future where intelligent writing tools are designed with care for user benefit, mitigating unintentional manipulations, improving awareness of perceptual effects, and fostering authentic and meaningful human expression.

\begin{acks}
We are grateful to Andrew Chen for his assistance in crafting the figures for this paper. We also thank the members of the PICL Lab---including James Eschrich, Claire Tian, and William Goss---as well as the broader Interactive Computing community at the University of Illinois for their insights, discussions, and support.
\end{acks}


\bibliographystyle{ACM-Reference-Format}




\begin{thebibliography}{128}


\ifx \showCODEN    \undefined \def \showCODEN     #1{\unskip}     \fi
\ifx \showDOI      \undefined \def \showDOI       #1{#1}\fi
\ifx \showISBNx    \undefined \def \showISBNx     #1{\unskip}     \fi
\ifx \showISBNxiii \undefined \def \showISBNxiii  #1{\unskip}     \fi
\ifx \showISSN     \undefined \def \showISSN      #1{\unskip}     \fi
\ifx \showLCCN     \undefined \def \showLCCN      #1{\unskip}     \fi
\ifx \shownote     \undefined \def \shownote      #1{#1}          \fi
\ifx \showarticletitle \undefined \def \showarticletitle #1{#1}   \fi
\ifx \showURL      \undefined \def \showURL       {\relax}        \fi
\providecommand\bibfield[2]{#2}
\providecommand\bibinfo[2]{#2}
\providecommand\natexlab[1]{#1}
\providecommand\showeprint[2][]{arXiv:#2}

\bibitem[Agarwal et~al\mbox{.}(2023)]%
        {agarwal23llm}
\bibfield{author}{\bibinfo{person}{Megha Agarwal}, \bibinfo{person}{Asfandyar
  Qureshi}, \bibinfo{person}{Nikhil Sardana}, \bibinfo{person}{Linden Li},
  \bibinfo{person}{Julian Quevedo}, {and} \bibinfo{person}{Daya Khudia}.}
  \bibinfo{year}{2023}\natexlab{}.
\newblock \bibinfo{title}{LLM Inference Performance Engineering: Best
  Practices}.
\newblock
\newblock
\urldef\tempurl%
\url{https://www.databricks.com/blog/llm-inference-performance-engineering-best-practices}
\showURL{%
\tempurl}


\bibitem[Agre(2014)]%
        {agre14toward}
\bibfield{author}{\bibinfo{person}{Philip~E Agre}.}
  \bibinfo{year}{2014}\natexlab{}.
\newblock \showarticletitle{Toward a critical technical practice: Lessons
  learned in trying to reform AI}.
\newblock In \bibinfo{booktitle}{\emph{Social science, technical systems, and
  cooperative work}}. \bibinfo{publisher}{Psychology Press},
  \bibinfo{pages}{131--157}.
\newblock


\bibitem[Amabile(1983)]%
        {amabile83social}
\bibfield{author}{\bibinfo{person}{Teresa~M Amabile}.}
  \bibinfo{year}{1983}\natexlab{}.
\newblock \showarticletitle{The social psychology of creativity: A componential
  conceptualization.}
\newblock \bibinfo{journal}{\emph{Journal of Personality and Social
  Psychology}}  \bibinfo{volume}{45(2)} (\bibinfo{year}{1983}),
  \bibinfo{pages}{357--376}.
\newblock


\bibitem[Anthropic(2024)]%
        {AnthropicClaude}
\bibfield{author}{\bibinfo{person}{Anthropic}.}
  \bibinfo{year}{2024}\natexlab{}.
\newblock \bibinfo{title}{Claude}.
\newblock
\newblock
\urldef\tempurl%
\url{https://claude.ai/}
\showURL{%
\tempurl}


\bibitem[Arnold et~al\mbox{.}(2020)]%
        {arnold20predictive}
\bibfield{author}{\bibinfo{person}{Kenneth~C. Arnold}, \bibinfo{person}{Krysta
  Chauncey}, {and} \bibinfo{person}{Krzysztof~Z. Gajos}.}
  \bibinfo{year}{2020}\natexlab{}.
\newblock \showarticletitle{Predictive text encourages predictable writing}. In
  \bibinfo{booktitle}{\emph{Proceedings of the 25th International Conference on
  Intelligent User Interfaces}} (Cagliari, Italy) \emph{(\bibinfo{series}{IUI
  '20})}. \bibinfo{publisher}{Association for Computing Machinery},
  \bibinfo{address}{New York, NY, USA}, \bibinfo{pages}{128–138}.
\newblock
\showISBNx{9781450371186}
\urldef\tempurl%
\url{https://doi.org/10.1145/3377325.3377523}
\showDOI{\tempurl}


\bibitem[Ayres and Martin{\'a}s(2005)]%
        {ayres05on}
\bibfield{author}{\bibinfo{person}{R.U. Ayres} {and} \bibinfo{person}{K.
  Martin{\'a}s}.} \bibinfo{year}{2005}\natexlab{}.
\newblock \bibinfo{booktitle}{\emph{On the Reappraisal of Microeconomics:
  Economic Growth and Change in a Material World}}.
\newblock \bibinfo{publisher}{Edward Elgar}, \bibinfo{address}{Cheltenham,
  United Kingdom}.
\newblock
\showISBNx{9781845422721}
\showLCCN{2005048435}
\urldef\tempurl%
\url{https://books.google.com/books?id=ksxK7J95IF8C}
\showURL{%
\tempurl}


\bibitem[Bambach(2003)]%
        {bambach03leonardo}
\bibfield{author}{\bibinfo{person}{Carmen Bambach}, \bibinfo{person}{Rachel Stern}, \bibinfo{person}{Alison Manges}, {and} \bibinfo{person}{Metropolitan Museum of Art (New York, N.Y.)}.}
  \bibinfo{year}{2003}\natexlab{}.
\newblock \bibinfo{booktitle}{\emph{Leonardo Da Vinci Master Draftsman: Catalogue to an Exhibition at The Metropolitan Museum of Art, New York 2003}}.
\newblock \bibinfo{publisher}{Metropolitan Museum of Art}.
\newblock
\showISBNx{9781588390332}
\showLCCN{02191234}
\urldef\tempurl%
\url{https://books.google.com/books?id=QwQxDJMKRE4C}
\showDOI{\tempurl}


\bibitem[Bazerman et~al\mbox{.}(1994)]%
        {bazerman94systems}
\bibfield{author}{\bibinfo{person}{Charles Bazerman} {et~al\mbox{.}}}
  \bibinfo{year}{1994}\natexlab{}.
\newblock \showarticletitle{Systems of genres and the enactment of social
  intentions}.
\newblock \bibinfo{journal}{\emph{Genre and the new rhetoric}}
  \bibinfo{volume}{79101} (\bibinfo{year}{1994}),
  \bibinfo{pages}{9780203393277--14}.
\newblock


\bibitem[Berkenkotter(2001)]%
        {berkenkotter01genre}
\bibfield{author}{\bibinfo{person}{Carol Berkenkotter}.}
  \bibinfo{year}{2001}\natexlab{}.
\newblock \showarticletitle{Genre Systems at Work: DSM-IV and Rhetorical
  Recontextualization in Psychotherapy Paperwork}.
\newblock \bibinfo{journal}{\emph{Written Communication}} \bibinfo{volume}{18},
  \bibinfo{number}{3} (\bibinfo{year}{2001}), \bibinfo{pages}{326--349}.
\newblock
\urldef\tempurl%
\url{https://doi.org/10.1177/0741088301018003004}
\showDOI{\tempurl}
\showeprint{https://doi.org/10.1177/0741088301018003004}


\bibitem[Biermann et~al\mbox{.}(2022)]%
        {biermann22from}
\bibfield{author}{\bibinfo{person}{Oloff~C. Biermann}, \bibinfo{person}{Ning~F.
  Ma}, {and} \bibinfo{person}{Dongwook Yoon}.} \bibinfo{year}{2022}\natexlab{}.
\newblock \showarticletitle{From Tool to Companion: Storywriters Want AI
  Writers to Respect Their Personal Values and Writing Strategies}. In
  \bibinfo{booktitle}{\emph{Designing Interactive Systems Conference}} (Virtual
  Event, Australia) \emph{(\bibinfo{series}{DIS '22})}.
  \bibinfo{publisher}{Association for Computing Machinery},
  \bibinfo{address}{New York, NY, USA}, \bibinfo{pages}{1209–1227}.
\newblock
\showISBNx{9781450393584}
\urldef\tempurl%
\url{https://doi.org/10.1145/3532106.3533506}
\showDOI{\tempurl}


\bibitem[Bolukbasi et~al\mbox{.}(2016)]%
        {bolukbasi16man}
\bibfield{author}{\bibinfo{person}{Tolga Bolukbasi}, \bibinfo{person}{Kai-Wei
  Chang}, \bibinfo{person}{James~Y Zou}, \bibinfo{person}{Venkatesh Saligrama},
  {and} \bibinfo{person}{Adam~T Kalai}.} \bibinfo{year}{2016}\natexlab{}.
\newblock \showarticletitle{Man is to Computer Programmer as Woman is to
  Homemaker? Debiasing Word Embeddings}. In \bibinfo{booktitle}{\emph{Advances
  in Neural Information Processing Systems}},
  \bibfield{editor}{\bibinfo{person}{D.~Lee}, \bibinfo{person}{M.~Sugiyama},
  \bibinfo{person}{U.~Luxburg}, \bibinfo{person}{I.~Guyon}, {and}
  \bibinfo{person}{R.~Garnett}} (Eds.), Vol.~\bibinfo{volume}{29}.
  \bibinfo{publisher}{Curran Associates, Inc.}
\newblock
\urldef\tempurl%
\url{https://proceedings.neurips.cc/paper_files/paper/2016/file/a486cd07e4ac3d270571622f4f316ec5-Paper.pdf}
\showURL{%
\tempurl}


\bibitem[Brand(1987)]%
        {brand87why}
\bibfield{author}{\bibinfo{person}{Alice~G. Brand}.}
  \bibinfo{year}{1987}\natexlab{}.
\newblock \showarticletitle{The Why of Cognition: Emotion and the Writing
  Process}.
\newblock \bibinfo{journal}{\emph{College Composition and Communication}}
  \bibinfo{volume}{38}, \bibinfo{number}{4} (\bibinfo{year}{1987}),
  \bibinfo{pages}{436--443}.
\newblock
\showISSN{0010096X}
\urldef\tempurl%
\url{http://www.jstor.org/stable/357637}
\showURL{%
\tempurl}


\bibitem[Braun and Clarke(2006)]%
        {braun2006TA}
\bibfield{author}{\bibinfo{person}{Virginia Braun} {and}
  \bibinfo{person}{Victoria Clarke}.} \bibinfo{year}{2006}\natexlab{}.
\newblock \showarticletitle{Using thematic analysis in psychology}.
\newblock \bibinfo{journal}{\emph{Qualitative Research in Psychology}}
  \bibinfo{volume}{3}, \bibinfo{number}{2} (\bibinfo{year}{2006}),
  \bibinfo{pages}{77--101}.
\newblock
\urldef\tempurl%
\url{https://doi.org/10.1191/1478088706qp063oa}
\showDOI{\tempurl}


\bibitem[Cahill et~al\mbox{.}(2021)]%
        {cahill21supporting}
\bibfield{author}{\bibinfo{person}{Aoife Cahill}, \bibinfo{person}{James
  Bruno}, \bibinfo{person}{James Ramey}, \bibinfo{person}{Gilmar
  Ayala~Meneses}, \bibinfo{person}{Ian Blood}, \bibinfo{person}{Florencia
  Tolentino}, \bibinfo{person}{Tamar Lavee}, {and} \bibinfo{person}{Slava
  Andreyev}.} \bibinfo{year}{2021}\natexlab{}.
\newblock \showarticletitle{Supporting Spanish Writers using Automated
  Feedback}. In \bibinfo{booktitle}{\emph{Proceedings of the 2021 Conference of
  the North American Chapter of the Association for Computational Linguistics:
  Human Language Technologies: Demonstrations}},
  \bibfield{editor}{\bibinfo{person}{Avi Sil} {and}
  \bibinfo{person}{Xi~Victoria Lin}} (Eds.). \bibinfo{publisher}{Association
  for Computational Linguistics}, \bibinfo{address}{Online},
  \bibinfo{pages}{116--124}.
\newblock
\urldef\tempurl%
\url{https://doi.org/10.18653/v1/2021.naacl-demos.14}
\showDOI{\tempurl}


\bibitem[Cameron(2024)]%
        {llmleaderboard}
\bibfield{author}{\bibinfo{person}{George Cameron}.}
  \bibinfo{year}{2024}\natexlab{}.
\newblock \bibinfo{title}{LLM Leaderboard - Comparison of GPT-4o, Llama 3,
  Mistral, Gemini and over 30 models}.
\newblock
\newblock
\urldef\tempurl%
\url{https://artificialanalysis.ai/leaderboards/models}
\showURL{%
\tempurl}


\bibitem[Card et~al\mbox{.}(1991)]%
        {card91information}
\bibfield{author}{\bibinfo{person}{Stuart~K. Card}, \bibinfo{person}{George~G.
  Robertson}, {and} \bibinfo{person}{Jock~D. Mackinlay}.}
  \bibinfo{year}{1991}\natexlab{}.
\newblock \showarticletitle{The information visualizer, an information
  workspace}. In \bibinfo{booktitle}{\emph{Proceedings of the SIGCHI Conference
  on Human Factors in Computing Systems}} (New Orleans, Louisiana, USA)
  \emph{(\bibinfo{series}{CHI '91})}. \bibinfo{publisher}{Association for
  Computing Machinery}, \bibinfo{address}{New York, NY, USA},
  \bibinfo{pages}{181–186}.
\newblock
\showISBNx{0897913833}
\urldef\tempurl%
\url{https://doi.org/10.1145/108844.108874}
\showDOI{\tempurl}


\bibitem[Carrera and Lee(2022)]%
        {carrera22watch}
\bibfield{author}{\bibinfo{person}{Dashiel Carrera} {and}
  \bibinfo{person}{Sang~Won Lee}.} \bibinfo{year}{2022}\natexlab{}.
\newblock \showarticletitle{Watch Me Write: Exploring the Effects of Revealing
  Creative Writing Process through Writing Replay}. In
  \bibinfo{booktitle}{\emph{Proceedings of the 14th Conference on Creativity
  and Cognition}} (Venice, Italy) \emph{(\bibinfo{series}{C\&C '22})}.
  \bibinfo{publisher}{Association for Computing Machinery},
  \bibinfo{address}{New York, NY, USA}, \bibinfo{pages}{146–160}.
\newblock
\showISBNx{9781450393270}
\urldef\tempurl%
\url{https://doi.org/10.1145/3527927.3532806}
\showDOI{\tempurl}


\bibitem[Chen et~al\mbox{.}(2023)]%
        {chen23accelerating}
\bibfield{author}{\bibinfo{person}{Charlie Chen}, \bibinfo{person}{Sebastian
  Borgeaud}, \bibinfo{person}{Geoffrey Irving}, \bibinfo{person}{Jean-Baptiste
  Lespiau}, \bibinfo{person}{Laurent Sifre}, {and} \bibinfo{person}{John
  Jumper}.} \bibinfo{year}{2023}\natexlab{}.
\newblock \bibinfo{title}{Accelerating Large Language Model Decoding with
  Speculative Sampling}.
\newblock
\newblock
\showeprint[arxiv]{2302.01318}~[cs.CL]
\urldef\tempurl%
\url{https://arxiv.org/abs/2302.01318}
\showURL{%
\tempurl}


\bibitem[Chen and Milligan(2023)]%
        {quill}
\bibfield{author}{\bibinfo{person}{Jason Chen} {and} \bibinfo{person}{Byron
  Milligan}.} \bibinfo{year}{2023}\natexlab{}.
\newblock \bibinfo{title}{quilljs.com}.
\newblock
\newblock


\bibitem[Clark et~al\mbox{.}(2018)]%
        {clark18creative}
\bibfield{author}{\bibinfo{person}{Elizabeth Clark},
  \bibinfo{person}{Anne~Spencer Ross}, \bibinfo{person}{Chenhao Tan},
  \bibinfo{person}{Yangfeng Ji}, {and} \bibinfo{person}{Noah~A. Smith}.}
  \bibinfo{year}{2018}\natexlab{}.
\newblock \showarticletitle{Creative Writing with a Machine in the Loop: Case
  Studies on Slogans and Stories}. In \bibinfo{booktitle}{\emph{23rd
  International Conference on Intelligent User Interfaces}} (Tokyo, Japan)
  \emph{(\bibinfo{series}{IUI '18})}. \bibinfo{publisher}{Association for
  Computing Machinery}, \bibinfo{address}{New York, NY, USA},
  \bibinfo{pages}{329–340}.
\newblock
\showISBNx{9781450349451}
\urldef\tempurl%
\url{https://doi.org/10.1145/3172944.3172983}
\showDOI{\tempurl}


\bibitem[Cohn et~al\mbox{.}(2024)]%
        {cohn24believing}
\bibfield{author}{\bibinfo{person}{Michelle Cohn}, \bibinfo{person}{Mahima
  Pushkarna}, \bibinfo{person}{Gbolahan~O. Olanubi}, \bibinfo{person}{Joseph~M.
  Moran}, \bibinfo{person}{Daniel Padgett}, \bibinfo{person}{Zion Mengesha},
  {and} \bibinfo{person}{Courtney Heldreth}.} \bibinfo{year}{2024}\natexlab{}.
\newblock \showarticletitle{Believing Anthropomorphism: Examining the Role of
  Anthropomorphic Cues on Trust in Large Language Models}. In
  \bibinfo{booktitle}{\emph{Extended Abstracts of the CHI Conference on Human
  Factors in Computing Systems}} (Honolulu, HI, USA)
  \emph{(\bibinfo{series}{CHI EA '24})}. \bibinfo{publisher}{Association for
  Computing Machinery}, \bibinfo{address}{New York, NY, USA}, Article
  \bibinfo{articleno}{54}, \bibinfo{numpages}{15}~pages.
\newblock
\showISBNx{9798400703317}
\urldef\tempurl%
\url{https://doi.org/10.1145/3613905.3650818}
\showDOI{\tempurl}


\bibitem[Collier(1983)]%
        {collier83word}
\bibfield{author}{\bibinfo{person}{Richard~M. Collier}.}
  \bibinfo{year}{1983}\natexlab{}.
\newblock \showarticletitle{The Word Processor and Revision Strategies}.
\newblock \bibinfo{journal}{\emph{College Composition and Communication}}
  \bibinfo{volume}{34}, \bibinfo{number}{2} (\bibinfo{year}{1983}),
  \bibinfo{pages}{149--155}.
\newblock
\showISSN{0010096X}
\urldef\tempurl%
\url{http://www.jstor.org/stable/357402}
\showURL{%
\tempurl}


\bibitem[Commission(2014)]%
        {fcc14closed}
\bibfield{author}{\bibinfo{person}{Federal~Communications Commission}.}
  \bibinfo{year}{2014}\natexlab{}.
\newblock \bibinfo{title}{Closed Captioning of Video Programming}.
\newblock , \bibinfo{numpages}{20--23}~pages.
\newblock
\urldef\tempurl%
\url{https://docs.fcc.gov/public/attachments/FCC-14-12A1.pdf}
\showURL{%
\tempurl}
\newblock
\shownote{FCC-14-12A1}.


\bibitem[Conde et~al\mbox{.}(2024)]%
        {conde24speed}
\bibfield{author}{\bibinfo{person}{Javier Conde}, \bibinfo{person}{Miguel
  González}, \bibinfo{person}{Pedro Reviriego}, \bibinfo{person}{Zhen Gao},
  \bibinfo{person}{Shanshan Liu}, {and} \bibinfo{person}{Fabrizio Lombardi}.}
  \bibinfo{year}{2024}\natexlab{}.
\newblock \showarticletitle{Speed and Conversational Large Language Models: Not
  All Is About Tokens per Second}.
\newblock \bibinfo{journal}{\emph{Computer}} \bibinfo{volume}{57},
  \bibinfo{number}{8} (\bibinfo{year}{2024}), \bibinfo{pages}{74--80}.
\newblock
\urldef\tempurl%
\url{https://doi.org/10.1109/MC.2024.3399384}
\showDOI{\tempurl}


\bibitem[Corporation(2024)]%
        {bbc24subtitle}
\bibfield{author}{\bibinfo{person}{British~Broadcasting Corporation}.}
  \bibinfo{year}{2024}\natexlab{}.
\newblock \bibinfo{title}{Subtitle Guidelines}.
\newblock
\newblock
\urldef\tempurl%
\url{https://www.bbc.co.uk/accessibility/forproducts/guides/subtitles/}
\showURL{%
\tempurl}
\newblock
\shownote{Version 1.2.3}.


\bibitem[Cundy and Ermon(2024)]%
        {cundy24sequencematch}
\bibfield{author}{\bibinfo{person}{Chris Cundy} {and} \bibinfo{person}{Stefano
  Ermon}.} \bibinfo{year}{2024}\natexlab{}.
\newblock \bibinfo{title}{SequenceMatch: Imitation Learning for Autoregressive
  Sequence Modelling with Backtracking}.
\newblock
\newblock
\showeprint[arxiv]{2306.05426}~[cs.LG]
\urldef\tempurl%
\url{https://arxiv.org/abs/2306.05426}
\showURL{%
\tempurl}


\bibitem[Dalsgaard(2017)]%
        {dalsgaard17instruments}
\bibfield{author}{\bibinfo{person}{Peter Dalsgaard}.}
  \bibinfo{year}{2017}\natexlab{}.
\newblock \showarticletitle{Instruments of inquiry: Understanding the nature
  and role of tools in design}.
\newblock \bibinfo{journal}{\emph{International Journal of Design}}
  \bibinfo{volume}{11}, \bibinfo{number}{1} (\bibinfo{year}{2017}),
  \bibinfo{pages}{21--33}.
\newblock


\bibitem[Dang et~al\mbox{.}(2022)]%
        {hai22beyond}
\bibfield{author}{\bibinfo{person}{Hai Dang}, \bibinfo{person}{Karim
  Benharrak}, \bibinfo{person}{Florian Lehmann}, {and} \bibinfo{person}{Daniel
  Buschek}.} \bibinfo{year}{2022}\natexlab{}.
\newblock \showarticletitle{Beyond Text Generation: Supporting Writers with
  Continuous Automatic Text Summaries}. In
  \bibinfo{booktitle}{\emph{Proceedings of the 35th Annual ACM Symposium on
  User Interface Software and Technology}} (Bend, OR, USA)
  \emph{(\bibinfo{series}{UIST '22})}. \bibinfo{publisher}{Association for
  Computing Machinery}, \bibinfo{address}{New York, NY, USA}, Article
  \bibinfo{articleno}{98}, \bibinfo{numpages}{13}~pages.
\newblock
\showISBNx{9781450393201}
\urldef\tempurl%
\url{https://doi.org/10.1145/3526113.3545672}
\showDOI{\tempurl}


\bibitem[Dave and Russell(2010)]%
        {dave10drafting}
\bibfield{author}{\bibinfo{person}{Anish~M. Dave} {and}
  \bibinfo{person}{David~R. Russell}.} \bibinfo{year}{2010}\natexlab{}.
\newblock \showarticletitle{Drafting and Revision Using Word Processing by
  Undergraduate Student Writers: Changing Conceptions and Practices}.
\newblock \bibinfo{journal}{\emph{Research in the Teaching of English}}
  \bibinfo{volume}{44}, \bibinfo{number}{4} (\bibinfo{year}{2010}),
  \bibinfo{pages}{406--434}.
\newblock
\showISSN{0034527X}
\urldef\tempurl%
\url{http://www.jstor.org/stable/25704888}
\showURL{%
\tempurl}


\bibitem[Dong and Xie(2024)]%
        {dong24large}
\bibfield{author}{\bibinfo{person}{Haiwei Dong} {and} \bibinfo{person}{Shuang
  Xie}.} \bibinfo{year}{2024}\natexlab{}.
\newblock \bibinfo{title}{Large Language Models (LLMs): Deployment, Tokenomics
  and Sustainability}.
\newblock
\newblock
\showeprint[arxiv]{2405.17147}~[cs.MM]
\urldef\tempurl%
\url{https://arxiv.org/abs/2405.17147}
\showURL{%
\tempurl}


\bibitem[Draxler et~al\mbox{.}(2024)]%
        {draxler24ai}
\bibfield{author}{\bibinfo{person}{Fiona Draxler}, \bibinfo{person}{Anna
  Werner}, \bibinfo{person}{Florian Lehmann}, \bibinfo{person}{Matthias Hoppe},
  \bibinfo{person}{Albrecht Schmidt}, \bibinfo{person}{Daniel Buschek}, {and}
  \bibinfo{person}{Robin Welsch}.} \bibinfo{year}{2024}\natexlab{}.
\newblock \showarticletitle{The AI ghostwriter effect: When users do not
  perceive ownership of AI-generated text but self-declare as authors}.
\newblock \bibinfo{journal}{\emph{ACM Transactions on Computer-Human
  Interaction}} \bibinfo{volume}{31}, \bibinfo{number}{2}
  (\bibinfo{year}{2024}), \bibinfo{pages}{1--40}.
\newblock


\bibitem[Elkin et~al\mbox{.}(2021)]%
        {elkin21uist}
\bibfield{author}{\bibinfo{person}{Lisa~A. Elkin}, \bibinfo{person}{Matthew
  Kay}, \bibinfo{person}{James~J. Higgins}, {and} \bibinfo{person}{Jacob~O.
  Wobbrock}.} \bibinfo{year}{2021}\natexlab{}.
\newblock \showarticletitle{An Aligned Rank Transform Procedure for Multifactor
  Contrast Tests}. In \bibinfo{booktitle}{\emph{The 34th Annual ACM Symposium
  on User Interface Software and Technology}} (Virtual Event, USA)
  \emph{(\bibinfo{series}{UIST '21})}. \bibinfo{publisher}{Association for
  Computing Machinery}, \bibinfo{address}{New York, NY, USA},
  \bibinfo{pages}{754–768}.
\newblock
\showISBNx{9781450386357}
\urldef\tempurl%
\url{https://doi.org/10.1145/3472749.3474784}
\showDOI{\tempurl}


\bibitem[Faul et~al\mbox{.}(2007)]%
        {gpower3}
\bibfield{author}{\bibinfo{person}{Franz Faul}, \bibinfo{person}{Edgar
  Erdfelder}, \bibinfo{person}{Albert-Georg Lang}, {and} \bibinfo{person}{Axel
  Buchner}.} \bibinfo{year}{2007}\natexlab{}.
\newblock \showarticletitle{G*Power 3: A flexible statistical power analysis
  program for the social, behavioral, and biomedical sciences}.
\newblock \bibinfo{journal}{\emph{Behavior Research Methods}}
  \bibinfo{volume}{39}, \bibinfo{number}{2} (\bibinfo{year}{2007}),
  \bibinfo{pages}{175--191}.
\newblock


\bibitem[Flower and Hayes(1980)]%
        {flower80rhetorical}
\bibfield{author}{\bibinfo{person}{Linda Flower} {and} \bibinfo{person}{John~R.
  Hayes}.} \bibinfo{year}{1980}\natexlab{}.
\newblock \showarticletitle{The Cognition of Discovery: Defining a Rhetorical
  Problem}.
\newblock \bibinfo{journal}{\emph{College Composition and Communication}}
  \bibinfo{volume}{31}, \bibinfo{number}{1} (\bibinfo{year}{1980}),
  \bibinfo{pages}{21--32}.
\newblock
\showISSN{0010096X}
\urldef\tempurl%
\url{http://www.jstor.org/stable/356630}
\showURL{%
\tempurl}


\bibitem[Flower and Hayes(1981)]%
        {flower81cognitive}
\bibfield{author}{\bibinfo{person}{Linda Flower} {and} \bibinfo{person}{John~R.
  Hayes}.} \bibinfo{year}{1981}\natexlab{}.
\newblock \showarticletitle{A Cognitive Process Theory of Writing}.
\newblock \bibinfo{journal}{\emph{College Composition and Communication}}
  \bibinfo{volume}{32}, \bibinfo{number}{4} (\bibinfo{year}{1981}),
  \bibinfo{pages}{365--387}.
\newblock
\showISSN{0010096X}
\urldef\tempurl%
\url{http://www.jstor.org/stable/356600}
\showURL{%
\tempurl}


\bibitem[Gabriel et~al\mbox{.}(2015)]%
        {gabriel15inkwell}
\bibfield{author}{\bibinfo{person}{Richard~P. Gabriel}, \bibinfo{person}{Jilin
  Chen}, {and} \bibinfo{person}{Jeffrey Nichols}.}
  \bibinfo{year}{2015}\natexlab{}.
\newblock \showarticletitle{InkWell: A Creative Writer's Creative Assistant}.
  In \bibinfo{booktitle}{\emph{Proceedings of the 2015 ACM SIGCHI Conference on
  Creativity and Cognition}} (Glasgow, United Kingdom)
  \emph{(\bibinfo{series}{C\&C '15})}. \bibinfo{publisher}{Association for
  Computing Machinery}, \bibinfo{address}{New York, NY, USA},
  \bibinfo{pages}{93–102}.
\newblock
\showISBNx{9781450335980}
\urldef\tempurl%
\url{https://doi.org/10.1145/2757226.2757229}
\showDOI{\tempurl}


\bibitem[George and Mallery(2016)]%
        {george16ibm}
\bibfield{author}{\bibinfo{person}{Darren George} {and} \bibinfo{person}{Paul Mallery}.} \bibinfo{year}{2016}\natexlab{}.
\newblock \bibinfo{booktitle}{\emph{IBM SPSS Statistics 23 Step by Step: A Simple Guide and Reference}}.
\newblock \bibinfo{publisher}{Routledge, Taylor \& Francis Group}.
\newblock


\bibitem[Gero et~al\mbox{.}(2022a)]%
        {gero22design}
\bibfield{author}{\bibinfo{person}{Katy Gero}, \bibinfo{person}{Alex
  Calderwood}, \bibinfo{person}{Charlotte Li}, {and} \bibinfo{person}{Lydia
  Chilton}.} \bibinfo{year}{2022}\natexlab{a}.
\newblock \showarticletitle{A Design Space for Writing Support Tools Using a
  Cognitive Process Model of Writing}. In \bibinfo{booktitle}{\emph{Proceedings
  of the First Workshop on Intelligent and Interactive Writing Assistants
  (In2Writing 2022)}}, \bibfield{editor}{\bibinfo{person}{Ting-Hao~'Kenneth'
  Huang}, \bibinfo{person}{Vipul Raheja}, \bibinfo{person}{Dongyeop Kang},
  \bibinfo{person}{John Joon~Young Chung}, \bibinfo{person}{Daniel Gissin},
  \bibinfo{person}{Mina Lee}, {and} \bibinfo{person}{Katy~Ilonka Gero}} (Eds.).
  \bibinfo{publisher}{Association for Computational Linguistics},
  \bibinfo{address}{Dublin, Ireland}, \bibinfo{pages}{11--24}.
\newblock
\urldef\tempurl%
\url{https://doi.org/10.18653/v1/2022.in2writing-1.2}
\showDOI{\tempurl}


\bibitem[Gero and Chilton(2019)]%
        {gero19metaphoria}
\bibfield{author}{\bibinfo{person}{Katy~Ilonka Gero} {and}
  \bibinfo{person}{Lydia~B. Chilton}.} \bibinfo{year}{2019}\natexlab{}.
\newblock \showarticletitle{Metaphoria: An Algorithmic Companion for Metaphor
  Creation}. In \bibinfo{booktitle}{\emph{Proceedings of the 2019 CHI
  Conference on Human Factors in Computing Systems}} (Glasgow, Scotland Uk)
  \emph{(\bibinfo{series}{CHI '19})}. \bibinfo{publisher}{Association for
  Computing Machinery}, \bibinfo{address}{New York, NY, USA},
  \bibinfo{pages}{1–12}.
\newblock
\showISBNx{9781450359702}
\urldef\tempurl%
\url{https://doi.org/10.1145/3290605.3300526}
\showDOI{\tempurl}


\bibitem[Gero et~al\mbox{.}(2022b)]%
        {gero22sparks}
\bibfield{author}{\bibinfo{person}{Katy~Ilonka Gero}, \bibinfo{person}{Vivian
  Liu}, {and} \bibinfo{person}{Lydia Chilton}.}
  \bibinfo{year}{2022}\natexlab{b}.
\newblock \showarticletitle{Sparks: Inspiration for Science Writing using
  Language Models}. In \bibinfo{booktitle}{\emph{Proceedings of the 2022 ACM
  Designing Interactive Systems Conference}} (Virtual Event, Australia)
  \emph{(\bibinfo{series}{DIS '22})}. \bibinfo{publisher}{Association for
  Computing Machinery}, \bibinfo{address}{New York, NY, USA},
  \bibinfo{pages}{1002–1019}.
\newblock
\showISBNx{9781450393584}
\urldef\tempurl%
\url{https://doi.org/10.1145/3532106.3533533}
\showDOI{\tempurl}


\bibitem[Gero et~al\mbox{.}(2023)]%
        {gero23social}
\bibfield{author}{\bibinfo{person}{Katy~Ilonka Gero}, \bibinfo{person}{Tao
  Long}, {and} \bibinfo{person}{Lydia~B Chilton}.}
  \bibinfo{year}{2023}\natexlab{}.
\newblock \showarticletitle{Social Dynamics of AI Support in Creative Writing}.
  In \bibinfo{booktitle}{\emph{Proceedings of the 2023 CHI Conference on Human
  Factors in Computing Systems}} (Hamburg, Germany) \emph{(\bibinfo{series}{CHI
  '23})}. \bibinfo{publisher}{Association for Computing Machinery},
  \bibinfo{address}{New York, NY, USA}, Article \bibinfo{articleno}{245},
  \bibinfo{numpages}{15}~pages.
\newblock
\showISBNx{9781450394215}
\urldef\tempurl%
\url{https://doi.org/10.1145/3544548.3580782}
\showDOI{\tempurl}


\bibitem[Goel and Pirolli(1992)]%
        {goel92structure}
\bibfield{author}{\bibinfo{person}{Vinod Goel} {and} \bibinfo{person}{Peter
  Pirolli}.} \bibinfo{year}{1992}\natexlab{}.
\newblock \showarticletitle{The structure of Design Problem Spaces}.
\newblock \bibinfo{journal}{\emph{Cognitive Science}} \bibinfo{volume}{16},
  \bibinfo{number}{3} (\bibinfo{year}{1992}), \bibinfo{pages}{395--429}.
\newblock
\urldef\tempurl%
\url{https://doi.org/10.1207/s15516709cog1603\_3}
\showDOI{\tempurl}
\showeprint{https://onlinelibrary.wiley.com/doi/pdf/10.1207/s15516709cog1603\_3}


\bibitem[Google(2024)]%
        {GoogleGemini}
\bibfield{author}{\bibinfo{person}{Google}.} \bibinfo{year}{2024}\natexlab{}.
\newblock \bibinfo{title}{Gemini}.
\newblock
\newblock
\urldef\tempurl%
\url{https://gemini.google.com/}
\showURL{%
\tempurl}


\bibitem[Gross et~al\mbox{.}(2017)]%
        {gross17technical}
\bibfield{author}{\bibinfo{person}{A. Gross}, \bibinfo{person}{A. Hamlin},
  \bibinfo{person}{B. Merck}, \bibinfo{person}{C. Rubio}, \bibinfo{person}{J.
  Naas}, \bibinfo{person}{M. Savage}, {and} \bibinfo{person}{M. DeSilva}.}
  \bibinfo{year}{2017}\natexlab{}.
\newblock \bibinfo{booktitle}{\emph{Technical Writing}}.
\newblock \bibinfo{publisher}{Open Oregon Educational Resources}.
\newblock
\showISBNx{9781636350660}
\urldef\tempurl%
\url{https://books.google.com/books?id=pxKMzQEACAAJ}
\showURL{%
\tempurl}


\bibitem[Gupta and Yu(2024)]%
        {sudowrite}
\bibfield{author}{\bibinfo{person}{Amit Gupta} {and} \bibinfo{person}{James
  Yu}.} \bibinfo{year}{2024}\natexlab{}.
\newblock \bibinfo{title}{sudowrite.com}.
\newblock
\newblock
\urldef\tempurl%
\url{https://www.sudowrite.com/}
\showURL{%
\tempurl}


\bibitem[Hill et~al\mbox{.}(1991)]%
        {hill91revising}
\bibfield{author}{\bibinfo{person}{Charles~A. Hill}, \bibinfo{person}{David~L.
  Wallace}, {and} \bibinfo{person}{Christina Haas}.}
  \bibinfo{year}{1991}\natexlab{}.
\newblock \showarticletitle{Revising on-line: Computer technologies and the
  revising process}.
\newblock \bibinfo{journal}{\emph{Computers and Composition}}
  \bibinfo{volume}{9}, \bibinfo{number}{1} (\bibinfo{year}{1991}),
  \bibinfo{pages}{83--109}.
\newblock
\showISSN{8755-4615}
\urldef\tempurl%
\url{https://doi.org/10.1016/8755-4615(91)80040-K}
\showDOI{\tempurl}


\bibitem[Hong et~al\mbox{.}(2021a)]%
        {hong21selfdriving}
\bibfield{author}{\bibinfo{person}{Joo-Wha Hong}, \bibinfo{person}{Ignacio
  Cruz}, {and} \bibinfo{person}{Dmitri Williams}.}
  \bibinfo{year}{2021}\natexlab{a}.
\newblock \showarticletitle{AI, you can drive my car: How we evaluate human
  drivers vs. self-driving cars}.
\newblock \bibinfo{journal}{\emph{Computers in Human Behavior}}
  \bibinfo{volume}{125} (\bibinfo{year}{2021}), \bibinfo{pages}{106944}.
\newblock
\showISSN{0747-5632}
\urldef\tempurl%
\url{https://doi.org/10.1016/j.chb.2021.106944}
\showDOI{\tempurl}


\bibitem[Hong et~al\mbox{.}(2021b)]%
        {hong21are}
\bibfield{author}{\bibinfo{person}{Joo~Wha Hong}, \bibinfo{person}{Qiyao Peng},
  {and} \bibinfo{person}{Dmitri Williams}.} \bibinfo{year}{2021}\natexlab{b}.
\newblock \showarticletitle{Are you ready for artificial Mozart and Skrillex?
  An experiment testing expectancy violation theory and AI music}.
\newblock \bibinfo{journal}{\emph{New Media \& Society}} \bibinfo{volume}{23},
  \bibinfo{number}{7} (\bibinfo{year}{2021}), \bibinfo{pages}{1920--1935}.
\newblock
\urldef\tempurl%
\url{https://doi.org/10.1177/1461444820925798}
\showDOI{\tempurl}
\showeprint{https://doi.org/10.1177/1461444820925798}


\bibitem[Hui and Sprouse(2023)]%
        {hui23lettersmith}
\bibfield{author}{\bibinfo{person}{Julie Hui} {and}
  \bibinfo{person}{Michelle~L. Sprouse}.} \bibinfo{year}{2023}\natexlab{}.
\newblock \showarticletitle{Lettersmith: Scaffolding Written Professional
  Communication Among College Students}. In
  \bibinfo{booktitle}{\emph{Proceedings of the 2023 CHI Conference on Human
  Factors in Computing Systems}} (Hamburg, Germany) \emph{(\bibinfo{series}{CHI
  '23})}. \bibinfo{publisher}{Association for Computing Machinery},
  \bibinfo{address}{New York, NY, USA}, Article \bibinfo{articleno}{703},
  \bibinfo{numpages}{17}~pages.
\newblock
\showISBNx{9781450394215}
\urldef\tempurl%
\url{https://doi.org/10.1145/3544548.3581029}
\showDOI{\tempurl}


\bibitem[Hui et~al\mbox{.}(2018)]%
        {hui18introassist}
\bibfield{author}{\bibinfo{person}{Julie~S. Hui}, \bibinfo{person}{Darren
  Gergle}, {and} \bibinfo{person}{Elizabeth~M. Gerber}.}
  \bibinfo{year}{2018}\natexlab{}.
\newblock \showarticletitle{IntroAssist: A Tool to Support Writing Introductory
  Help Requests}. In \bibinfo{booktitle}{\emph{Proceedings of the 2018 CHI
  Conference on Human Factors in Computing Systems}} (Montreal QC, Canada)
  \emph{(\bibinfo{series}{CHI '18})}. \bibinfo{publisher}{Association for
  Computing Machinery}, \bibinfo{address}{New York, NY, USA},
  \bibinfo{pages}{1–13}.
\newblock
\showISBNx{9781450356206}
\urldef\tempurl%
\url{https://doi.org/10.1145/3173574.3173596}
\showDOI{\tempurl}


\bibitem[Hunt and Sampson(1998)]%
        {hunt88self}
\bibfield{author}{\bibinfo{person}{Celia Hunt} {and} \bibinfo{person}{Fiona
  Sampson}.} \bibinfo{year}{1998}\natexlab{}.
\newblock \bibinfo{booktitle}{\emph{The Self on the Page: Theory and Practice
  of Creative Writing in Personal Development}}.
\newblock \bibinfo{publisher}{Jessica Kingsley Publishers}.
\newblock
\showISBNx{978-1-85302-470-2}
\showLCCN{19981806}


\bibitem[Hwang et~al\mbox{.}(2019)]%
        {hwang19when}
\bibfield{author}{\bibinfo{person}{Sun~Young Hwang}, \bibinfo{person}{Negar
  Khojasteh}, {and} \bibinfo{person}{Susan~R. Fussell}.}
  \bibinfo{year}{2019}\natexlab{}.
\newblock \showarticletitle{When Delayed in a Hurry: Interpretations of
  Response Delays in Time-Sensitive Instant Messaging}.
\newblock \bibinfo{journal}{\emph{Proc. ACM Hum.-Comput. Interact.}}
  \bibinfo{volume}{3}, \bibinfo{number}{GROUP}, Article
  \bibinfo{articleno}{234} (\bibinfo{date}{dec} \bibinfo{year}{2019}),
  \bibinfo{numpages}{20}~pages.
\newblock
\urldef\tempurl%
\url{https://doi.org/10.1145/3361115}
\showDOI{\tempurl}


\bibitem[Ippolito et~al\mbox{.}(2022)]%
        {ippolito22creative}
\bibfield{author}{\bibinfo{person}{Daphne Ippolito}, \bibinfo{person}{Ann
  Yuan}, \bibinfo{person}{Andy Coenen}, {and} \bibinfo{person}{Sehmon Burnam}.}
  \bibinfo{year}{2022}\natexlab{}.
\newblock \bibinfo{title}{Creative Writing with an AI-Powered Writing
  Assistant: Perspectives from Professional Writers}.
\newblock
\newblock
\urldef\tempurl%
\url{https://doi.org/10.48550/ARXIV.2211.05030}
\showDOI{\tempurl}


\bibitem[Jakesch et~al\mbox{.}(2023)]%
        {jakesch23cowriting}
\bibfield{author}{\bibinfo{person}{Maurice Jakesch}, \bibinfo{person}{Advait
  Bhat}, \bibinfo{person}{Daniel Buschek}, \bibinfo{person}{Lior Zalmanson},
  {and} \bibinfo{person}{Mor Naaman}.} \bibinfo{year}{2023}\natexlab{}.
\newblock \showarticletitle{Co-Writing with Opinionated Language Models Affects
  Users’ Views}. In \bibinfo{booktitle}{\emph{Proceedings of the 2023 CHI
  Conference on Human Factors in Computing Systems}} (Hamburg, Germany)
  \emph{(\bibinfo{series}{CHI '23})}. \bibinfo{publisher}{Association for
  Computing Machinery}, \bibinfo{address}{New York, NY, USA}, Article
  \bibinfo{articleno}{111}, \bibinfo{numpages}{15}~pages.
\newblock
\showISBNx{9781450394215}
\urldef\tempurl%
\url{https://doi.org/10.1145/3544548.3581196}
\showDOI{\tempurl}


\bibitem[Jensema et~al\mbox{.}(1996)]%
        {jensema96closed}
\bibfield{author}{\bibinfo{person}{Carl Jensema}, \bibinfo{person}{Ralph
  McCann}, {and} \bibinfo{person}{Scott Ramsey}.}
  \bibinfo{year}{1996}\natexlab{}.
\newblock \showarticletitle{Closed-Captioned Television Presentation Speed and
  Vocabulary}.
\newblock \bibinfo{journal}{\emph{American Annals of the Deaf}}
  \bibinfo{volume}{141}, \bibinfo{number}{4} (\bibinfo{year}{1996}),
  \bibinfo{pages}{284--292}.
\newblock


\bibitem[Karsten(2014)]%
        {karsten14writing}
\bibfield{author}{\bibinfo{person}{Andrea Karsten}.}
  \bibinfo{year}{2014}\natexlab{}.
\newblock \showarticletitle{Writing: Movements of the self}.
\newblock \bibinfo{journal}{\emph{Theory \& Psychology}} \bibinfo{volume}{24},
  \bibinfo{number}{4} (\bibinfo{year}{2014}), \bibinfo{pages}{479--503}.
\newblock
\urldef\tempurl%
\url{https://doi.org/10.1177/0959354314541020}
\showDOI{\tempurl}
\showeprint{https://doi.org/10.1177/0959354314541020}


\bibitem[Khadpe et~al\mbox{.}(2020)]%
        {khadpe20conceptual}
\bibfield{author}{\bibinfo{person}{Pranav Khadpe}, \bibinfo{person}{Ranjay
  Krishna}, \bibinfo{person}{Li Fei-Fei}, \bibinfo{person}{Jeffrey~T. Hancock},
  {and} \bibinfo{person}{Michael~S. Bernstein}.}
  \bibinfo{year}{2020}\natexlab{}.
\newblock \showarticletitle{Conceptual Metaphors Impact Perceptions of Human-AI
  Collaboration}.
\newblock \bibinfo{journal}{\emph{Proc. ACM Hum.-Comput. Interact.}}
  \bibinfo{volume}{4}, \bibinfo{number}{CSCW2}, Article
  \bibinfo{articleno}{163} (\bibinfo{date}{oct} \bibinfo{year}{2020}),
  \bibinfo{numpages}{26}~pages.
\newblock
\urldef\tempurl%
\url{https://doi.org/10.1145/3415234}
\showDOI{\tempurl}


\bibitem[Khurana et~al\mbox{.}(2024)]%
        {khurana24why}
\bibfield{author}{\bibinfo{person}{Anjali Khurana}, \bibinfo{person}{Hariharan
  Subramonyam}, {and} \bibinfo{person}{Parmit~K Chilana}.}
  \bibinfo{year}{2024}\natexlab{}.
\newblock \showarticletitle{Why and When LLM-Based Assistants Can Go Wrong:
  Investigating the Effectiveness of Prompt-Based Interactions for Software
  Help-Seeking}. In \bibinfo{booktitle}{\emph{Proceedings of the 29th
  International Conference on Intelligent User Interfaces}} (Greenville, SC,
  USA) \emph{(\bibinfo{series}{IUI '24})}. \bibinfo{publisher}{Association for
  Computing Machinery}, \bibinfo{address}{New York, NY, USA},
  \bibinfo{pages}{288–303}.
\newblock
\showISBNx{9798400705083}
\urldef\tempurl%
\url{https://doi.org/10.1145/3640543.3645200}
\showDOI{\tempurl}


\bibitem[Kim et~al\mbox{.}(2023)]%
        {kim23cells}
\bibfield{author}{\bibinfo{person}{Tae~Soo Kim}, \bibinfo{person}{Yoonjoo Lee},
  \bibinfo{person}{Minsuk Chang}, {and} \bibinfo{person}{Juho Kim}.}
  \bibinfo{year}{2023}\natexlab{}.
\newblock \showarticletitle{Cells, Generators, and Lenses: Design Framework for
  Object-Oriented Interaction with Large Language Models}. In
  \bibinfo{booktitle}{\emph{Proceedings of the 36th Annual ACM Symposium on
  User Interface Software and Technology}} (San Francisco, CA, USA)
  \emph{(\bibinfo{series}{UIST '23})}. \bibinfo{publisher}{Association for
  Computing Machinery}, \bibinfo{address}{New York, NY, USA}, Article
  \bibinfo{articleno}{4}, \bibinfo{numpages}{18}~pages.
\newblock
\showISBNx{9798400701320}
\urldef\tempurl%
\url{https://doi.org/10.1145/3586183.3606833}
\showDOI{\tempurl}


\bibitem[Kincaid(1975)]%
        {kincaid75derivation}
\bibfield{author}{\bibinfo{person}{JP Kincaid}.}
  \bibinfo{year}{1975}\natexlab{}.
\newblock \showarticletitle{Derivation of new readability formulas (automated
  readability index, fog count and flesch reading ease formula) for navy
  enlisted personnel}.
\newblock \bibinfo{journal}{\emph{Chief of Naval Technical Training}}
  (\bibinfo{year}{1975}).
\newblock


\bibitem[Kirschenbaum(2016)]%
        {kirschenbaum16track}
\bibfield{author}{\bibinfo{person}{M.G. Kirschenbaum}.}
  \bibinfo{year}{2016}\natexlab{}.
\newblock \bibinfo{booktitle}{\emph{Track Changes: A Literary History of Word
  Processing}}.
\newblock \bibinfo{publisher}{Harvard University Press}.
\newblock
\showISBNx{9780674417076}
\showLCCN{2015041450}
\urldef\tempurl%
\url{https://books.google.com/books?id=viy7CwAAQBAJ}
\showURL{%
\tempurl}


\bibitem[Lee and Moray(1992)]%
        {lee92trust}
\bibfield{author}{\bibinfo{person}{John Lee} {and} \bibinfo{person}{Neville
  Moray}.} \bibinfo{year}{1992}\natexlab{}.
\newblock \showarticletitle{Trust, control strategies and allocation of
  function in human-machine systems}.
\newblock \bibinfo{journal}{\emph{Ergonomics}} \bibinfo{volume}{35},
  \bibinfo{number}{10} (\bibinfo{year}{1992}), \bibinfo{pages}{1243--1270}.
\newblock
\urldef\tempurl%
\url{https://doi.org/10.1080/00140139208967392}
\showDOI{\tempurl}
\showeprint{https://doi.org/10.1080/00140139208967392}
\newblock
\shownote{PMID: 1516577}.


\bibitem[Lee et~al\mbox{.}(2024)]%
        {lee2024design}
\bibfield{author}{\bibinfo{person}{Mina Lee}, \bibinfo{person}{Katy~Ilonka
  Gero}, \bibinfo{person}{John Joon~Young Chung},
  \bibinfo{person}{Simon~Buckingham Shum}, \bibinfo{person}{Vipul Raheja},
  \bibinfo{person}{Hua Shen}, \bibinfo{person}{Subhashini Venugopalan},
  \bibinfo{person}{Thiemo Wambsganss}, \bibinfo{person}{David Zhou},
  \bibinfo{person}{Emad~A. Alghamdi}, \bibinfo{person}{Tal August},
  \bibinfo{person}{Avinash Bhat}, \bibinfo{person}{Madiha~Zahrah Choksi},
  \bibinfo{person}{Senjuti Dutta}, \bibinfo{person}{Jin~L.C. Guo},
  \bibinfo{person}{Md~Naimul Hoque}, \bibinfo{person}{Yewon Kim},
  \bibinfo{person}{Simon Knight}, \bibinfo{person}{Seyed~Parsa Neshaei},
  \bibinfo{person}{Antonette Shibani}, \bibinfo{person}{Disha Shrivastava},
  \bibinfo{person}{Lila Shroff}, \bibinfo{person}{Agnia Sergeyuk},
  \bibinfo{person}{Jessi Stark}, \bibinfo{person}{Sarah Sterman},
  \bibinfo{person}{Sitong Wang}, \bibinfo{person}{Antoine Bosselut},
  \bibinfo{person}{Daniel Buschek}, \bibinfo{person}{Joseph~Chee Chang},
  \bibinfo{person}{Sherol Chen}, \bibinfo{person}{Max Kreminski},
  \bibinfo{person}{Joonsuk Park}, \bibinfo{person}{Roy Pea},
  \bibinfo{person}{Eugenia Ha~Rim Rho}, \bibinfo{person}{Zejiang Shen}, {and}
  \bibinfo{person}{Pao Siangliulue}.} \bibinfo{year}{2024}\natexlab{}.
\newblock \showarticletitle{A Design Space for Intelligent and Interactive
  Writing Assistants}. In \bibinfo{booktitle}{\emph{Proceedings of the CHI
  Conference on Human Factors in Computing Systems}} (Honolulu, HI, USA)
  \emph{(\bibinfo{series}{CHI '24})}. \bibinfo{publisher}{Association for
  Computing Machinery}, \bibinfo{address}{New York, NY, USA}, Article
  \bibinfo{articleno}{1054}, \bibinfo{numpages}{35}~pages.
\newblock
\showISBNx{9798400703300}
\urldef\tempurl%
\url{https://doi.org/10.1145/3613904.3642697}
\showDOI{\tempurl}


\bibitem[Lee et~al\mbox{.}(2022)]%
        {lee22coauthor}
\bibfield{author}{\bibinfo{person}{Mina Lee}, \bibinfo{person}{Percy Liang},
  {and} \bibinfo{person}{Qian Yang}.} \bibinfo{year}{2022}\natexlab{}.
\newblock \showarticletitle{CoAuthor: Designing a Human-AI Collaborative
  Writing Dataset for Exploring Language Model Capabilities}. In
  \bibinfo{booktitle}{\emph{Proceedings of the 2022 CHI Conference on Human
  Factors in Computing Systems}} (New Orleans, LA, USA)
  \emph{(\bibinfo{series}{CHI '22})}. \bibinfo{publisher}{Association for
  Computing Machinery}, \bibinfo{address}{New York, NY, USA}, Article
  \bibinfo{articleno}{388}, \bibinfo{numpages}{19}~pages.
\newblock
\showISBNx{9781450391573}
\urldef\tempurl%
\url{https://doi.org/10.1145/3491102.3502030}
\showDOI{\tempurl}


\bibitem[Leong and Selinger(2019)]%
        {leong19robot}
\bibfield{author}{\bibinfo{person}{Brenda Leong} {and} \bibinfo{person}{Evan
  Selinger}.} \bibinfo{year}{2019}\natexlab{}.
\newblock \showarticletitle{Robot Eyes Wide Shut: Understanding Dishonest
  Anthropomorphism}. In \bibinfo{booktitle}{\emph{Proceedings of the Conference
  on Fairness, Accountability, and Transparency}} (Atlanta, GA, USA)
  \emph{(\bibinfo{series}{FAT* '19})}. \bibinfo{publisher}{Association for
  Computing Machinery}, \bibinfo{address}{New York, NY, USA},
  \bibinfo{pages}{299–308}.
\newblock
\showISBNx{9781450361255}
\urldef\tempurl%
\url{https://doi.org/10.1145/3287560.3287591}
\showDOI{\tempurl}


\bibitem[Liu et~al\mbox{.}(2024)]%
        {liu24andes}
\bibfield{author}{\bibinfo{person}{Jiachen Liu}, \bibinfo{person}{Zhiyu Wu},
  \bibinfo{person}{Jae-Won Chung}, \bibinfo{person}{Fan Lai},
  \bibinfo{person}{Myungjin Lee}, {and} \bibinfo{person}{Mosharaf Chowdhury}.}
  \bibinfo{year}{2024}\natexlab{}.
\newblock \bibinfo{title}{Andes: Defining and Enhancing Quality-of-Experience
  in LLM-Based Text Streaming Services}.
\newblock
\newblock
\showeprint[arxiv]{2404.16283}~[cs.DC]
\urldef\tempurl%
\url{https://arxiv.org/abs/2404.16283}
\showURL{%
\tempurl}


\bibitem[Liu et~al\mbox{.}(2023)]%
        {liu23modeling}
\bibfield{author}{\bibinfo{person}{Xingyu~"Bruce" Liu}, \bibinfo{person}{Jun
  Zhang}, \bibinfo{person}{Leonardo Ferrer}, \bibinfo{person}{Susan Xu},
  \bibinfo{person}{Vikas Bahirwani}, \bibinfo{person}{Boris Smus},
  \bibinfo{person}{Alex Olwal}, {and} \bibinfo{person}{Ruofei Du}.}
  \bibinfo{year}{2023}\natexlab{}.
\newblock \showarticletitle{Modeling and Improving Text Stability in Live
  Captions}. In \bibinfo{booktitle}{\emph{Extended Abstracts of the 2023 CHI
  Conference on Human Factors in Computing Systems}} (Hamburg, Germany)
  \emph{(\bibinfo{series}{CHI EA '23})}. \bibinfo{publisher}{Association for
  Computing Machinery}, \bibinfo{address}{New York, NY, USA}, Article
  \bibinfo{articleno}{208}, \bibinfo{numpages}{9}~pages.
\newblock
\showISBNx{9781450394222}
\urldef\tempurl%
\url{https://doi.org/10.1145/3544549.3585609}
\showDOI{\tempurl}


\bibitem[Long and Magerko(2020)]%
        {long20what}
\bibfield{author}{\bibinfo{person}{Duri Long} {and} \bibinfo{person}{Brian
  Magerko}.} \bibinfo{year}{2020}\natexlab{}.
\newblock \showarticletitle{What is AI Literacy? Competencies and Design
  Considerations}. In \bibinfo{booktitle}{\emph{Proceedings of the 2020 CHI
  Conference on Human Factors in Computing Systems}} (Honolulu, HI, USA)
  \emph{(\bibinfo{series}{CHI '20})}. \bibinfo{publisher}{Association for
  Computing Machinery}, \bibinfo{address}{New York, NY, USA},
  \bibinfo{pages}{1–16}.
\newblock
\showISBNx{9781450367080}
\urldef\tempurl%
\url{https://doi.org/10.1145/3313831.3376727}
\showDOI{\tempurl}


\bibitem[Maeda and Quan-Haase(2024)]%
        {maeda24when}
\bibfield{author}{\bibinfo{person}{Takuya Maeda} {and} \bibinfo{person}{Anabel
  Quan-Haase}.} \bibinfo{year}{2024}\natexlab{}.
\newblock \showarticletitle{When Human-AI Interactions Become Parasocial:
  Agency and Anthropomorphism in Affective Design}. In
  \bibinfo{booktitle}{\emph{Proceedings of the 2024 ACM Conference on Fairness,
  Accountability, and Transparency}} (Rio de Janeiro, Brazil)
  \emph{(\bibinfo{series}{FAccT '24})}. \bibinfo{publisher}{Association for
  Computing Machinery}, \bibinfo{address}{New York, NY, USA},
  \bibinfo{pages}{1068–1077}.
\newblock
\showISBNx{9798400704505}
\urldef\tempurl%
\url{https://doi.org/10.1145/3630106.3658956}
\showDOI{\tempurl}


\bibitem[Martinuzzi(2022)]%
        {martinuzzi22why}
\bibfield{author}{\bibinfo{person}{Bruna Martinuzzi}.}
  \bibinfo{year}{2022}\natexlab{}.
\newblock \bibinfo{title}{Why Talking Too Fast Can Hurt Your Message}.
\newblock
\newblock
\urldef\tempurl%
\url{https://www.americanexpress.com/en-us/business/trends-and-insights/articles/slow-down-why-speaking-too-fast-can-hurt-your-message/}
\showURL{%
\tempurl}


\bibitem[Mavrakakis and Chipman(2021)]%
        {mavrakakis21writing}
\bibfield{author}{\bibinfo{person}{Konstantinos Mavrakakis} {and}
  \bibinfo{person}{Abigail Chipman}.} \bibinfo{year}{2021}\natexlab{}.
\newblock \showarticletitle{Writing fiction; a phenomenological study of the
  creative writing experience of fiction writers}.
\newblock \bibinfo{journal}{\emph{Modern Psychological Studies}}
  \bibinfo{volume}{26}, \bibinfo{number}{2} (\bibinfo{year}{2021}),
  \bibinfo{pages}{1--35}.
\newblock


\bibitem[Mayer et~al\mbox{.}(1995)]%
        {mayer95integrative}
\bibfield{author}{\bibinfo{person}{Roger~C. Mayer}, \bibinfo{person}{James~H.
  Davis}, {and} \bibinfo{person}{F.~David Schoorman}.}
  \bibinfo{year}{1995}\natexlab{}.
\newblock \showarticletitle{An Integrative Model of Organizational Trust}.
\newblock \bibinfo{journal}{\emph{The Academy of Management Review}}
  \bibinfo{volume}{20}, \bibinfo{number}{3} (\bibinfo{year}{1995}),
  \bibinfo{pages}{709--734}.
\newblock
\showISSN{03637425}
\urldef\tempurl%
\url{http://www.jstor.org/stable/258792}
\showURL{%
\tempurl}


\bibitem[Meta(2024)]%
        {MetaAI}
\bibfield{author}{\bibinfo{person}{Meta}.} \bibinfo{year}{2024}\natexlab{}.
\newblock \bibinfo{title}{Meta AI}.
\newblock
\newblock
\urldef\tempurl%
\url{https://www.meta.ai/}
\showURL{%
\tempurl}


\bibitem[Miller(1984)]%
        {miller84genre}
\bibfield{author}{\bibinfo{person}{Carolyn~R. Miller}.}
  \bibinfo{year}{1984}\natexlab{}.
\newblock \showarticletitle{Genre as social action}.
\newblock \bibinfo{journal}{\emph{Quarterly Journal of Speech}}
  \bibinfo{volume}{70}, \bibinfo{number}{2} (\bibinfo{year}{1984}),
  \bibinfo{pages}{151--167}.
\newblock
\urldef\tempurl%
\url{https://doi.org/10.1080/00335638409383686}
\showDOI{\tempurl}
\showeprint{https://doi.org/10.1080/00335638409383686}


\bibitem[Murray(1991)]%
        {murray91all}
\bibfield{author}{\bibinfo{person}{Donald Murray}.}
  \bibinfo{year}{1991}\natexlab{}.
\newblock \showarticletitle{All Writing Is Autobiography}.
\newblock \bibinfo{journal}{\emph{College Composition and Communication}}
  \bibinfo{volume}{42}, \bibinfo{number}{1} (\bibinfo{year}{1991}),
  \bibinfo{pages}{66--74}.
\newblock
\showISSN{0010096X}
\urldef\tempurl%
\url{http://www.jstor.org/stable/357540}
\showURL{%
\tempurl}


\bibitem[Nass and Moon(2000)]%
        {nass00machines}
\bibfield{author}{\bibinfo{person}{Clifford Nass} {and}
  \bibinfo{person}{Youngme Moon}.} \bibinfo{year}{2000}\natexlab{}.
\newblock \showarticletitle{Machines and Mindlessness: Social Responses to
  Computers}.
\newblock \bibinfo{journal}{\emph{Journal of Social Issues}}
  \bibinfo{volume}{56}, \bibinfo{number}{1} (\bibinfo{year}{2000}),
  \bibinfo{pages}{81--103}.
\newblock
\urldef\tempurl%
\url{https://doi.org/10.1111/0022-4537.00153}
\showDOI{\tempurl}
\showeprint{https://spssi.onlinelibrary.wiley.com/doi/pdf/10.1111/0022-4537.00153}


\bibitem[Newell(1990)]%
        {newell90unified}
\bibfield{author}{\bibinfo{person}{Allen Newell}.}
  \bibinfo{year}{1990}\natexlab{}.
\newblock \bibinfo{booktitle}{\emph{Unified theories of cognition}}.
\newblock \bibinfo{publisher}{Harvard University Press},
  \bibinfo{address}{USA}.
\newblock
\showISBNx{0674920996}


\bibitem[Ni(2019)]%
        {ni19do}
\bibfield{author}{\bibinfo{person}{Preston Ni}.}
  \bibinfo{year}{2019}\natexlab{}.
\newblock \bibinfo{title}{Do You Talk Too Fast? How to Slow Down}.
\newblock
\newblock
\urldef\tempurl%
\url{https://www.psychologytoday.com/us/blog/communication-success/201911/do-you-talk-too-fast-how-to-slow-down}
\showURL{%
\tempurl}


\bibitem[Nielsen(1993)]%
        {nielsen93response}
\bibfield{author}{\bibinfo{person}{Jakob Nielsen}.}
  \bibinfo{year}{1993}\natexlab{}.
\newblock \bibinfo{title}{Response Times: The 3 Important Limits}.
\newblock
\newblock
\urldef\tempurl%
\url{https://www.nngroup.com/articles/response-times-3-important-limits/}
\showURL{%
\tempurl}


\bibitem[Nielsen(1994)]%
        {nielsen94usability}
\bibfield{author}{\bibinfo{person}{Jakob Nielsen}.}
  \bibinfo{year}{1994}\natexlab{}.
\newblock \bibinfo{booktitle}{\emph{Usability Engineering}}.
\newblock \bibinfo{publisher}{Morgan Kaufmann Publishers Inc.},
  \bibinfo{address}{San Francisco, CA, USA}.
\newblock
\showISBNx{9780080520292}


\bibitem[Niles et~al\mbox{.}({[n.\,d.]})]%
        {niles14randomized}
\bibfield{author}{\bibinfo{person}{Andrea~N. Niles}, \bibinfo{person}{Kate
  E.~Byrne Haltom}, \bibinfo{person}{Catherine~M. Mulvenna},
  \bibinfo{person}{Matthew~D. Lieberman}, {and} \bibinfo{person}{Annette~L.
  Stanton}.} \bibinfo{year}{[n.\,d.]}\natexlab{}.
\newblock \showarticletitle{Randomized controlled trial of expressive writing
  for psychological and physical health: the moderating role of emotional
  expressivity.}
\newblock  \bibinfo{volume}{27}, \bibinfo{number}{1}
  (\bibinfo{year}{[n.\,d.]}), \bibinfo{pages}{1--17}.
\newblock
\showISSN{1477-2205 1061-5806}
\urldef\tempurl%
\url{https://doi.org/10.1080/10615806.2013.802308}
\showDOI{\tempurl}
\newblock
\shownote{Place: England}.


\bibitem[Norman(2010)]%
        {norman10likert}
\bibfield{author}{\bibinfo{person}{Geoff Norman}.}
  \bibinfo{year}{2010}\natexlab{}.
\newblock \showarticletitle{Likert scales, levels of measurement and the
  “laws” of statistics}.
\newblock \bibinfo{journal}{\emph{Advances in health sciences education}}
  \bibinfo{volume}{15} (\bibinfo{year}{2010}), \bibinfo{pages}{625--632}.
\newblock


\bibitem[Novikova et~al\mbox{.}(2018)]%
        {novikova18rankME}
\bibfield{author}{\bibinfo{person}{Jekaterina Novikova},
  \bibinfo{person}{Ond{\v{r}}ej Du{\v{s}}ek}, {and} \bibinfo{person}{Verena
  Rieser}.} \bibinfo{year}{2018}\natexlab{}.
\newblock \showarticletitle{RankME: Reliable human ratings for natural language
  generation}.
\newblock \bibinfo{journal}{\emph{arXiv preprint arXiv:1803.05928}}
  (\bibinfo{year}{2018}).
\newblock


\bibitem[NVIDIA(2024)]%
        {nvidia24metrics}
\bibfield{author}{\bibinfo{person}{NVIDIA}.} \bibinfo{year}{2024}\natexlab{}.
\newblock \bibinfo{title}{NIM for LLM Benchmarking Guide | Metrics}.
\newblock
\newblock
\urldef\tempurl%
\url{https://docs.nvidia.com/nim/benchmarking/llm/latest/metrics.html}
\showURL{%
\tempurl}


\bibitem[OpenAI(2024a)]%
        {ChatGPT}
\bibfield{author}{\bibinfo{person}{OpenAI}.} \bibinfo{year}{2024}\natexlab{a}.
\newblock \bibinfo{title}{ChatGPT}.
\newblock
\newblock
\urldef\tempurl%
\url{https://chat.openai.com}
\showURL{%
\tempurl}


\bibitem[OpenAI(2024b)]%
        {ChatGPTAPIStreaming}
\bibfield{author}{\bibinfo{person}{OpenAI}.} \bibinfo{year}{2024}\natexlab{b}.
\newblock \bibinfo{title}{Streaming}.
\newblock
\newblock
\urldef\tempurl%
\url{https://platform.openai.com/docs/api-reference/streaming}
\showURL{%
\tempurl}


\bibitem[OpenAI(2024c)]%
        {openai24what}
\bibfield{author}{\bibinfo{person}{OpenAI}.} \bibinfo{year}{2024}\natexlab{c}.
\newblock \bibinfo{title}{What are tokens and how to count them?}
\newblock
\newblock
\urldef\tempurl%
\url{https://help.openai.com/en/articles/4936856-what-are-tokens-and-how-to-count-them}
\showURL{%
\tempurl}


\bibitem[Peng et~al\mbox{.}(2020)]%
        {peng20exploring}
\bibfield{author}{\bibinfo{person}{Zhenhui Peng}, \bibinfo{person}{Qingyu Guo},
  \bibinfo{person}{Ka~Wing Tsang}, {and} \bibinfo{person}{Xiaojuan Ma}.}
  \bibinfo{year}{2020}\natexlab{}.
\newblock \showarticletitle{Exploring the Effects of Technological Writing
  Assistance for Support Providers in Online Mental Health Community}. In
  \bibinfo{booktitle}{\emph{Proceedings of the 2020 CHI Conference on Human
  Factors in Computing Systems}} (Honolulu, HI, USA)
  \emph{(\bibinfo{series}{CHI '20})}. \bibinfo{publisher}{Association for
  Computing Machinery}, \bibinfo{address}{New York, NY, USA},
  \bibinfo{pages}{1–15}.
\newblock
\showISBNx{9781450367080}
\urldef\tempurl%
\url{https://doi.org/10.1145/3313831.3376695}
\showDOI{\tempurl}


\bibitem[Piepho(2018)]%
        {piepho18letters}
\bibfield{author}{\bibinfo{person}{Hans-Peter Piepho}.}
  \bibinfo{year}{2018}\natexlab{}.
\newblock \showarticletitle{Letters in Mean Comparisons: What They Do and
  Don’t Mean}.
\newblock \bibinfo{journal}{\emph{Agronomy Journal}} \bibinfo{volume}{110},
  \bibinfo{number}{2} (\bibinfo{year}{2018}), \bibinfo{pages}{431--434}.
\newblock
\urldef\tempurl%
\url{https://doi.org/10.2134/agronj2017.10.0580}
\showDOI{\tempurl}
\showeprint{https://acsess.onlinelibrary.wiley.com/doi/pdf/10.2134/agronj2017.10.0580}


\bibitem[Prior(2013)]%
        {prior13writing}
\bibfield{author}{\bibinfo{person}{Paul Prior}.}
  \bibinfo{year}{2013}\natexlab{}.
\newblock \bibinfo{booktitle}{\emph{Writing/disciplinarity: A sociohistoric
  account of literate activity in the academy}}.
\newblock \bibinfo{publisher}{Routledge}.
\newblock


\bibitem[Raja~Parasuraman and Singh(1993)]%
        {parasuraman93performance}
\bibfield{author}{\bibinfo{person}{Robert~Molloy Raja~Parasuraman} {and}
  \bibinfo{person}{Indramani~L. Singh}.} \bibinfo{year}{1993}\natexlab{}.
\newblock \showarticletitle{Performance Consequences of Automation-Induced
  'Complacency'}.
\newblock \bibinfo{journal}{\emph{The International Journal of Aviation
  Psychology}} \bibinfo{volume}{3}, \bibinfo{number}{1} (\bibinfo{year}{1993}),
  \bibinfo{pages}{1--23}.
\newblock
\urldef\tempurl%
\url{https://doi.org/10.1207/s15327108ijap0301\_1}
\showDOI{\tempurl}
\showeprint{https://doi.org/10.1207/s15327108ijap0301\_1}


\bibitem[Rayner et~al\mbox{.}(2016)]%
        {rayner16so}
\bibfield{author}{\bibinfo{person}{Keith Rayner}, \bibinfo{person}{Elizabeth~R.
  Schotter}, \bibinfo{person}{Michael E.~J. Masson}, \bibinfo{person}{Mary~C.
  Potter}, {and} \bibinfo{person}{Rebecca Treiman}.}
  \bibinfo{year}{2016}\natexlab{}.
\newblock \showarticletitle{So Much to Read, So Little Time: How Do We Read,
  and Can Speed Reading Help?}
\newblock \bibinfo{journal}{\emph{Psychological Science in the Public
  Interest}} \bibinfo{volume}{17}, \bibinfo{number}{1} (\bibinfo{year}{2016}),
  \bibinfo{pages}{4--34}.
\newblock
\urldef\tempurl%
\url{https://doi.org/10.1177/1529100615623267}
\showDOI{\tempurl}
\showeprint{https://doi.org/10.1177/1529100615623267}
\newblock
\shownote{PMID: 26769745}.


\bibitem[Readable({[n.\,d.]})]%
        {readableflesch}
\bibfield{author}{\bibinfo{person}{Readable}.}
  \bibinfo{year}{[n.\,d.]}\natexlab{}.
\newblock \bibinfo{title}{Flesch Reading Ease and the Flesch Kincaid Grade
  Level}.
\newblock
\newblock
\urldef\tempurl%
\url{https://readable.com/readability/flesch-reading-ease-flesch-kincaid-grade-level/}
\showURL{%
\tempurl}


\bibitem[Research(2023)]%
        {microsoft23how}
\bibfield{author}{\bibinfo{person}{Microsoft Research}.}
  \bibinfo{year}{2023}\natexlab{}.
\newblock \bibinfo{title}{How to Evaluate LLMs: A Complete Metric Framework}.
\newblock
\newblock
\urldef\tempurl%
\url{https://www.microsoft.com/en-us/research/group/experimentation-platform-exp/articles/how-to-evaluate-llms-a-complete-metric-framework/}
\showURL{%
\tempurl}


\bibitem[Richard L.~Street et~al\mbox{.}(1983)]%
        {street83influence}
\bibfield{author}{\bibinfo{person}{JR Richard L.~Street},
  \bibinfo{person}{Robert~M. Brady}, {and} \bibinfo{person}{William~B.
  Putman}.} \bibinfo{year}{1983}\natexlab{}.
\newblock \showarticletitle{The Influence of Speech Rate Stereotypes and Rate
  Similarity or Listeners' Evaluations of Speakers}.
\newblock \bibinfo{journal}{\emph{Journal of Language and Social Psychology}}
  \bibinfo{volume}{2}, \bibinfo{number}{1} (\bibinfo{year}{1983}),
  \bibinfo{pages}{37--56}.
\newblock
\urldef\tempurl%
\url{https://doi.org/10.1177/0261927X8300200103}
\showDOI{\tempurl}
\showeprint{https://doi.org/10.1177/0261927X8300200103}


\bibitem[Robson(2022)]%
        {robson22reflective}
\bibfield{author}{\bibinfo{person}{Ian Robson}.}
  \bibinfo{year}{2022}\natexlab{}.
\newblock \bibinfo{booktitle}{\emph{The Reflective Leader : Reflexivity in
  Practice}}.
\newblock \bibinfo{publisher}{Emerald Publishing Limited},
  \bibinfo{address}{Bingley}.
\newblock
\showISBNx{978-1-83982-555-2}


\bibitem[Roemmele and Gordon(2015)]%
        {roemmele15creative}
\bibfield{author}{\bibinfo{person}{Melissa Roemmele} {and}
  \bibinfo{person}{Andrew~S. Gordon}.} \bibinfo{year}{2015}\natexlab{}.
\newblock \showarticletitle{Creative Help: A Story Writing Assistant}. In
  \bibinfo{booktitle}{\emph{Interactive {Storytelling}}}
  \emph{(\bibinfo{series}{Lecture Notes in Computer Science})},
  \bibfield{editor}{\bibinfo{person}{Henrik Schoenau-Fog},
  \bibinfo{person}{Luis~Emilio Bruni}, \bibinfo{person}{Sandy Louchart}, {and}
  \bibinfo{person}{Sarune Baceviciute}} (Eds.). \bibinfo{publisher}{Springer
  International Publishing}, \bibinfo{address}{Cham}, \bibinfo{pages}{81--92}.
\newblock
\showISBNx{978-3-319-27036-4}
\urldef\tempurl%
\url{https://doi.org/10.1007/978-3-319-27036-4_8}
\showDOI{\tempurl}


\bibitem[Roemmele and Gordon(2018)]%
        {roemmele18automated}
\bibfield{author}{\bibinfo{person}{Melissa Roemmele} {and}
  \bibinfo{person}{Andrew~S. Gordon}.} \bibinfo{year}{2018}\natexlab{}.
\newblock \showarticletitle{Automated Assistance for Creative Writing with an
  RNN Language Model}. In \bibinfo{booktitle}{\emph{Proceedings of the 23rd
  International Conference on Intelligent User Interfaces Companion}}
  \emph{(\bibinfo{series}{IUI '18 Companion})}. \bibinfo{publisher}{Association
  for Computing Machinery}, \bibinfo{address}{New York, NY, USA},
  \bibinfo{pages}{1--2}.
\newblock
\showISBNx{978-1-4503-5571-1}
\urldef\tempurl%
\url{https://doi.org/10.1145/3180308.3180329}
\showDOI{\tempurl}


\bibitem[Rosenhead(1996)]%
        {rosenhead96whats}
\bibfield{author}{\bibinfo{person}{Jonathan Rosenhead}.}
  \bibinfo{year}{1996}\natexlab{}.
\newblock \showarticletitle{What's the Problem? An Introduction to Problem
  Structuring Methods}.
\newblock \bibinfo{journal}{\emph{Interfaces}} \bibinfo{volume}{26},
  \bibinfo{number}{6} (\bibinfo{date}{December} \bibinfo{year}{1996}),
  \bibinfo{pages}{117–131}.
\newblock
\showISSN{0092-2102}
\urldef\tempurl%
\url{https://doi.org/10.1287/inte.26.6.117}
\showDOI{\tempurl}


\bibitem[Rousseau et~al\mbox{.}(1998)]%
        {rousseau98introduction}
\bibfield{author}{\bibinfo{person}{Denise~M. Rousseau}, \bibinfo{person}{Sim~B.
  Sitkin}, \bibinfo{person}{Ronald~S. Burt}, {and} \bibinfo{person}{Colin
  Camerer}.} \bibinfo{year}{1998}\natexlab{}.
\newblock \showarticletitle{Introduction to Special Topic Forum: Not so
  Different after All: A Cross-Discipline View of Trust}.
\newblock \bibinfo{journal}{\emph{The Academy of Management Review}}
  \bibinfo{volume}{23}, \bibinfo{number}{3} (\bibinfo{year}{1998}),
  \bibinfo{pages}{393--404}.
\newblock
\showISSN{03637425}
\urldef\tempurl%
\url{http://www.jstor.org/stable/259285}
\showURL{%
\tempurl}


\bibitem[Saisubramanian et~al\mbox{.}(2021)]%
        {saisubramanian21understanding}
\bibfield{author}{\bibinfo{person}{Sandhya Saisubramanian},
  \bibinfo{person}{Shannon~C. Roberts}, {and} \bibinfo{person}{Shlomo
  Zilberstein}.} \bibinfo{year}{2021}\natexlab{}.
\newblock \showarticletitle{Understanding User Attitudes Towards Negative Side
  Effects of AI Systems}. In \bibinfo{booktitle}{\emph{Extended Abstracts of
  the 2021 CHI Conference on Human Factors in Computing Systems}} (Yokohama,
  Japan) \emph{(\bibinfo{series}{CHI EA '21})}. \bibinfo{publisher}{Association
  for Computing Machinery}, \bibinfo{address}{New York, NY, USA}, Article
  \bibinfo{articleno}{368}, \bibinfo{numpages}{6}~pages.
\newblock
\showISBNx{9781450380959}
\urldef\tempurl%
\url{https://doi.org/10.1145/3411763.3451654}
\showDOI{\tempurl}


\bibitem[Sarkar(2023)]%
        {sarkar23enough}
\bibfield{author}{\bibinfo{person}{Advait Sarkar}.}
  \bibinfo{year}{2023}\natexlab{}.
\newblock \showarticletitle{Enough With “Human-AI Collaboration”}. In
  \bibinfo{booktitle}{\emph{Extended Abstracts of the 2023 CHI Conference on
  Human Factors in Computing Systems}} (Hamburg, Germany)
  \emph{(\bibinfo{series}{CHI EA '23})}. \bibinfo{publisher}{Association for
  Computing Machinery}, \bibinfo{address}{New York, NY, USA}, Article
  \bibinfo{articleno}{415}, \bibinfo{numpages}{8}~pages.
\newblock
\showISBNx{9781450394222}
\urldef\tempurl%
\url{https://doi.org/10.1145/3544549.3582735}
\showDOI{\tempurl}


\bibitem[Saumya(2024)]%
        {saumya24how}
\bibfield{author}{\bibinfo{person}{Saumya}.} \bibinfo{year}{2024}\natexlab{}.
\newblock \bibinfo{title}{How (and why) to implement streaming in your LLM
  application}.
\newblock
\newblock
\urldef\tempurl%
\url{https://blog.kusho.ai/how-and-why-to-implement-streaming-in-your-llm-application/}
\showURL{%
\tempurl}


\bibitem[Schmitt and Buschek(2021)]%
        {schmitt21characterchat}
\bibfield{author}{\bibinfo{person}{Oliver Schmitt} {and}
  \bibinfo{person}{Daniel Buschek}.} \bibinfo{year}{2021}\natexlab{}.
\newblock \showarticletitle{CharacterChat: Supporting the Creation of Fictional
  Characters through Conversation and Progressive Manifestation with a
  Chatbot}. In \bibinfo{booktitle}{\emph{Proceedings of the 13th Conference on
  Creativity and Cognition}} (Virtual Event, Italy)
  \emph{(\bibinfo{series}{C\&C '21})}. \bibinfo{publisher}{Association for
  Computing Machinery}, \bibinfo{address}{New York, NY, USA}, Article
  \bibinfo{articleno}{10}, \bibinfo{numpages}{10}~pages.
\newblock
\showISBNx{9781450383769}
\urldef\tempurl%
\url{https://doi.org/10.1145/3450741.3465253}
\showDOI{\tempurl}


\bibitem[Sch{\"o}n(1983)]%
        {schon1983reflective}
\bibfield{author}{\bibinfo{person}{Donald Sch{\"o}n}.}
  \bibinfo{year}{1983}\natexlab{}.
\newblock \bibinfo{booktitle}{\emph{The Reflective Practitioner}}.
\newblock \bibinfo{publisher}{Basic Books, Perseus Books Group},
  \bibinfo{address}{London}.
\newblock


\bibitem[Schraeder(2019)]%
        {schraeder19public}
\bibfield{author}{\bibinfo{person}{Terry~L. Schraeder}.}
  \bibinfo{year}{2019}\natexlab{}.
\newblock \showarticletitle{{Public Speaking and Presentation Skills}}.
\newblock In \bibinfo{booktitle}{\emph{{Physician Communication: Connecting
  with Patients, Peers, and the Public}}}. \bibinfo{publisher}{Oxford
  University Press}.
\newblock
\showISBNx{9780190882440}
\urldef\tempurl%
\url{https://doi.org/10.1093/med/9780190882440.003.0003}
\showDOI{\tempurl}
\showeprint{https://academic.oup.com/book/0/chapter/265380356/chapter-ag-pdf/44543567/book\_31716\_section\_265380356.ag.pdf}


\bibitem[Sears et~al\mbox{.}(1993)]%
        {sears93investigating}
\bibfield{author}{\bibinfo{person}{Andrew Sears}, \bibinfo{person}{Doreen
  Revis}, \bibinfo{person}{Janet Swatski}, \bibinfo{person}{Rob Crittenden},
  {and} \bibinfo{person}{Ben Shneiderman}.} \bibinfo{year}{1993}\natexlab{}.
\newblock \showarticletitle{Investigating touchscreen typing: the effect of
  keyboard size on typing speed}.
\newblock \bibinfo{journal}{\emph{Behaviour \& Information Technology}}
  \bibinfo{volume}{12}, \bibinfo{number}{1} (\bibinfo{year}{1993}),
  \bibinfo{pages}{17--22}.
\newblock
\urldef\tempurl%
\url{https://doi.org/10.1080/01449299308924362}
\showDOI{\tempurl}
\showeprint{https://doi.org/10.1080/01449299308924362}


\bibitem[Sheehan(1999)]%
        {sheehan99metaphor}
\bibfield{author}{\bibinfo{person}{Richard D.~Johnson Sheehan}.}
  \bibinfo{year}{1999}\natexlab{}.
\newblock \showarticletitle{Metaphor as Hermeneutic}.
\newblock \bibinfo{journal}{\emph{Rhetoric Society Quarterly}}
  \bibinfo{volume}{29}, \bibinfo{number}{2} (\bibinfo{year}{1999}),
  \bibinfo{pages}{47--64}.
\newblock
\showISSN{02773945}
\urldef\tempurl%
\url{http://www.jstor.org/stable/3886085}
\showURL{%
\tempurl}


\bibitem[Singh et~al\mbox{.}(2023)]%
        {singh22stolen}
\bibfield{author}{\bibinfo{person}{Nikhil Singh}, \bibinfo{person}{Guillermo
  Bernal}, \bibinfo{person}{Daria Savchenko}, {and} \bibinfo{person}{Elena~L.
  Glassman}.} \bibinfo{year}{2023}\natexlab{}.
\newblock \showarticletitle{Where to Hide a Stolen Elephant: Leaps in Creative
  Writing with Multimodal Machine Intelligence}.
\newblock \bibinfo{journal}{\emph{ACM Trans. Comput.-Hum. Interact.}}
  \bibinfo{volume}{30}, \bibinfo{number}{5}, Article \bibinfo{articleno}{68}
  (\bibinfo{date}{sep} \bibinfo{year}{2023}), \bibinfo{numpages}{57}~pages.
\newblock
\showISSN{1073-0516}
\urldef\tempurl%
\url{https://doi.org/10.1145/3511599}
\showDOI{\tempurl}


\bibitem[Skjuve et~al\mbox{.}(2023)]%
        {skjuve23user}
\bibfield{author}{\bibinfo{person}{Marita Skjuve}, \bibinfo{person}{Asbj\o{}rn
  F\o{}lstad}, {and} \bibinfo{person}{Petter~Bae Brandtzaeg}.}
  \bibinfo{year}{2023}\natexlab{}.
\newblock \showarticletitle{The User Experience of ChatGPT: Findings from a
  Questionnaire Study of Early Users}. In \bibinfo{booktitle}{\emph{Proceedings
  of the 5th International Conference on Conversational User Interfaces}}
  (Eindhoven, Netherlands) \emph{(\bibinfo{series}{CUI '23})}.
  \bibinfo{publisher}{Association for Computing Machinery},
  \bibinfo{address}{New York, NY, USA}, Article \bibinfo{articleno}{2},
  \bibinfo{numpages}{10}~pages.
\newblock
\showISBNx{9798400700149}
\urldef\tempurl%
\url{https://doi.org/10.1145/3571884.3597144}
\showDOI{\tempurl}


\bibitem[Smith et~al\mbox{.}(1975)]%
        {smith75effects}
\bibfield{author}{\bibinfo{person}{Bruce Smith}, \bibinfo{person}{Bruce Brown},
  \bibinfo{person}{William Strong}, {and} \bibinfo{person}{Alvin Rencher}.}
  \bibinfo{year}{1975}\natexlab{}.
\newblock \showarticletitle{Effects of Speech Rate on Personality Perception}.
\newblock \bibinfo{journal}{\emph{Language and speech}}  \bibinfo{volume}{18}
  (\bibinfo{date}{04} \bibinfo{year}{1975}), \bibinfo{pages}{145--52}.
\newblock
\urldef\tempurl%
\url{https://doi.org/10.1177/002383097501800203}
\showDOI{\tempurl}


\bibitem[Sommers(1980)]%
        {sommers80revision}
\bibfield{author}{\bibinfo{person}{Nancy~I. Sommers}.}
  \bibinfo{year}{1980}\natexlab{}.
\newblock \showarticletitle{Revision Strategies of Student Writers and
  Experienced Adult Writers.}
\newblock \bibinfo{journal}{\emph{College Composition and Communication}}
  \bibinfo{volume}{31} (\bibinfo{year}{1980}), \bibinfo{pages}{378--387}.
\newblock
\urldef\tempurl%
\url{https://www.jstor.org/stable/356588}
\showURL{%
\tempurl}


\bibitem[Szarkowska and Gerber-Mor{\'o}n(2018)]%
        {szarkowska18viewers}
\bibfield{author}{\bibinfo{person}{Agnieszka Szarkowska} {and}
  \bibinfo{person}{Olivia Gerber-Mor{\'o}n}.} \bibinfo{year}{2018}\natexlab{}.
\newblock \showarticletitle{Viewers can keep up with fast subtitles: Evidence
  from eye movements}.
\newblock \bibinfo{journal}{\emph{PloS one}} \bibinfo{volume}{13},
  \bibinfo{number}{6} (\bibinfo{year}{2018}), \bibinfo{pages}{e0199331}.
\newblock


\bibitem[Tian et~al\mbox{.}(2024)]%
        {tian24designing}
\bibfield{author}{\bibinfo{person}{Mingyan~Claire Tian}, \bibinfo{person}{James
  Eschrich}, {and} \bibinfo{person}{Sarah Sterman}.}
  \bibinfo{year}{2024}\natexlab{}.
\newblock \showarticletitle{Designing AI with Metaphors: Leveraging Ambiguity
  and Defamiliarization to Support Design Creativity}. In
  \bibinfo{booktitle}{\emph{Proceedings of the 16th Conference on Creativity \&
  Cognition}} (Chicago, IL, USA) \emph{(\bibinfo{series}{C\&C '24})}.
  \bibinfo{publisher}{Association for Computing Machinery},
  \bibinfo{address}{New York, NY, USA}, \bibinfo{pages}{537–541}.
\newblock
\showISBNx{9798400704857}
\urldef\tempurl%
\url{https://doi.org/10.1145/3635636.3664250}
\showDOI{\tempurl}


\bibitem[Tombaugh et~al\mbox{.}(1985)]%
        {tombaugh85effect}
\bibfield{author}{\bibinfo{person}{Jo~W. Tombaugh}, \bibinfo{person}{Michael~D.
  Arkin}, {and} \bibinfo{person}{Richard~F. Dillion}.}
  \bibinfo{year}{1985}\natexlab{}.
\newblock \showarticletitle{The effect of VDU text-presentation rate on reading
  comprehension and reading speed}. In \bibinfo{booktitle}{\emph{Proceedings of
  the SIGCHI Conference on Human Factors in Computing Systems}} (San Francisco,
  California, USA) \emph{(\bibinfo{series}{CHI '85})}.
  \bibinfo{publisher}{Association for Computing Machinery},
  \bibinfo{address}{New York, NY, USA}, \bibinfo{pages}{1–6}.
\newblock
\showISBNx{0897911490}
\urldef\tempurl%
\url{https://doi.org/10.1145/317456.317457}
\showDOI{\tempurl}


\bibitem[Turkle(1984)]%
        {turkle84second}
\bibfield{author}{\bibinfo{person}{Sherry Turkle}.}
  \bibinfo{year}{1984}\natexlab{}.
\newblock \bibinfo{booktitle}{\emph{The second self: computers and the human
  spirit}}.
\newblock \bibinfo{publisher}{Simon \& Schuster, Inc.}, \bibinfo{address}{USA}.
\newblock
\showISBNx{0671468480}


\bibitem[Vanneste and Puranam(2024)]%
        {vanneste24artificial}
\bibfield{author}{\bibinfo{person}{Bart~S. Vanneste} {and}
  \bibinfo{person}{Phanish Puranam}.} \bibinfo{year}{2024}\natexlab{}.
\newblock \showarticletitle{Artificial Intelligence, Trust, and Perceptions of
  Agency}.
\newblock \bibinfo{journal}{\emph{Academy of Management Review}}
  \bibinfo{volume}{0}, \bibinfo{number}{ja} (\bibinfo{year}{2024}),
  \bibinfo{pages}{amr.2022.0041}.
\newblock
\urldef\tempurl%
\url{https://doi.org/10.5465/amr.2022.0041}
\showDOI{\tempurl}
\showeprint{https://doi.org/10.5465/amr.2022.0041}


\bibitem[Wambsganss and Niklaus(2022)]%
        {wanbsganss22modeling}
\bibfield{author}{\bibinfo{person}{Thiemo Wambsganss} {and}
  \bibinfo{person}{Christina Niklaus}.} \bibinfo{year}{2022}\natexlab{}.
\newblock \showarticletitle{Modeling Persuasive Discourse to Adaptively Support
  Students{'} Argumentative Writing}. In \bibinfo{booktitle}{\emph{Proceedings
  of the 60th Annual Meeting of the Association for Computational Linguistics
  (Volume 1: Long Papers)}}, \bibfield{editor}{\bibinfo{person}{Smaranda
  Muresan}, \bibinfo{person}{Preslav Nakov}, {and} \bibinfo{person}{Aline
  Villavicencio}} (Eds.). \bibinfo{publisher}{Association for Computational
  Linguistics}, \bibinfo{address}{Dublin, Ireland},
  \bibinfo{pages}{8748--8760}.
\newblock
\urldef\tempurl%
\url{https://doi.org/10.18653/v1/2022.acl-long.599}
\showDOI{\tempurl}


\bibitem[Wang et~al\mbox{.}(2017)]%
        {wang17why}
\bibfield{author}{\bibinfo{person}{Dakuo Wang}, \bibinfo{person}{Haodan Tan},
  {and} \bibinfo{person}{Tun Lu}.} \bibinfo{year}{2017}\natexlab{}.
\newblock \showarticletitle{Why Users Do Not Want to Write Together When They
  Are Writing Together: Users' Rationales for Today's Collaborative Writing
  Practices}.
\newblock \bibinfo{journal}{\emph{Proc. ACM Hum.-Comput. Interact.}}
  \bibinfo{volume}{1}, \bibinfo{number}{CSCW}, Article \bibinfo{articleno}{107}
  (\bibinfo{date}{dec} \bibinfo{year}{2017}), \bibinfo{numpages}{18}~pages.
\newblock
\urldef\tempurl%
\url{https://doi.org/10.1145/3134742}
\showDOI{\tempurl}


\bibitem[Weber et~al\mbox{.}(2023)]%
        {weber23structured}
\bibfield{author}{\bibinfo{person}{Florian Weber}, \bibinfo{person}{Thiemo
  Wambsganss}, \bibinfo{person}{Seyed~Parsa Neshaei}, {and}
  \bibinfo{person}{Matthias Soellner}.} \bibinfo{year}{2023}\natexlab{}.
\newblock \showarticletitle{Structured Persuasive Writing Support in Legal
  Education: A Model and Tool for {G}erman Legal Case Solutions}. In
  \bibinfo{booktitle}{\emph{Findings of the Association for Computational
  Linguistics: ACL 2023}}, \bibfield{editor}{\bibinfo{person}{Anna Rogers},
  \bibinfo{person}{Jordan Boyd-Graber}, {and} \bibinfo{person}{Naoaki Okazaki}}
  (Eds.). \bibinfo{publisher}{Association for Computational Linguistics},
  \bibinfo{address}{Toronto, Canada}, \bibinfo{pages}{2296--2313}.
\newblock
\urldef\tempurl%
\url{https://doi.org/10.18653/v1/2023.findings-acl.145}
\showDOI{\tempurl}


\bibitem[Wobbrock et~al\mbox{.}(2011)]%
        {wobbrock11aligned}
\bibfield{author}{\bibinfo{person}{Jacob~O. Wobbrock}, \bibinfo{person}{Leah
  Findlater}, \bibinfo{person}{Darren Gergle}, {and} \bibinfo{person}{James~J.
  Higgins}.} \bibinfo{year}{2011}\natexlab{}.
\newblock \showarticletitle{The aligned rank transform for nonparametric
  factorial analyses using only anova procedures}. In
  \bibinfo{booktitle}{\emph{Proceedings of the SIGCHI Conference on Human
  Factors in Computing Systems}} (Vancouver, BC, Canada)
  \emph{(\bibinfo{series}{CHI '11})}. \bibinfo{publisher}{Association for
  Computing Machinery}, \bibinfo{address}{New York, NY, USA},
  \bibinfo{pages}{143–146}.
\newblock
\showISBNx{9781450302289}
\urldef\tempurl%
\url{https://doi.org/10.1145/1978942.1978963}
\showDOI{\tempurl}


\bibitem[Woolridge et~al\mbox{.}(2024)]%
        {woolridge24do}
\bibfield{author}{\bibinfo{person}{Lyndsay~R. Woolridge},
  \bibinfo{person}{Amy-May Leach}, \bibinfo{person}{Chelsea Blake}, {and}
  \bibinfo{person}{Elizabeth Elliott}.} \bibinfo{year}{2024}\natexlab{}.
\newblock \showarticletitle{Do Accents Speak Louder Than Words? Perceptions of
  Linguistic Speech Characteristics on Deception Detection}.
\newblock \bibinfo{journal}{\emph{Journal of Language and Social Psychology}}
  \bibinfo{volume}{43}, \bibinfo{number}{2} (\bibinfo{year}{2024}),
  \bibinfo{pages}{195--223}.
\newblock
\urldef\tempurl%
\url{https://doi.org/10.1177/0261927X231209428}
\showDOI{\tempurl}
\showeprint{https://doi.org/10.1177/0261927X231209428}


\bibitem[Wu(2018)]%
        {wu18smart}
\bibfield{author}{\bibinfo{person}{Yonghui Wu}.}
  \bibinfo{year}{2018}\natexlab{}.
\newblock \bibinfo{title}{Smart Compose: Using Neural Networks to Help Write
  Emails}.
\newblock
\newblock
\urldef\tempurl%
\url{https://research.google/blog/smart-compose-using-neural-networks-to-help-write-emails/}
\showURL{%
\tempurl}


\bibitem[Yatani(2016)]%
        {yatani16effect}
\bibfield{author}{\bibinfo{person}{Koji Yatani}.}
  \bibinfo{year}{2016}\natexlab{}.
\newblock \bibinfo{booktitle}{\emph{Effect Sizes and Power Analysis in HCI}}.
\newblock \bibinfo{publisher}{Springer International Publishing},
  \bibinfo{address}{Cham}, \bibinfo{pages}{87--110}.
\newblock
\showISBNx{978-3-319-26633-6}
\urldef\tempurl%
\url{https://doi.org/10.1007/978-3-319-26633-6_5}
\showDOI{\tempurl}


\bibitem[Yuan et~al\mbox{.}(2022)]%
        {yuan22wordcraft}
\bibfield{author}{\bibinfo{person}{Ann Yuan}, \bibinfo{person}{Andy Coenen},
  \bibinfo{person}{Emily Reif}, {and} \bibinfo{person}{Daphne Ippolito}.}
  \bibinfo{year}{2022}\natexlab{}.
\newblock \showarticletitle{Wordcraft: Story Writing With Large Language
  Models}. In \bibinfo{booktitle}{\emph{27th International Conference on
  Intelligent User Interfaces}} \emph{(\bibinfo{series}{{IUI} '22})}.
  \bibinfo{publisher}{Association for Computing Machinery},
  \bibinfo{address}{New York, NY, USA}, \bibinfo{pages}{841--852}.
\newblock
\showISBNx{978-1-4503-9144-3}
\urldef\tempurl%
\url{https://doi.org/10.1145/3490099.3511105}
\showDOI{\tempurl}


\bibitem[Yurrita et~al\mbox{.}(2023)]%
        {yurrita23disentangling}
\bibfield{author}{\bibinfo{person}{Mireia Yurrita}, \bibinfo{person}{Tim
  Draws}, \bibinfo{person}{Agathe Balayn}, \bibinfo{person}{Dave Murray-Rust},
  \bibinfo{person}{Nava Tintarev}, {and} \bibinfo{person}{Alessandro Bozzon}.}
  \bibinfo{year}{2023}\natexlab{}.
\newblock \showarticletitle{Disentangling Fairness Perceptions in Algorithmic
  Decision-Making: the Effects of Explanations, Human Oversight, and
  Contestability}. In \bibinfo{booktitle}{\emph{Proceedings of the 2023 CHI
  Conference on Human Factors in Computing Systems}} (Hamburg, Germany)
  \emph{(\bibinfo{series}{CHI '23})}. \bibinfo{publisher}{Association for
  Computing Machinery}, \bibinfo{address}{New York, NY, USA}, Article
  \bibinfo{articleno}{134}, \bibinfo{numpages}{21}~pages.
\newblock
\showISBNx{9781450394215}
\urldef\tempurl%
\url{https://doi.org/10.1145/3544548.3581161}
\showDOI{\tempurl}


\bibitem[Zhang et~al\mbox{.}(2023)]%
        {zhang23visar}
\bibfield{author}{\bibinfo{person}{Zheng Zhang}, \bibinfo{person}{Jie Gao},
  \bibinfo{person}{Ranjodh~Singh Dhaliwal}, {and} \bibinfo{person}{Toby Jia-Jun
  Li}.} \bibinfo{year}{2023}\natexlab{}.
\newblock \showarticletitle{VISAR: A Human-AI Argumentative Writing Assistant
  with Visual Programming and Rapid Draft Prototyping}. In
  \bibinfo{booktitle}{\emph{Proceedings of the 36th Annual ACM Symposium on
  User Interface Software and Technology}} (San Francisco, CA, USA)
  \emph{(\bibinfo{series}{UIST '23})}. \bibinfo{publisher}{Association for
  Computing Machinery}, \bibinfo{address}{New York, NY, USA}, Article
  \bibinfo{articleno}{5}, \bibinfo{numpages}{30}~pages.
\newblock
\showISBNx{9798400701320}
\urldef\tempurl%
\url{https://doi.org/10.1145/3586183.3606800}
\showDOI{\tempurl}


\bibitem[Zhao et~al\mbox{.}(2017)]%
        {zhao17men}
\bibfield{author}{\bibinfo{person}{Jieyu Zhao}, \bibinfo{person}{Tianlu Wang},
  \bibinfo{person}{Mark Yatskar}, \bibinfo{person}{Vicente Ordonez}, {and}
  \bibinfo{person}{Kai-Wei Chang}.} \bibinfo{year}{2017}\natexlab{}.
\newblock \showarticletitle{Men Also Like Shopping: Reducing Gender Bias
  Amplification using Corpus-level Constraints}. In
  \bibinfo{booktitle}{\emph{Proceedings of the 2017 Conference on Empirical
  Methods in Natural Language Processing}},
  \bibfield{editor}{\bibinfo{person}{Martha Palmer}, \bibinfo{person}{Rebecca
  Hwa}, {and} \bibinfo{person}{Sebastian Riedel}} (Eds.).
  \bibinfo{publisher}{Association for Computational Linguistics},
  \bibinfo{address}{Copenhagen, Denmark}, \bibinfo{pages}{2979--2989}.
\newblock
\urldef\tempurl%
\url{https://doi.org/10.18653/v1/D17-1323}
\showDOI{\tempurl}


\bibitem[Zhou and Sterman(2023)]%
        {zhou23creative}
\bibfield{author}{\bibinfo{person}{David Zhou} {and} \bibinfo{person}{Sarah
  Sterman}.} \bibinfo{year}{2023}\natexlab{}.
\newblock \showarticletitle{Creative Struggle: Arguing for the Value of
  Difficulty in Supporting Ownership and Self-Expression in Creative Writing}.
  In \bibinfo{booktitle}{\emph{Proceedings of the Second Workshop on
  Intelligent and Interactive Writing Assistants (In2Writing 2023)}}.
\newblock
\urldef\tempurl%
\url{https://cdn.glitch.global/d058c114-3406-43be-8a3c-d3afff35eda2/paper11_2023.pdf}
\showURL{%
\tempurl}


\bibitem[Zhou and Sterman(2024)]%
        {zhou24aillude}
\bibfield{author}{\bibinfo{person}{David Zhou} {and} \bibinfo{person}{Sarah
  Sterman}.} \bibinfo{year}{2024}\natexlab{}.
\newblock \showarticletitle{Ai.llude: Investigating Rewriting AI-Generated Text
  to Support Creative Expression}. In \bibinfo{booktitle}{\emph{Proceedings of
  the 16th Conference on Creativity \& Cognition}} (Chicago, IL, USA)
  \emph{(\bibinfo{series}{C\&C '24})}. \bibinfo{publisher}{Association for
  Computing Machinery}, \bibinfo{address}{New York, NY, USA},
  \bibinfo{pages}{241–254}.
\newblock
\showISBNx{9798400704857}
\urldef\tempurl%
\url{https://doi.org/10.1145/3635636.3656187}
\showDOI{\tempurl}







\end{thebibliography}

\appendix


\section{Exploratory Analysis}
\label{sec:exploratory}

We performed an exploratory analysis to better understand our findings by noting any significant but unforeseen results. 
Note that these results are not confirmatory, as hypotheses were not declared in the main study plan before beginning the user study.

We collected participants' history of using AI tools, attitude towards AI, writing experience, and demographic data.
\begin{enumerate}
    \item \textit{Past Experience with AI:} Personal experience with AI can shape future interactions with writing tools. These items were adapted from the two measures of prior experience in Draxler et al. \cite{draxler24ai}.
    \item \textit{Attitude Towards AI:} Overall perceptions of AI can influence how users perceive AI tools. These items include the perceived usefulness of AI  \cite{hong21selfdriving}, self-confidence in explaining AI, \cite{hong21selfdriving}, and assessment of AI creativity \cite{hong21are}; all questions were adapted from Draxler et al. \cite{draxler24ai}.
    \item \textit{Past Writing Experience:} We collected data on professional writing frequency (five-point Likert scale), personal writing frequency (five-point Likert scale), and whether the participant had been paid before to write (boolean).
    \item \textit{Demographics:} We collected age (continuous) and level of education (categorical) to search for correlations. 
\end{enumerate}

We conducted exploratory analyses using ANOVAs to note unforeseen trends and applied the same Bonferroni correction to the significance threshold. 

\subsection{Past Writing Experience with AI}
As \underline{reading comfort} ($\mathbf{H_1}$) can be influenced by past AI writing tool usage, we conducted further investigation on three external variables. We found \textit{frequency of AI writing tool usage} to have a statistically significant but small effect ($F_{4,2965}=5.265, p<0.001, \eta^2_p<0.01$) on \underline{reading comfort}, as well as whether or not a participant had \textit{previously been paid to write} ($F_{1,2968}=12.466, p<0.001, \eta^2_p<0.01$). We did not find evidence of \textit{specific AI tool usage}, such as ChatGPT or Sudowrite, having significant effect. We also did not find the \textit{frequency of professional} or \textit{frequency of personal writing} affecting \underline{reading comfort}.

Post-hoc analysis for \textit{frequency of AI writing tool usage} reveals that participants that \textit{use AI writing tools everyday} find text generation more \underline{comfortable} than participants that \textit{use AI writing tools less often} ($p\leq0.001)$. This finding is in line with our expectations, as users who use AI writing tools frequently, as part of their daily workflows, likely feel more accustomed to reading AI generated text.

Additionally, we found past writing experience with AI to increase \underline{perceptions of trust} ($\mathbf{H_5}$) ($F_{4,2965}=8.059, p<0.001, \eta^2_p<0.01$). Specifically, we found participants who \textit{use AI writing tools daily} or \textit{weekly} to perceive greater \underline{trust} than participants who \textit{have never used AI writing tools} ($p<0.001$), have used them \textit{a few times} ($p<0.001$), or \textit{monthly} ($p<0.005$). This aligns with intuition, as users who use AI writing tools more often would likely find them to be trustworthy and reliable.

\subsection{Perceived Usefulness of AI}
Secondly, attitude towards the usefulness of AI can play a role in perceptions of AI tools. Specifically, we saw \textit{perceived helpfulness AI} have a significant, but small effect on \underline{comfort} ($\mathbf{H_1}$) ($F_{4,2965}=14.617, p<0.001, \eta^2_p=0.01$) and \underline{perception of humanness} ($\mathbf{H_3}$) ($F_{4,2965}=6.164, p<0.001, \eta^2_p<0.01$). We found that participants who strongly agree with the statement \textit{AI is helpful} ($p\leq0.003$) judged text appearance overall to be more \underline{comfortable to read} and the AI tool more \underline{human-like}.

We did not find evidence that perceptions of AI being a \textit{positive force in the world} have a significant effect on the aforementioned dependent variables. We also did not see any significant effect of \textit{perceived need to use AI} have a significant effect on \underline{comfort} and \underline{humanness}. We suggest that future work may look into connections between various metaphorical interpretations of AI with perceived comfort and tool preferences, particularly agentic and collaboration metaphors.

\subsection{Assessment of AI Creativity}
Finally, we assessed the potential impact of attitudes towards AI creativity on the perception of AI tools. Here, we saw the belief that \textit{AI can be creative on its own} having a significant but small effect on \underline{reading comfort} ($\mathbf{H_1}$) ($F_{4,2965}=5.746, p<0.001, \eta^2_p<0.01$), on \underline{judgement of humanness} ($\mathbf{H_3}$) $(F_{4,2965}=5.263, p<0.001, \eta^2_p<0.01)$, as well as \underline{trustworthiness} ($F_{4,2965}=4.037, p=0.003, \eta^2_p<0.01$).

After conducting post-hoc analysis on \underline{comfort}, we found all pair-wise comparisons ($p<0.001$) except one (neutral and disagree, $p=0.999$) to have a significant positive effect on \underline{comfort}. In other words, a more positive outlook of AI creativity almost always led to a greater feeling of \underline{comfort}. We found similar evidence of the same relation for \underline{perceived quality of AI text} when conducting all pair-wise comparisons ($p\leq0.010$) except one (neutral and disagree, $p=0.901$); for \underline{perception of humanness} when conducting all pair-wise comparisons ($p<0.001$) except two (neutral and disagree, $p=0.343$; agree and neutral, $p=0.068$); and for \underline{perception of trustworthiness} when conducting all pair-wise comparisons ($p<0.001$) except one (neutral and disagree, $p=0.998$). Our results suggest that strong attitudes towards \textit{AI creativity} effect perceived output quality and regard for the AI tool.

Future work might look into perceptions of AI creativity within artistic and AI-centered communities, as observing these communities can inform future community-designed AI tools.




\end{document}